\documentclass[10pt,compsoc]{IEEEtran}

\makeatletter \def\mdseries@tt{m} \makeatother 

\usepackage{wrapfig}
\usepackage{graphicx}
\usepackage{indentfirst}
\usepackage{booktabs} 
\usepackage{graphicx}
\graphicspath{./figures}
\usepackage{colortbl}
\usepackage{float}
\usepackage{amsmath}
\newcommand\norm[1]{\left\lVert#1\right\rVert}
\usepackage{rotating}

\newcommand{\added}[1]{\textcolor{black}{{#1}}}

\newcommand{\Revise}[1]{\textcolor{black}{{#1}}}

\usepackage[backgroundcolor=white,bordercolor=black,linecolor=black]{todonotes}

\usepackage[singlespacing]{setspace}
\usepackage{threeparttable}
\usepackage{multirow}
\usepackage{subcaption}
\usepackage[utf8]{inputenc}
\usepackage[english]{babel}
\usepackage{amsthm}
\usepackage{changepage} 
\usepackage{mathtools}
\usepackage{textgreek}
\usepackage{tcolorbox}
\usepackage[compact]{titlesec}     
\usepackage{amsmath}
\usepackage{tikz}

\usepackage{float}
\usepackage{wrapfig}
\usetikzlibrary{shapes,positioning,arrows,calc}

\DeclareFontFamily{OT1}{pzc}{}
\DeclareFontShape{OT1}{pzc}{m}{it}{<-> s * [1.1] pzcmi7t}{}
\DeclareMathAlphabet{\mathpzc}{OT1}{pzc}{m}{it}
\theoremstyle{definition}
\DeclareMathAlphabet\mathbfcal{OMS}{cmsy}{b}{n}
\usepackage{centernot}
\usepackage{enumitem}
\makeatletter
\def\algbackskip{\hskip-\ALG@thistlm}

\newcommand{\ygg@basicalert}[2]{\fbox{\bfseries\sffamily\scriptsize#1}{\sf\small$\blacktriangleright$\textit{#2}$\blacktriangleleft$}}
\newcommand{\manel}[1]{\ygg@basicalert{Manel}{#1}}

\usepackage{xcolor}
\usepackage[ruled, vlined, linesnumbered]{algorithm2e}
\newcommand{\nonl}{\renewcommand{\nl}{\let\nl\oldnl}}

\SetKwInput{KwInput}{Input}                
\SetKwInput{KwOutput}{Output} 

\SetAlFnt{\small\sffamily}

\SetCommentSty{mycommfont}

\usepackage{amsmath, listings, amsthm, amssymb, xspace}

\usepackage{hyperref}
\hypersetup{colorlinks,allcolors=black}

\theoremstyle{remark}

\author{Zohreh Aghababaeyan, Manel Abdellatif, Lionel Briand, Ramesh S, and Mojtaba Bagherzadeh

\IEEEcompsocitemizethanks{\IEEEcompsocthanksitem Zohreh Aghababaeyan is with the School of EECS, University of Ottawa, Ottawa, Canada. \protect\\
E-mail: zagha052@uottawa.ca
\IEEEcompsocthanksitem Manel Abdellatif is with the Software and Information Technology Engineering Department, École de Technologie Supérieure, Montreal, Canada. She contributed to this work mainly during her postdoctoral fellowship at the School of EECS, University of Ottawa, Ottawa, Canada.\protect\\
E-mail: Manel.abdellatif@etsmtl.ca
\IEEEcompsocthanksitem Lionel Briand is with the School of EECS, University of Ottawa, Ottawa, Canada, and also with the SnT Centre for Security, Reliability and Trust, University of Luxembourg, Luxembourg.\protect\\
E-mail: Lbriand@uottawa.ca
\IEEEcompsocthanksitem Mojtaba Bagherzadeh is with the School of EECS, University of Ottawa, Ottawa, Canada.\protect\\
E-mail: Mbagherz@cisco.com
\IEEEcompsocthanksitem Ramesh S is with the Department of Research and Development, General Motors, Warren, MI, USA.\protect\\
E-mail: Ramesh.s@gm.com
}}

\begin{document}

\title{Black-Box Testing of Deep Neural Networks through Test Case Diversity}

\IEEEtitleabstractindextext{%

\begin{abstract}
    
Deep Neural Networks (DNNs) have been extensively used in many areas including image processing, medical diagnostics and autonomous driving. However, DNNs can exhibit erroneous behaviours that may lead to critical errors, especially when used in safety-critical systems. Inspired by testing techniques for traditional software systems, researchers have proposed neuron coverage criteria, as an analogy to source code coverage, to guide the testing of DNNs. Despite very active research on DNN coverage, several recent studies have questioned the usefulness of such criteria in guiding DNN testing. Further, from a practical standpoint, these criteria are white-box as they require access to the internals or training data of DNNs, which is often not feasible or convenient. \Revise{Measuring such coverage requires executing DNNs with candidate inputs to guide testing, which is not an option in many practical contexts.} 

\Revise{In this paper, we investigate diversity metrics as an alternative to white-box coverage criteria. For the previously mentioned reasons, we require such metrics to be black-box and not rely on the execution and outputs of DNNs under test.} To this end, we first select and adapt three diversity metrics and study, in a controlled manner, their capacity to measure actual diversity in input sets. We then analyze their statistical association with fault detection using \added{four} datasets and \added{five} DNNs. We further compare diversity with state-of-the-art white-box coverage criteria. \Revise{As a mechanism to enable such analysis, we also propose a novel way to estimate fault detection in DNNs.} 

Our experiments show that relying on the diversity of image features embedded in test input sets is a more reliable indicator than coverage criteria to effectively guide DNN testing. Indeed, we found that one of our selected black-box diversity metrics far outperforms existing coverage criteria in terms of fault-revealing capability and computational time.
Results also confirm the suspicions that state-of-the-art coverage criteria are not adequate to guide the construction of test input sets to detect as many faults as possible using natural inputs.   
\end{abstract}    
}

\maketitle 

\section{Introduction} 
\label{Sec:Introduction}

Over the last decade, Deep Neural Networks (DNNs) have achieved successful performance in many domains, such as image processing~\cite{yang2019survey,giusti2013fast}, medical diagnostics~\cite{mallick2019brain,rajinikanth2020customized,debelee2020deep}, speech recognition~\cite{pan2012investigation} and autonomous driving~\cite{sallab2017deep,stocco2020misbehaviour}. 
Similar to traditional software components, DNN models often exhibit erroneous behaviours that may lead to potentially critical errors. Therefore, like traditional software, DNNs need to be tested effectively to ensure their reliability and safety.

In the software testing context, code coverage criteria (e.g. branch coverage, statement coverage) are used to guide the generation of test cases and assess the completeness of test suites~\cite{cai2005effect}. While full coverage does not ensure functional correctness, high coverage increases stakeholders’ confidence in the testing results because it triggers more code execution paths. Inspired by code coverage, several coverage criteria have been introduced to measure the adequacy of test data in the context of DNNs~\cite{kim2019guiding,gerasimou2020importance,pei2017deepxplore,Ma2018DeepGaugeMT,sun2019structural}. Neuron coverage measures the extent to which neurons in a DNN are activated based on certain input data. Intuitively, test inputs with  higher neuron coverage are desirable. However, reaching high neuron coverage with a few test inputs is usually easy to achieve~\cite{sun2019structural,sekhon2019towards} and the usefulness of such coverage is therefore questionable.  
Furthermore, defining coverage in DNNs is not as straightforward as testing traditional software because in the latter the code logic is explicit but in DNNs that logic is not represented explicitly.
Although more sophisticated coverage criteria have been proposed, several articles have criticized the use of such coverage to guide the testing of DNN models~\cite{li2019structural,dong2019there,chen2020deep}.

In traditional software systems, testers rely on coverage metrics because they assume that (1) inputs covering the same part of the source code are homogeneous (i.e. either all or none of these inputs trigger a failure), and (2) the inputs used in testing should be diverse to ensure high coverage~\cite{li2019structural}. 
However, these assumptions break down in DNN testing because (1) unlike code coverage, neuron coverage does not fully exercise the implicit logic embedded in DNNs; (2) the homogeneity assumption is broken with adversarial inputs; and (3) increasing the diversity of inputs does not necessarily increase DNN coverage~\cite{li2019structural}. Further, most coverage studies rely on adversarial inputs to validate their proposed criteria~\cite{pei2017deepxplore,Ma2018DeepGaugeMT,sun2019structural,kim2019guiding,gerasimou2020importance}. However, these inputs are mostly unrealistic and used to study the robustness of the DNN model instead of its accuracy.
While state-of-the-art coverage criteria have been largely validated with artificial inputs generated based on adversarial methods, their claimed sensitivity to adversarial inputs does not necessarily mean that they relate to the fault detection capability of natural test input sets. This is confirmed by various studies~\cite{li2019structural,chen2020deep} that have failed to find a significant correlation between coverage and the number of misclassified inputs in a natural test input set, despite a positive correlation in the presence of adversarial test inputs. Consequently, coverage criteria may be ineffective in guiding DNN testing to increase the fault-detection capability of natural test input sets. Further, another study~\cite{dong2019there} found that retraining DNN models with new input sets that improve coverage does not increase the robustness of the model to adversarial attacks.

Furthermore, coverage criteria require full access to the internals of the DNN state or training data, both of which are often not available to testers, especially when the DNN model is proprietary and provided by a third party. Thus, in our project, we focus on black-box input diversity metrics to provide guidance on how to assess test suites or select test cases for DNNs. We target diversity because it has been successfully used in testing software systems~\cite{biagiola2019diversity,hemmati2015prioritizing,feldt2016test}. Intuitively, relying on diverse test inputs should increase the exploration of the fault space and thus increase the fault detection capability of a given test input set. 
\Revise{Further, we target black-box metrics that do not require executing test inputs on the DNN under test since this is a strong practical impediment in many application contexts, such as when dealing with large models and large databases of unlabeled inputs. We also target black-box metrics that are model-independent and do not rely on the  outputs of DNNs under test because they cannot be trusted when the models are not accurate~\cite{li2019boosting}.
Based on these requirements, we propose and investigate black-box diversity metrics for DNNs that rely on inputs' features, investigate their relationships with coverage metrics and analyze their association with fault detection. In other words, this paper focuses on the fundamental assumptions related to the relationship between testing criteria (i.e. coverage and diversity metrics) and faults in DNNs.} 
However, this paper does not investigate how these testing criteria might be used for specific testing scenarios such as the selection, minimization or generation of test sets. Nonetheless, investigating the relationship between DNN faults and testing criteria is an essential step for selecting proper criteria, independent of any specific purpose.

In traditional software systems, some of the inputs causing failures are usually very close to each other~\cite{chen2010adaptive,chen2004adaptive}. Similarly, it has been observed that many mispredicted inputs in DNNs fail due to the same causes~\cite{fahmy2021supporting}. 
Counting such inputs to assess the fault detection capability of a test suite is therefore misleading. However, the notion of fault, though rather straightforward in regular software, is elusive in DNNs. For this reason, we rely on a clustering-based fault estimation approach to group similar mispredicted inputs based on their features and misprediction behaviour~\cite{fahmy2021supporting}. We assume that each cluster corresponds to a fault because similar mispredicted inputs belonging to the same cluster are assumed to be mispredicted for similar reasons. 
To assess test suites for DNNs, we use and adapt three diversity metrics. As we evaluate datasets composed of images, commonly used as inputs in many DNNs (e.g. the perception layer of cyber-physical systems), we rely on a feature extraction model to extract features from images that will be  used to compute the diversity of test input sets. We evaluate the selected metrics in terms of their capability to measure actual diversity based on extracted features. \added{We then analyze their associations with fault detection in DNNs using four widely used datasets and five different DNN models.} We further study state-of-the-art white-box coverage metrics and their associations with diversity and fault detection.

Based on our experiments, we show that diversity metrics, and geometric diversity (GD)~\cite{kulesza2012determinantal} in particular, though black-box and without the use of any DNN internal information, far outperform existing coverage criteria in terms of fault-revealing capability and computational time. We also show that state-of-the-art coverage metrics are not correlated to faults or diversity in natural test input sets.

Overall, the main contributions of our paper are as follows:

\begin{itemize}
    \item We propose and study the use of black-box diversity metrics to guide the testing of DNN models. We show that geometric diversity is the best option to guide the testing of DNN models because it is positively correlated to faults in subsets.   
    
    \item We introduce and validate a clustering-based approach to estimate faults in DNNs as test input sets typically contain many similar mispredicted inputs caused by the same problems in the DNN model. We explain why this is a requirement to evaluate any test set evaluation criterion. 
    \item We study state-of-the-art coverage criteria and show that there is no correlation between coverage and faults in DNN models. Further, coverage is not correlated with diversity in input sets. Our results  question the reliability of coverage, as it is currently defined, to guide DNN testing if the objective is to detect as many faults as possible.

\end{itemize}

The remainder of the paper is structured as follows. Section~\ref{Sec:Approach} presents our approach and describes the selected diversity metrics. Section~\ref{Sec:Evaluation} presents our empirical evaluation and results. Section~\ref{Sec:Discussions} discusses the implications of our results and our recommendations for guiding the testing of DNN models. Section~\ref{sec:Threats} describes the threats to the validity of our study. Sections~\ref{Sec:RW} and \ref{Sec:Conclusion} contrast our work with related work and conclude the paper, respectively.

\section{Approach}
\label{Sec:Approach}

A central problem in software testing, especially when test oracles (verdicts) are not automated, is the selection of a small set of test cases that sufficiently exercise a software system. Intuitively, testers should select a set of diverse test cases because selecting similar test cases does not bring extra benefits to fault detection. \Revise{In this paper, we study diversity metrics with the ultimate aim of using them to guide DNN testing, relying on the best diversity metric in both the capacity to uncover erroneous behavior and computational complexity. We target black-box diversity metrics that are model-independent because we cannot rely on DNNs output when the models are not accurate~\cite{li2019structural}.
We also target black-box metrics that do not require executing the model with all inputs because it would impede their application when working with large models and datasets.  
Therefore, we use and adapt three diversity metrics that have been applied widely in other contexts and are based on inputs.} We rely on a feature extraction model to extract features from images that we use to compute diversity. In section~\ref{Sec:Evaluation}, we will first evaluate the selected metrics in terms of their capability to measure the actual diversity of a test input set. Then we will study their relationships with state-of-the-art white-box coverage metrics and analyze their associations with fault detection in DNNs. 

In this section, we describe the feature extraction method and the diversity metrics that we used, and detail the evaluation process in the following section.  

\subsection{Feature Extraction} \label{SubSec:FeatureExtraction}

For diversity to account for the content of images, we need to extract features from each input image in the test input set. Consequently, we rely on VGG-16~\cite{simonyan2014very}, which is one of the most-used and accurate state-of-the-art feature extraction models~\cite{mousser2019deep,kaur2019automated}. It is a pre-trained convolutional neural network model and consists of 16 weight layers, including 13 convolutional layers with a filter size of 3×3, and three fully connected layers. The model is trained on ImageNet~\footnote{https://image-net.org/index.php}, which is a dataset of over 14 million labeled images belonging to 22,000 categories. 

We use VGG-16 to extract the features of images. A feature is an activation value  on the layer after the last convolutional layer of the VGG-16 model. A set of features can characterize semantic elements such as shapes and colors.  We extract the features in the test input set $S$ and build the related feature matrix \Revise{\textit{Vs}} where (1) each row of the matrix corresponds to the feature vector of an input in the test set, and (2) each column corresponds to a feature.

After generating the feature matrix, we normalize it by applying \textit{Min-Max normalization} per feature, which is one of the most common and simple ways to normalize data. For each feature in \Revise{\textit{Vs}}, the maximum and minimum values of that feature are transformed to one and zero, respectively, and every other value is transformed to a real value between zero and one.
The \textit{Min-Max normalization}  is defined as follows.
For every feature  in the feature matrix \Revise{\textit{Vs}} where $j \in [1,2,...,m]$ and $m$ is the number of features, the normalized feature \Revise{\textit{Vs$_j'$}} is calculated as follows: 

\begin{equation}\label{Eq:MinMaxNormalization}
 Vs_j'(i) = \frac{Vs_j(i) - min(Vs_j)}{max(Vs_j)-min(Vs_j)}
\end{equation}

We normalize the feature matrix (1) to make the computation of the selected diversity metrics more scalable, and (2) to eliminate the dominance effect of features with large value ranges. 

\subsection{Diversity metrics}

In this section, we describe the selected diversity metrics: Geometric Diversity~\cite{kulesza2012determinantal,gong2019diversity}, Normalized Compression Distance~\cite{cohen2014normalized,feldt2016test}, and Standard Deviation.

\Revise{We chose these metrics based on the following criteria. First, we targeted diversity metrics that measure diversity within a subset. We did not consider metrics that measure diversity in relation to another subset (e.g. Kullback-Leibler~\cite{hershey2007approximating}, Jensen-Shannon divergence~\cite{chen2021comparison}). Second, we selected diversity metrics that can be applied to our datasets, specifically targeting metrics that can be applied to images. Third, we selected diversity metrics that do not depend on the DNN model under test and do not require the execution of this model with all inputs. Finally, we  targeted diversity metrics that are widely used in a variety of other application contexts. For instance, the geometric diversity metric has been used in a variety of machine learning applications such as the selection of training sets with the Determinantal Point Process method ~\cite{kulesza2012determinantal},  data summarization~\cite{lin2012learning} and data clustering~\cite{kang2013fast,xu2016scalable}. Furthermore, the standard deviation metric is considered to be a common diversity metric that has been successfully applied in different contexts to measure text and image similarity~\cite{nafchi2016mean}.
The Normalized Compression Distance metric has been  employed in many application domains such as image processing~\cite{cohen2014normalized}, security~\cite{borbely2016normalized} and clustering~\cite{cilibrasi2005clustering,cohen2014normalized}. This metric supports any type of input and has been used recently  to guide the selection of diverse input tests for regular software systems~\cite{feldt2016test}.}

In this section, we will describe each of these metrics and discuss their strengths and limitations. 

\subsubsection{Geometric Diversity}

The geometric diversity metric measures the diversity of the selected inputs~\cite{kulesza2012determinantal}. As mentioned previously, this metric is widely used to select diverse input sets with the Determinantal Point Process (DPP) method~\cite{kulesza2012determinantal,gong2019diversity}. DPP is applied to guide the selection of diverse subsets from a fixed ground set~\cite{elfeki2019gdpp} and has been used in a variety of machine learning applications for images~\cite{kulesza2012determinantal}, videos~\cite{gong2014diverse}, documents~\cite{lin2012learning}, recommendation systems~\cite{zhou2010solving} and sensor placement~\cite{krause2008near}. The key characteristic of DPP is that including one item makes including other similar items less likely (i.e. a DPP assigns a greater probability to subsets of items that are diverse). Thus, a DPP value of a subset indicates its diversity, where the higher this value, the more diverse the subset. The key component in DPP is geometric diversity that measures the diversity of an input set in terms of the (hyper)volume spanned by the input feature vectors (feature matrix).

\paragraph{\textbf{Definition}}

\begin{figure}[ht]
\includegraphics[width = \columnwidth]{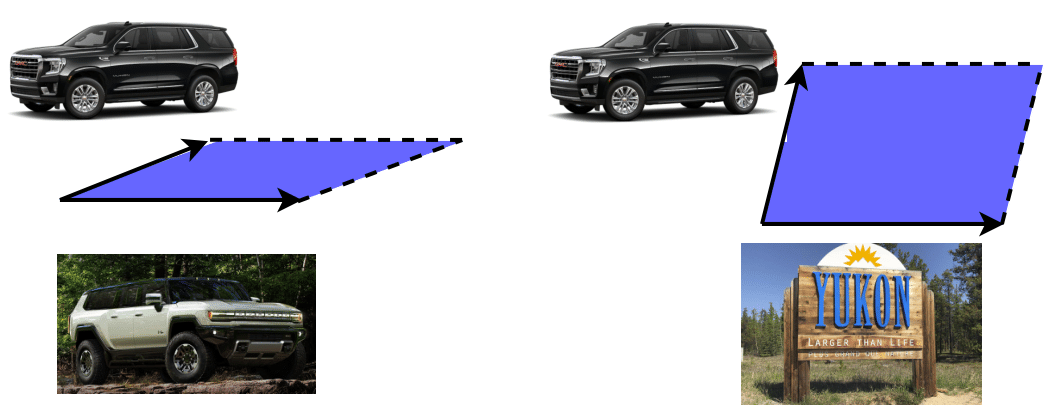}

\centering
\caption{Illustration of the geometric diversity metric}
\label{fig:ExampleGeometricDiversity}
\end{figure}

The geometric diversity \textit{G(.)} is defined as follows. Given a dataset $X$, a number of inputs $n$, a number of features $m$,  and  feature vectors $V \in R^{n*m}$,  the geometric diversity of a subset $S {\subseteq} X $ is defined as: 

\begin{equation}\label{Eq:GeometricDiversity}
 G(S) = det(Vs * Vs^{T} )  
\end{equation}

which corresponds to the squared volume of the parallelepiped spanned by the rows of \textit{Vs}, since they correspond to vectors in the feature space. The larger the volume, the more diverse $S$ is in the feature space, as illustrated in Figure~\ref{fig:ExampleGeometricDiversity}. It is expected that very different (similar) images result in very different (similar) feature vectors.

\paragraph{\textbf{Calculation}}

Geometric diversity takes as input the feature matrix of the test input set, as generated using the feature extraction model. Because geometric diversity relies on the calculation of the determinant of a matrix, we need to handle several challenges related to such processing.

\textbf{Determinant Overflow.} The determinant is likely to run into overflow when we work with large feature matrices. The main cause of the problem is that the determinant value is too large to be represented by a real number. To overcome this problem, we follow the recommendations of Celis \textit{et al}.~\cite{celis2018fair} and use the logarithm of this value rather than the determinant itself. We also overcome the determinant overflow problem by using the normalized feature matrix, \Revise{\textit{Vs'}} thus making the geometric diversity computation more scalable.

\textbf{Mathematical Limitations.} If a matrix contains at least two linearly dependent vectors, its determinant will be equal to zero. Consequently, we cannot calculate the geometric diversity score of an input set that contains duplicate inputs. The feature extraction model predicts features for each test input. If the feature values are the same for two test inputs, we have duplicate inputs. This means the two test images are redundant in terms of this feature extraction model. We therefore have to delete redundant inputs before calculating the diversity score. This kind of pre-processing is acceptable in our context because (1) duplicate inputs that do not add any value to our testing model, and (2) we aim to test the DNN model with a diverse input set to detect faults.

Further, the maximum subset size for which we calculate GD must be less than the number of the features in \Revise{\textit{Vs}}.
This is also due to the mathematical limitations of the determinant and the rank of matrices.

\textit{\textbf{Proof:} } In linear algebra, the rank of a matrix $A$  of size $n*m$ refers to the number of linearly independent rows or columns in the matrix. Consequently, $Rank(A_{n*m}) <= min(n,m)$, where $n$ is the number of lines in the matrix $A$, and $m$ is the number of columns.
Consider a square matrix $B$ of size $n*n$. By definition, If $Rank(B)<n$ then $Det(B)=0$.
Let us assume that $B= A* A^T$.
By definition $Rank(B)= Rank(A * A^T)= Rank(A)$.
If $n > m$ then  $Rank(B)\leq m <n$. As a result $Det(B)=Det(A* A^T) = 0$.

\Revise{To mitigate this mathematical limitation, we can select one of the internal layers of the feature extraction model where the number of linearly independent features is equal to or greater than the size of the subset. We propose to use the deepest hidden layer, which provides enough features because, as noted by Bengio \textit{et al.}~\cite{bengio2013better,kim2019guiding}, deeper layers represent higher-level features of the input. 
Specifically, we can select hidden layers that are possible candidates for feature extraction because their number of linearly independent features is greater than or equal to the size of the subset. 
From these candidates, we then select the deepest hidden layer because it is likely to contain the most semantically significant and helpful features for characterizing an input.}

\subsubsection{Normalized Compression Distance }

The Normalized Compression Distance (NCD) is a similarity metric based on the Kolmogorov complexity~\cite{kolmogorov1965three} and information distance~\cite{bennett1998information} where we measure the information required to transform one object into another to assess the similarity between these objects. Because of the complexity in calculating the Kolmogorov complexity, we approximate it by using real-world compressors~\cite{cohen2014normalized,li2004similarity}. This leads to the normalized compression distance~\cite{cilibrasi2005clustering}, which has been extended by Cohen \textit{et al}.~\cite{cohen2014normalized} to support the calculation of multisets' similarity. 

\paragraph{\textbf{Definition}}

The NCD metric for a multiset $S$ is calculated via an intermediate measure $NCD_1$~\cite{cohen2014normalized,feldt2008searching, feldt2016test}: 

\begin{equation}\label{Eq:NCD1}
\color{black} NCD_1(S) = \frac{C(S)- min_{s \in S}\{ C(s) \}}{max_{s \in S}\{ C(S \backslash \{ s \}) \} } 
\end{equation}
\begin{equation}\label{Eq:NCD}
NCD(S) = max\Big\{ NCD_1(S), max_{Y \subset S}\{ NCD(Y)\} \Big\}
\end{equation}
\Revise{where $C(S)$ denotes the length of S after compression~\cite{cohen2014normalized,feldt2016test}.This metric is interpreted by Cohen \textit{et al.}~\cite{cohen2014normalized} as follows. For example, if a multiset S of strings (inputs) of about 1,000,000 bits each have pairwise information distances of 1,000 bits between each pair of inputs, then those strings can be considered relatively similar. If, on the other hand, a multiset S contains strings of about 1,200 bits each, and each pair of strings in S has a pairwise information distance of 1,000 bits, then we can conclude that the inputs in S are quite diverse~\cite{cohen2014normalized}.} NCD supports any type of input (e.g text, images, execution traces) and has many applications, such as in pattern recognition~\cite{coltuc2018use,kocsor2006application}, clustering~\cite{cohen2014normalized,cilibrasi2005clustering}, security~\cite{borbely2016normalized} and measuring the diversity of test sets~\cite{feldt2016test,henard2016comparing}.

\paragraph{\textbf{Calculation}}
We have re-implemented the NCD metric for multisets based on the original paper~\cite{feldt2016test}. NCD takes the normalized feature matrix of an input set and measures its diversity score. It takes values in the range $[0,1]$. The more diverse the input set, the larger the NCD score. However, one limitation of this metric is its high computational cost, such that its application on large input sets becomes prohibitive~\cite{cohen2014normalized,feldt2016test}. NCD is highly sensitive to the used compression tool~\cite{feldt2016test}. 
Different compression tools determine various performance aspects of NCD, such as computation time, used memory and compression distance. Following the recommendations of existing papers on NCD ~\cite{feldt2016test,cohen2014normalized,borbely2016normalized}, we tried different compression tools like \textit{Lzm}, \textit{Bzip2} and \textit{Zlip}. We tested their efficiency in computational cost and correctness in generating  diversity scores. We evaluated the correctness of the diversity scores by controlling the actual diversity of input sets in terms of features and compared the corresponding NCD scores. We compared the NCD score of input sets with similar images to other sets with different images. The NCD score was expected to increase when the input set was more diverse in terms of features.  
The best results were obtained with \textit{Bzip2}, which we used in our experiments.

\subsubsection{Standard Deviation}

Standard deviation (STD) is a statistical measure of how far from the mean a group of data points is, determined by calculating the square root of the variance. 

\paragraph{\textbf{Definition}}

STD is a straightforward measure of the diversity of a test input set based on the statistical variation of the inputs' features. We define the STD metric as the norm of the standard deviation of each feature in the test input set. Formally, we define the STD of an input set $S$ of size $n$ as follows:

\begin{equation}\label{Eq:STD} \color{black}
STD(S) = \norm {  \biggl(\sqrt {\sum_{i=1}^n \frac{Vs_{i,j} - \mu_j}{n} },  1 \leq j \leq m\biggr)} 
\end{equation}

where \textit{Vs} is the feature matrix of the input set $S$, $m$ is the number of features, and $\mu_j$ is the mean value of feature $j$ in \textit{Vs}. 

\paragraph{\textbf{Calculation}}

To calculate STD for an input set $S$, we first extract the feature matrix for $S$ and normalize it. Then we calculate the norm of the standard deviations of each feature in the matrix to measure the diversity of the input set. The higher the STD, the more diverse the input set. One of the limitations of the standard deviation is its dependence on the mean, which introduces unwanted bias in some cases. To explain this, we use two same-size subsets, $A$ and $B$, where (1) in subset $A$ we have two sets of similar inputs and these two sets are far from each other in the features space, and (2) in subset $B$ all inputs are different from one another. The variance of the inputs in subset $A$ with respect to the mean could be larger than the one in subset $B$. In such a case, $STD(A)$ would be larger than $STD(B)$ though subset $B$ is more diverse than $A$, as the latter only contains two truly distinct groups of inputs.

\section{Empirical Evaluation}
\label{Sec:Evaluation}

This section describes the empirical evaluation of our approach, including research questions, datasets, DNN models, experiments and results. 

\begin{table*}[ht]
    \centering
    \begin{tabular}{lp{8cm}p{5cm}l}
    \hline \hline
    \\Dataset  & Description & DNN Model & Accuracy  \\ \\\hline \hline

    \\MNIST & \raggedright Handwritten digit images composed of 60,000 images for training and 10,000 images for testing.  & LeNet-5 &    87.85\% \\ \\
    
    & & LeNet-1 &  84.5\% \\ \hline
    
    \\\added{Fashion-MNIST} & \raggedright \added{Grayscale images in 10 different classes of clothes composed of 60,000 images for training and 10,000 images for testing.}  & \added{LeNet-4} &    \added{88\%} \\  \hline 
    
    \\Cifar-10 & \raggedright Object recognition dataset in ten different classes composed of 50,000 images for training and 10,000 images for testing. & \raggedright A 12-layer ConvNet with max-pooling and dropout layers. & 82.93\%  \\ 
    
    \\ \\
    & &\added{ResNet20} &  \added{86\%} \\ \hline
    \\ \added{SVHN} & \raggedright \added{A real-world image dataset for recognizing house numbers obtained from Google Street View images. \\ It is composed of 73,257 training images and 26,032 testing images.}  & \added{LeNet-5} &    \added{88\%} \\  \hline

    \end{tabular}
    \caption{Datasets and models used for evaluation}
    \label{tab:Dataset}
\end{table*}

\subsection{Research Questions} \label{Sec:RQs}

Our empirical evaluation is designed to answer the following research questions.

\begin{itemize}

   \item \textbf{RQ1. To what extent do the selected diversity metrics measure actual diversity in input sets?} We want to assess, in a controlled way, the reliability of the selected diversity metrics for measuring the actual diversity of an input set in terms of the features the images contain. Only the metrics that reliably reflect changes in image diversity will be retained for the next research questions.

    \item \textbf{RQ2. How does diversity relate to fault detection?} Similar to other studies in different contexts~\cite{feldt2016test,bueno2007improving,leon2003comparison}, we aim to investigate the correlation between diversity and faults to assess whether diverse input sets lead to higher fault coverage. We do not investigate in this research question the correlation between diversity and the number of mispredicted inputs, as this is misleading. Many mispredictions result from the same problems in the DNN model and are therefore redundant. This is similar to failures in regular software. In classification problems, for example, guiding the selection of test inputs to maximize misprediction rates (the number of mispredicted inputs / total number of inputs) could thus be misleading. However, the notion of fault in DNN models is not as straightforward as it is in regular software, where we can identify statements responsible for failures. Therefore, to investigate this research question, we need to first define a mechanism to compare how effective test sets are in detecting faults in DNNs, so that we can then investigate the relationship between diversity and faults.

    \item \textbf{RQ3. How  does  coverage  relate  to fault detection?} Similar to diversity, we aim to assess the association between state-of-the-art coverage metrics and faults. This enables us to compare black-box diversity and white-box coverage in selecting test sets with high fault-revealing power. Note that recent studies questioned the use of coverage metrics to assess DNN test inputs~\cite{li2019structural,harel2020neuron}.
    \Revise{Most state-of-the-art coverage metrics strongly rely on artificial inputs generated based on adversarial methods~\cite{pei2017deepxplore,Ma2018DeepGaugeMT,sun2019structural,kim2019guiding,gerasimou2020importance}.}
    However, their positive correlation with the presence of adversarial inputs does not necessarily mean that they are efficient enough to reveal the fault detection capability of natural test input sets. Several studies~\cite{li2019structural,chen2020deep} failed to find a strong correlation between coverage and misprediction rates when using only natural input sets. Furthermore, coverage metrics showed poor performance in guiding the retraining of DNN models to improve the robustness of the model to adversarial attacks~\cite{dong2019there,yan2020correlations}. 
   Therefore, there is still no consensus on which coverage metrics are suitable for different DNN testing-related tasks such as test selection, minimization and generation. 
    
     \item \textbf{RQ4. How do diversity and coverage perform in  terms  of  computation  time?} We aim to compare the computation times of selected diversity and coverage metrics. Most importantly, we aim to study how these computation times scale as the sizes of the test sets increase. Excessive computation times may limit applicability, though what is acceptable depends on  the context.

     \item \textbf{RQ5. How does diversity relate to coverage?} Though diversity is black-box and therefore has inherent practical advantages, it is interesting to study the correlation between diversity and coverage to determine if they essentially capture the same thing. \Revise{Though this question can be answered indirectly by some of the previous questions (correlations are transitive), such correlation analysis can provide additional insights to explain and support previous results. }

\end{itemize}

\subsection{Subject Datasets and DL Models}

\added{Table~\ref{tab:Dataset} shows the characteristics of the datasets and models in our experiments. We used four common image recognition datasets, Cifar-10~\cite{Cifar10}, MNIST~\cite{deng2012mnist}, Fashion-MNIST~\cite{deng2012mnist} and SVHN~\cite{netzer2011reading}.
We use these datasets with five state-of-the-art DNN models: 12 layers Convolutional Neural Network (12-layer ConvNet), LeNet-1, LeNet-4, LeNet-5 and ResNet20.}

Cifar-10 contains 50,000 images for training and 10,000 for testing. These images belong to 10 different classes (e.g cats, dogs, trucks). 

We also used MNIST, which contains 70,000 images (60,000 for training and 10,000 for testing). Each of these images represents a handwritten digit and belongs to one of the 10 classes.
We included Fashion-MNIST, which contains grayscale images in 10 different classes of clothes. It is composed of 60,000 images for training and 10,000 images for testing.
Finally, we use SVHN, a real-world image dataset for recognizing house numbers. It contains 99,289 images where 73,257 are for training and 26,032 are for testing. 
 
For Cifar-10, we used a 12-layer ConvNet and ResNet20 that we trained for 50 and 100 epochs, respectively. For MNIST, we used the LeNet-1 and LeNet-5 models that we trained for 50 epochs. We trained the LeNet-4 model with Fashion-MNIST for 20 epochs. Finally, for SVHN, we used the LeNet-5 model, which we trained for 100 epochs. The different combinations of models and datasets, along with the models' accuracy, are detailed in Table~\ref{tab:Dataset}.

We selected these datasets and models because they are widely used in the literature~\cite{pei2017deepxplore,kim2019guiding,gerasimou2020importance,Ma2018DeepGaugeMT}. Further, all the inputs in the selected datasets are correctly labelled. These datasets and models are considered good baselines to observe key trends, as they offer a wide range of diverse inputs (in classes and domain concepts) and different models (in terms of internal architecture).

\subsection{Evaluation and Results}

Before addressing our research questions, one essential issue was how to count faults in DNNs. A misprediction implies the existence of a fault in the DNN. However, identifying faults is not as straightforward as in regular software, where faulty statements that cause failures can be identified. Nevertheless, estimating fault detection effectively is essential to compare coverage and diversity metrics. Simply comparing misprediction rates is misleading as many test inputs are typically mispredicted for the same reasons~\cite{fahmy2021supporting}. Typically, with regular software, a tester does not select input tests to maximize the failure rate (equivalent to the misprediction rate in our context) but rather wants to maximize the number of distinct detected faults. This should be not different with DNNs, where we want to detect the distinct causes of mispredictions.

\begin{figure}[h]
\centering
\includegraphics[scale=0.5]{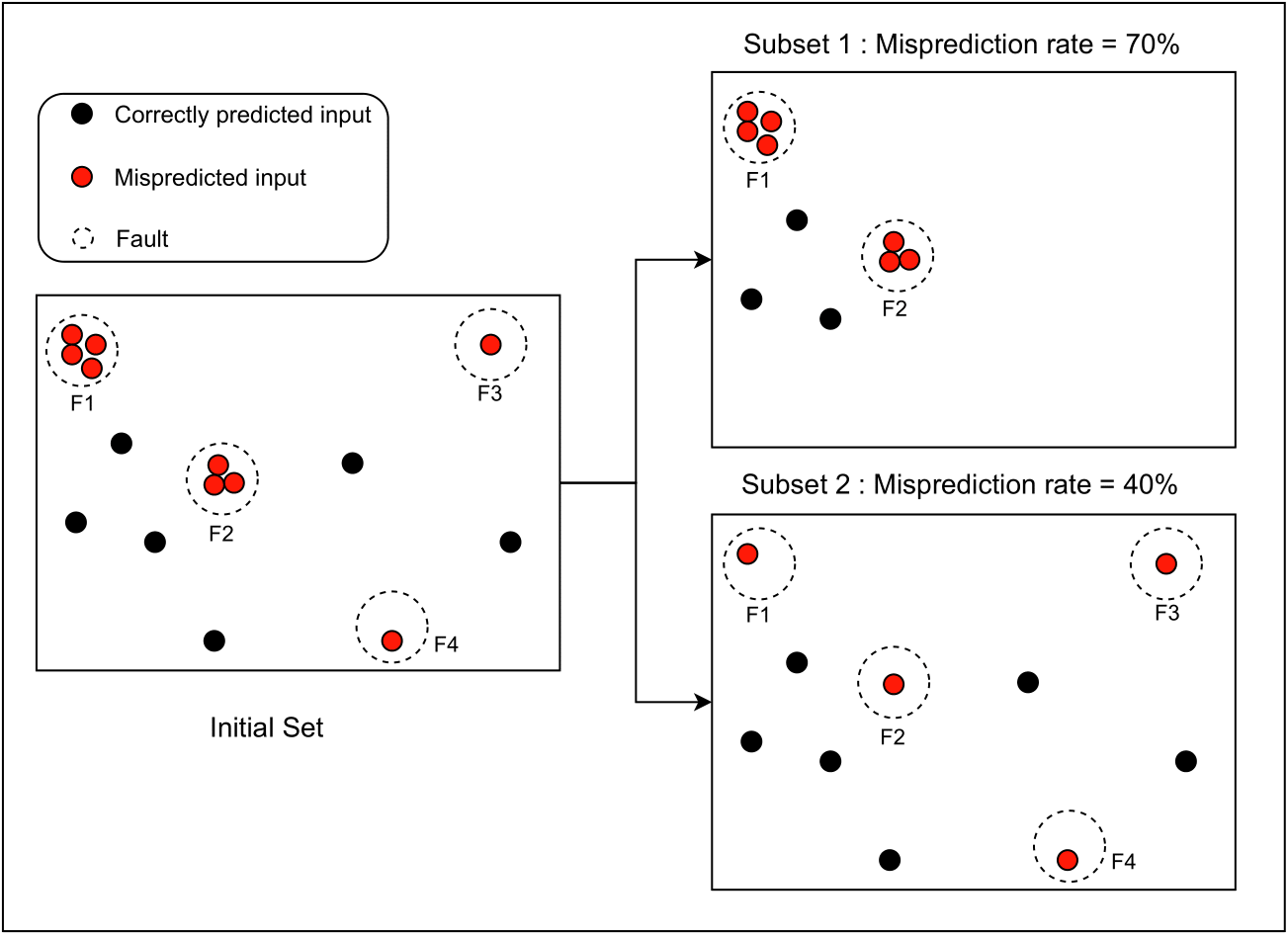}
\caption{Relying on misprediction rates is misleading}
\label{fig:FaultsFig}
\end{figure}

We illustrate this issue in Figure~\ref{fig:FaultsFig} where we represent an example of a test input set in a two-dimensional feature space. Black dots refer to the inputs correctly predicted by the DNN under test, and red dots represent the mispredicted ones. We select two subsets from the initial set and measure their corresponding misprediction rates. As shown in Figure~\ref{fig:FaultsFig}, subset 1 is less diverse than subset 2 but has a higher misprediction rate. However, some of the mispredicted inputs are very similar and somewhat redundant. 

As a result, it can be argued that subset 2 is more diverse than subset 1 and is more informative for testing the model because its mispredicted inputs potentially  reveal more faults in the DNN model.
In preliminary experiments (not included in this paper), we evaluated the computation of misprediction rates in test input sets and studied their correlation with diversity and coverage and found no statistically significant correlation for both diversity and coverage metrics. We suspected that accounting for numerous redundant test inputs affected our correlation analysis. \added{
In practice, selecting or generating test inputs that trigger failures (i.e mispredictions) is far more useful when these failures are diverse~\cite{zohdinasab2021deephyperion}. A test set that repeatedly exposes the same problem in the DNN model is a waste of computational resources, especially when we have a limited testing budget and a high labeling cost for testing data~\cite{zohdinasab2021deephyperion}. } This is why, similar to other studies comparing the effectiveness of test strategies with regular software, we want here to address the notion of faults detected in DNNs and study their association with diversity and coverage.

\subsubsection{Estimating Faults in DNNs}\label{Sec:Faults}

\begin{figure*}[ht]
\centering
\includegraphics[width=\textwidth]{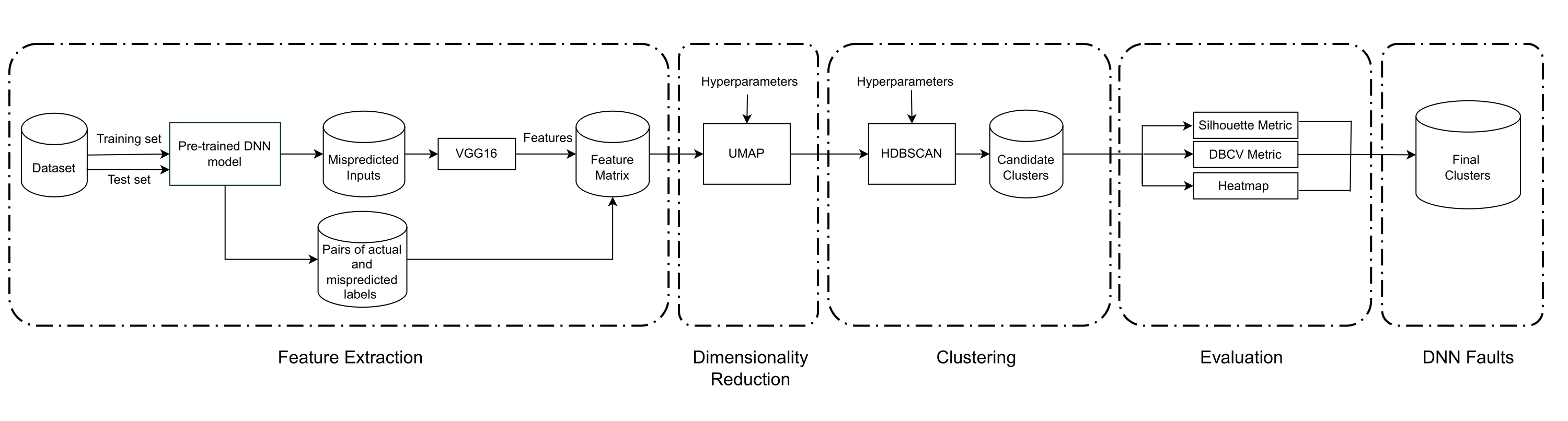}
\centering
\caption{Estimating faults in DNNs}
\label{fig:FaultsApproach}
\end{figure*}

Following a similar approach to the work of Fahmy \textit{et al}.~\cite{fahmy2021supporting} and Attaoui \textit{et al.}~\cite{attaoui2022black}, we rely on a clustering approach to group similar mispredicted inputs presenting a common set of characteristics that are plausible causes for mispredictions. We approximate the number of detected faults in a DNN through such clustering. Although many mispredicted test inputs are redundant and result from the same causes, we assume that test inputs belonging to different clusters are mispredicted due to distinct problems~\cite{fahmy2021supporting} in the DNN model. This is an approximation but a practical and plausible way to estimate and compare the number of detected faults across coverage and diversity strategies. Although faults can only be addressed by retraining in DNNs, as opposed to debugging, clusters nevertheless capture common causes for mispredictions and are thus comparable to faults in regular software. Figure~\ref{fig:FaultsApproach} depicts how faults are counted in DNNs and we describe below each step in detail.

\paragraph{Feature Extraction}  We start by training our model using the training dataset. We then run our pre-trained model on the test and training datasets to identify all mispredicted inputs. 
We \added{not only} use mispredicted inputs from the test set but also mispredicted inputs from the training dataset to extract the best clusters and estimate detected faults as accurately as possible. We rely on VGG16 to extract the mispredicted inputs' features and build the corresponding feature matrix as described in section~\ref{SubSec:FeatureExtraction}. \Revise{We add two extra features to the matrix from the DNN model to capture actual and mispredicted classes (labels) related to each misclassified input. }This adds information to the feature matrix about the misprediction behaviour of the model under test for each mispredicted input, which we believe builds better clusters to reflect common misprediction causes. 

\paragraph{Dimensionality Reduction} By definition, the number of input features for a dataset corresponds to its dimensionality. Low density in high-dimensional spaces makes it difficult, in general, for typical clustering algorithms to find a continuous boundary that separates the different clusters~\cite{joswiak2019dimensionality}. Therefore, employing dimensionality reduction techniques can help clustering algorithms make the inputs and their related clusters more distinguishable. 
Because we are working with high-dimensional inputs (512 features from the VGG model and two features from the DNN model), we rely on the Uniform Manifold Approximation and Projection (UMAP)~\cite{mcinnes2018umap} dimensionality reduction technique. We selected UMAP because several studies~\cite{diaz2021review,hozumi2021umap} have shown its effectiveness as a pre-processing step to boost the performance of clustering algorithms when compared to other state-of-the-art dimensionality reduction techniques, such as PCA~\cite{jolliffe2016principal} and t-SNE~\cite{van2008visualizing}. PCA is a linear dimensionality reduction technique that performs poorly on features with nonlinear relationships. To work with high-dimensionality data to obtain low-dimensionality and nonlinear manifolds, some nonlinear dimensionality reduction algorithms, such as UMAP and t-SNE, should be used~\cite{hozumi2021umap}. However, t-SNE is more computationally expensive than UMAP and PCA. It is used in practice for data visualization and data reduction to two or three dimensions. Furthermore, it involves hyperparameters that are not always easy to tune in order to get the best results. Therefore, we relied on UMAP for dimensionality reduction as an effective pre-processing step to boost the performance of density-based clustering. This will be used in the next step.

\paragraph{Clustering} After performing dimensionality reduction, we apply the HDBSCAN~\cite{mcinnes2017hdbscan} clustering algorithm to group mispredicted inputs that are similar and believed to result from the same causes (faults) in the DNN model. HDBSCAN is a density-based clustering algorithm where each dense region is considered a cluster and low-density regions are considered noise. In other words, it views clusters as areas of high density separated by areas of low density. Clusters found by HDBSCAN can be of any shape, as opposed to other types of clustering algorithms, such as  k-means or hierarchical clustering, which assume that clusters are convex-shaped. Each cluster is supposed to correspond to a fault (common problems) in the DNN model because their inputs are similar in terms of extracted features and actual and mispredicted classes. 

\begin{table*}[ht]
    \centering
    \small
    \begin{tabular}{llllllll}
    \hline \hline \\
    Dataset  & Model & \raggedright\#Misp. in training set & \raggedright \#Misp. in test set & Silh. & DBCV & \#Noisy test inputs & \#Clusters \\ \\ \hline \hline \\
    MNIST           & LeNet-5           & 8055 & 1215 & 0.64 & 0.68 & 58 & 85   \\ \hline \\
    MNIST           & LeNet-1           & 9754 & 1542 & 0.71 & 0.74 & 72 & 137  \\ \hline \\
    
    \added{Fashion-MNIST}   &  \added{LeNet-4}          &   \added{5636}    &  \added{1157}    &   \added{0.65}   &    \added{0.53}   &  \added{101}  &    \added{141}  \\ \hline \\
    Cifar-10        &  12-layer ConvNet & 1173 & 1707 & 0.71 & 0.62 & 56 & 187  \\ \hline \\
    \added{Cifar-10}       &  \added{ResNet20}         &   \added{2191}   &    \added{1384}  &     \added{0.75} &   \added{0.63}  &  \added{78}    &   \added{177}   \\ \hline \\

    \added{SVHN}           &  \added{LeNet-5}          &    \added{151}   &   \added{3009}    &   \added{0.69}   &    \added{0.59}   &  \added{213}  &   \added{147}   \\ \hline \\

    \end{tabular}
    \caption{Fault estimates across datasets and models}
    \label{tab:Cluster}
\end{table*}

\paragraph{Evaluation} 
As with any clustering algorithm, there are several hyperparameters to fine-tune to obtain the best clustering results.  Such hyperparameters include for example, the minimum distance that controls how tightly UMAP is allowed to pack points together, the number of neighbours to consider as locally connected in UMAP and the minimum size of clusters in HDBSCAN. We tried several hyperparameter configurations and selected the best configurations based on both manual and metric-based evaluations. For the latter, we relied on two standard metrics to evaluate the clusters, which are the Silhouette score~\cite{rousseeuw1987silhouettes} and the Density-Based Clustering Validation (DBCV)~\cite{moulavi2014density} metric. 

The Silhouette score is one of the state-of-the-art clustering evaluation metrics that compare inter- and intra-cluster distances. It varies between minus one and one. The closer to one, the better the clustering. A score near zero represents clusters with inputs very close to the decision boundary of the neighboring clusters. A negative score generally indicates that the inputs are assigned to the wrong clusters.

We also relied on the DBCV metric to evaluate the generated clusters. This metric is dedicated to density-based clustering algorithms and assesses clustering quality based on the relative density connection between pairs of inputs. It evaluates the within- and between-cluster density connectedness~\cite{ester1997density}. Similar to Silhouette, DBCV generates scores between -1 and 1~\cite{moulavi2014density}. High-density within clusters and low-density between clusters lead to high DBCV scores, indicating better clustering results.

\begin{figure}[ht]
\begin{minipage}{.5\columnwidth}
\centering
\includegraphics[scale=0.35]{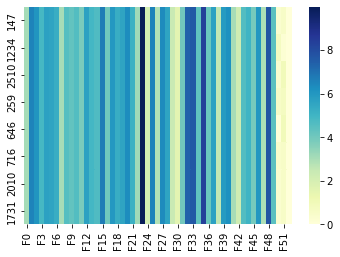}
\caption{Heatmap example \\ 1 related to a final cluster}
\label{Fig:HeatMap1}
\end{minipage}
\begin{minipage}{.5\columnwidth}
\centering
\includegraphics[scale=0.35]{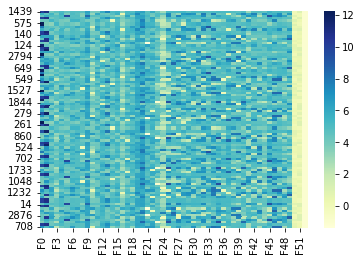}
\caption{Heatmap example 2  related to noisy cluster}
\label{Fig:HeatmapNoisy}
\end{minipage}

\begin{minipage}{.5\columnwidth}
\centering
\includegraphics[scale=0.35]{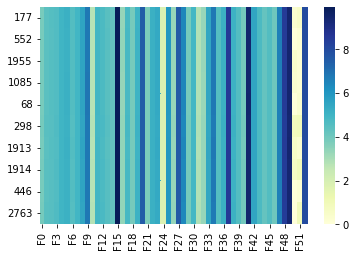}
\caption{\added{Heatmap example \\ 3  related to a final cluster}}
\label{Fig:HeatMap2}
\end{minipage}
\begin{minipage}{.5\columnwidth}
\centering
\includegraphics[scale=0.35]{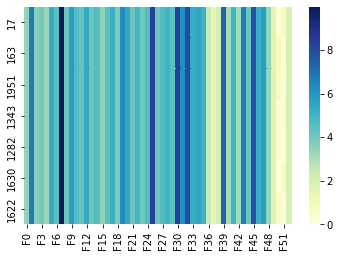}
\caption{\added{Heatmap example 4 related to a final cluster}}
\label{Fig:Heatmap3}
\end{minipage}
\end{figure}

We selected the configuration with the best Silhouette and DBCV scores. We further evaluated the generated clusters by performing a manual evaluation. First, we tried to check the content of the clusters to see whether their inputs were similar to or shared some features that might have led to mispredictions by the DNN model. Because of the large number of  mispredicted inputs, an exhaustive manual inspection of the clusters was impractical. Therefore, we relied on generating  the features' heatmaps related to each cluster to better visualize and assess the quality of the clusters. \added{Figures~\ref{Fig:HeatMap1},~\ref{Fig:HeatmapNoisy},~\ref{Fig:HeatMap2} and~\ref{Fig:Heatmap3} illustrate four representative examples of heatmaps where rows correspond to the inputs' ids in one cluster, columns refer to their features and colours encode the features' values. As we observe in Figures~\ref{Fig:HeatMap1},~\ref{Fig:HeatMap2} and~\ref{Fig:Heatmap3}, within a cluster, well-clustered inputs share common patterns in terms of the features' distribution while ill-clustered inputs (such as noisy inputs) do not, as is visible in Figure~\ref{Fig:HeatmapNoisy}. Based on our manual analysis of the final selected clusters, we observed that most of them look like the first three figures and share common patterns.
We therefore conclude that the mispredicted inputs inside each cluster are similar and share common characteristics (features), potentially causing mispredictions. }

Table~\ref{tab:Cluster} describes  the final clusters that we generated for the different datasets and models that we used in our experiments. 

We observe that the number of noisy inputs (inputs that do not belong to any cluster) is not large compared to the total number of mispredicted inputs. We decided to delete them from the sets of mispredicted inputs in all the following experiments because (1) they do not belong to any cluster and cannot therefore be associated with faults as we defined them, and (2) the minimum number of detected faults in the studied DNN models can thus be assumed to correspond to the number of clusters.

\begin{table*}[ht]
\centering
\begin{tabular}{|c|c||c|c|}
\hline
\hline
 \textbf{Model}    & \textbf{Faults $C_{i}$} & \begin{tabular}[c]{@{}c@{}}\textbf{Accuracy on the} \\ \textbf{retraining cluster $C_{i}$}  \end{tabular}   & \begin{tabular}[c]{@{}c@{}}\textbf{Average Accuracy on the} \\ \textbf{other clusters $C_{j \neq i}$}  \end{tabular}  \\ \hline \hline

 \multirow{10}{*}{\rotatebox{90}{\begin{tabular}[c]{@{}c@{}}12-layer \\ ConvNet \end{tabular}  }}  & Cluster 1  & 80\%  & 28.23\%\\ \cline{2-4}

                          & Cluster 2  & 60\% & 27.42\%\\ \cline{2-4} 
                          & Cluster 3  & 55\%  & 27.45\%\\ \cline{2-4}  
                          & Cluster 4 & 71.40\% &  27.06\%\\ 
                        \cline{2-4} 
                          & Cluster 5 & 66.70\%  & 27.40\%\\ \cline{2-4}  
                          & Cluster 6  & 60\% & 27.95\% \\
                          
                        \cline{2-4} 
                          & Cluster 7  & 58.33\%  &27.76\% \\ \cline{2-4}  
                         & Cluster 8  & 67.70\% & 28.60\% \\ 
                          
                        \cline{2-4} 
                          & Cluster 9  & 66.30\%  &27.74\% \\ \cline{2-4}  
                          & Cluster 10  & 61\% & 27.80\%   \\ \cline{2-4}  \hline \hline
\multicolumn{2}{|c||}{\textbf{Average}}   &\textbf{ 66.64\%} & \textbf{27.74\%} \\ \hline \hline
 \multirow{10}{*}{\rotatebox{90}{\begin{tabular}[c]{@{}c@{}}ResNet20 \end{tabular}  }}  & Cluster 1  & 77\%  & 30\%\\ \cline{2-4}

                          & Cluster 2  & 75\% & 29\%\\ \cline{2-4} 
                          & Cluster 3  & 74\%  & 28\%\\ \cline{2-4}  
                          & Cluster 4 & 62.5\% &  31\%\\ 
                        \cline{2-4} 
                          & Cluster 5 & 65\%  & 29\%\\ \cline{2-4}  
                          & Cluster 6  & 56.5\% & 28.70\% \\
                          
                        \cline{2-4} 
                          & Cluster 7  & 78.40\%  &30.3\% \\ \cline{2-4}  
                         & Cluster 8  & 75\% & 29.9\% \\ 
                          
                        \cline{2-4} 
                          & Cluster 9  & 62.5\%  &31\% \\ \cline{2-4}  
                          & Cluster 10  & 64\% & 31\%   \\ \cline{2-4}  \hline \hline
\multicolumn{2}{|c||}{\textbf{Average}}   & \textbf{69\%} & \textbf{29.80\%} \\

                          \hline \hline

\end{tabular}
\caption{\color{black} The results of faults validation experiment on 12-layer ConvNet and ResNet20. In each row, we retrained the model under test on 85\% of each cluster $C_{i}$. We report in the third column the accuracy of the retrained model on the remaining 15\% of cluster $C_{i}$. The last column refers to the average of accuracies of the retrained models over all other fault clusters $C_{j}$ ($j\neq i$) }
\label{Tab:Fault_Validation_Cifar10}
\end{table*}

\paragraph{Validation} As mentioned earlier, we followed an approach similar to existing work~\cite{fahmy2021supporting},~\cite{attaoui2022black} to estimate faults in DNNs. The authors conducted an empirical study on six DNNs to validate the clustering-based fault estimation approach. They retrained the DNNs using the original training set and subsets selected from each identified cluster (i.e. fault), which led to a significant improvement in the models' accuracy, thus demonstrating the usefulness of clustering. To further validate the identified faults (clusters) in our work, we followed a finer-grained validation method which aimed to prove that (1) inputs in the same cluster tend to be mispredicted due to the same fault, and (2) inputs belonging to different clusters are mispredicted because of distinct faults.

\noindent\textbf{\newline Inputs that belong to the same cluster are mispredicted due to the same fault in the DNN model.} If inputs within one cluster are mispredicted due to the same fault, retraining the model with a subset of the cluster should help fix the model with respect to that fault.
We verified this hypothesis for each fault-related cluster $C_i$ where $i \in [1,2,...,n]$ and n is the total number of the identified faults (clusters), by retraining the original DNN model under test with a retraining dataset consisting of the original training set and 85\% of randomly selected mispredicted inputs inside $C_i$. The original training set was reused to prevent any reduction in model accuracy~\cite{fahmy2021supporting,attaoui2022black,hu2022empirical}.
We then tested the retrained model on the remaining 15\% of inputs in $C_i$.  
We repeated this process five times (as we randomly selected inputs from cluster $C_i$) and measured the average accuracy of the retrained models when tested with 15\% of the remaining inputs from cluster $C_i$. If clusters included mispredictions caused by the same fault, the retraining process was expected to alleviate the cause of input mispredictions in $C_i$ and thus significantly improve the accuracy of the model for images in that particular cluster. We did not expect a perfect model with no mispredictions because we did not have a large number of mispredicted inputs in each cluster to retrain the model. 
Moreover, clustering was not expected to be perfect but only an approximation of faults. 
Because clusters did not have the same size and were not all large enough to enable this analysis, our analysis focused on the ten largest clusters across datasets and models. Due to the high computational expense of our experiments, we used two models, 12-layer ConvNet and ResNet20, to validate the fault-estimation method.  Table~\ref{Tab:Fault_Validation_Cifar10} shows the accuracy of the retrained 12-layer ConvNet and ResNet20 models when tested on the ten largest clusters in each of their corresponding datasets. 
By definition, the accuracy of the original model on all cluster images was zero because they were all mispredicted. As shown in Table~\ref{Tab:Fault_Validation_Cifar10}, there was significant improvement in the models' accuracy, with an average equal to 66.64\% in 12-layer ConvNet and 69\% in ResNet20.  

The results therefore suggested that test inputs belonging to the same cluster are indeed mispredicted due to the same faults, thus supporting the hypothesis underlying our method of counting faults.
\newline\textbf{\newline Inputs belonging to different clusters are mispredicted due to distinct faults.} If the clusters represent distinct faults in the DNN, retraining the model with a subset of a cluster $C_i$ should have a significant effect on the other images in $C_i$ when compared to images in other clusters.  
\newline Consequently, to validate that inputs belonging to different clusters are mispredicted due to distinct faults, we test each of the previously retrained DNN models for each cluster $C_i$ on the other clusters $C_j$ where $j \neq i$. We measure the average accuracy of each of the retrained DNN models over all remaining clusters and report the results in Table~\ref{Tab:Fault_Validation_Cifar10}. The retrained ResNet20 and 12-layer ConvNet models are significantly more accurate on the clusters for which they were retrained than on other clusters. Indeed, in 12-layer ConvNet, for example, the average accuracy for the latter is only 27.74\% compared to 66.64\% for the former. 
We therefore conclude that inputs belonging to different clusters tend to be mispredicted due to distinct faults. 
Although they are quite limited, we nevertheless observed improvements in model accuracy on clusters for which the model was not retrained. This can be expected because fixing one fault in the DNN model through retraining may also, to a more limited extent, fix other related faults, potentially improving the accuracy on other clusters. As acknowledged previously, our clustering is not perfect and although the obtained clusters' Silhouette and DBCV scores are high, they are not equal to one as shown in Table~\ref{tab:Cluster}.

\subsubsection{RQ1. To what extent do the selected diversity metrics measure actual diversity in input sets?}
To directly evaluate the capability of the selected metrics to measure diversity in input sets, we studied how diversity scores changed while varying, in a controlled manner, the number of image classes covered by the input sets. Classes characterize the content of images. For example, a set of images, sampled from the Cifar-10 dataset and containing the two classes \textit{Car} and \textit{Deer} is considered more diverse than a set containing only \textit{cars}. We assumed that diversity scores should increase with the number of classes that are present in an input set.

\begin{figure*}[ht]
\begin{minipage}{.3\textwidth}
\centering
\includegraphics[scale=0.4]{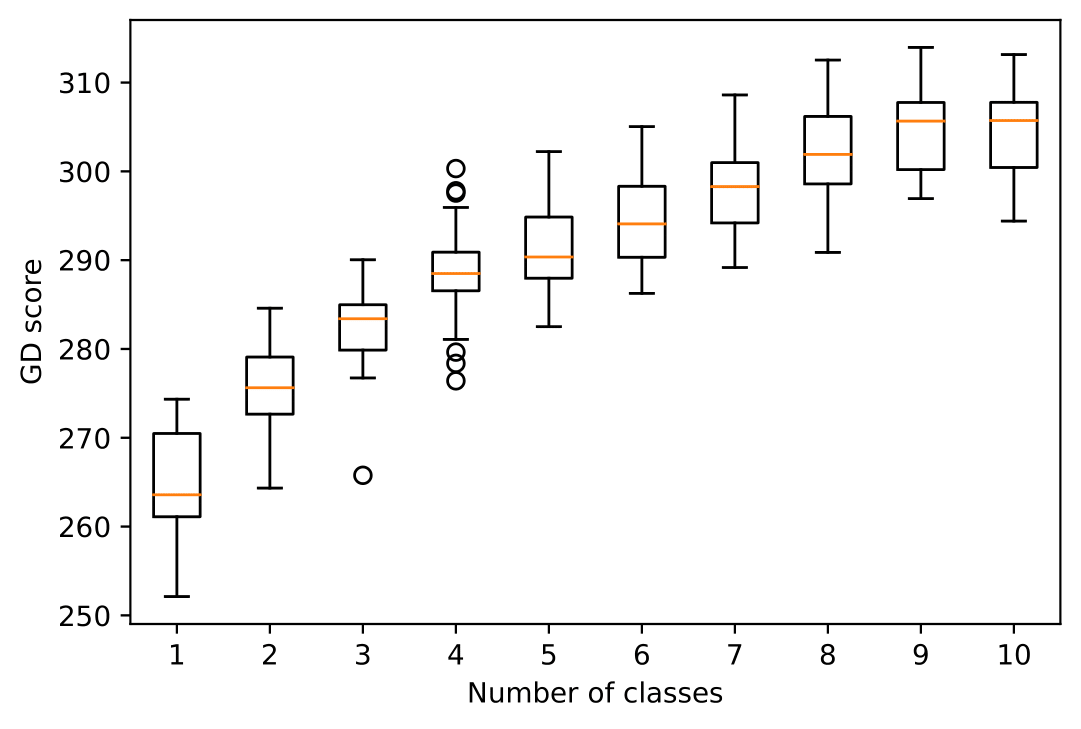}\\
(a) Evolution of GD on Cifar-10
\label{Fig:GDCifar}
\end{minipage}
\begin{minipage}{.3\textwidth}
\centering
\includegraphics[scale=0.4]{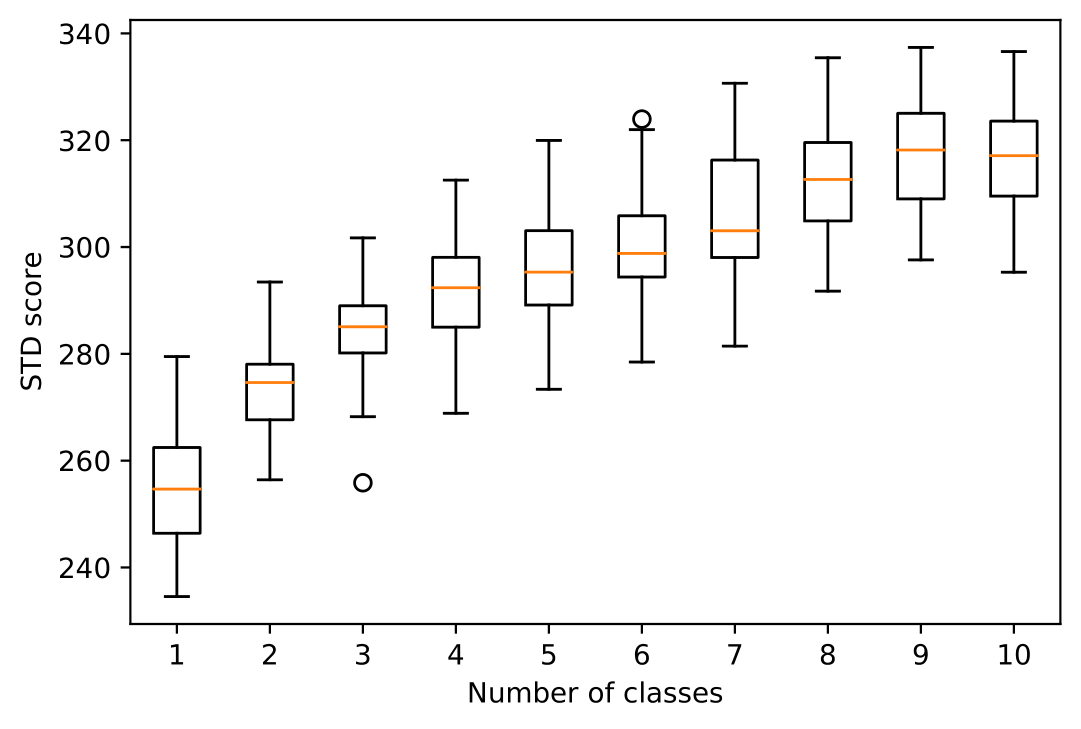}\\
(b) Evolution of STD on Cifar-10
\label{Fig:STDCifar}
\end{minipage}
\begin{minipage}{.3\textwidth}
\centering
\includegraphics[scale=0.4]{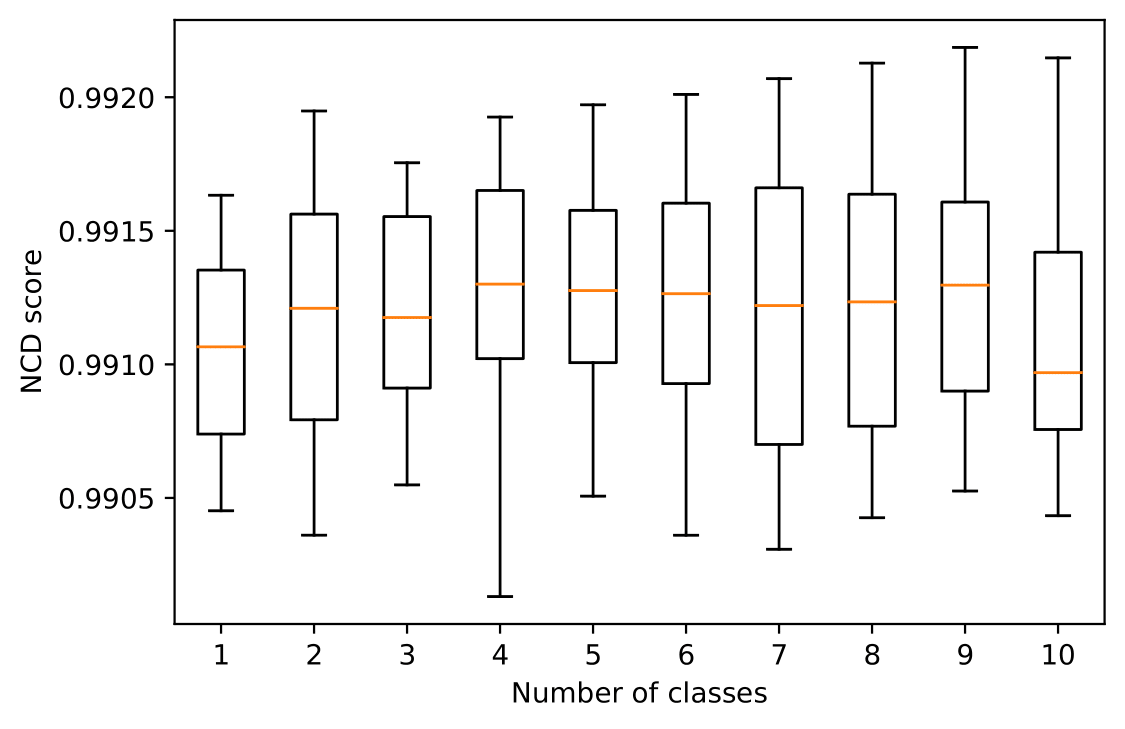}\\
(c) Evolution of NCD on Cifar-10
\label{Fig:NCDCifar}
\end{minipage}

\begin{minipage}{.3\textwidth}
\centering
\includegraphics[scale=0.4]{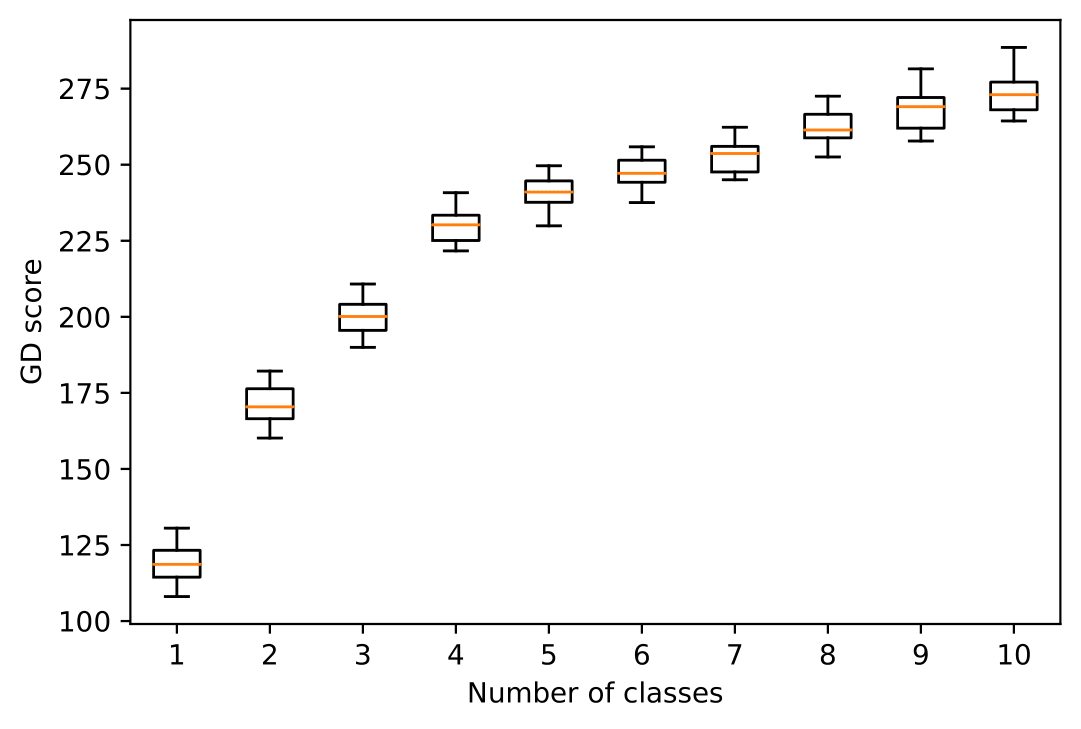}\\
(d) Evolution of GD on MNIST
\label{Fig:GDMnist}
\end{minipage}
\begin{minipage}{.3\textwidth}
\centering
\includegraphics[scale=0.4]{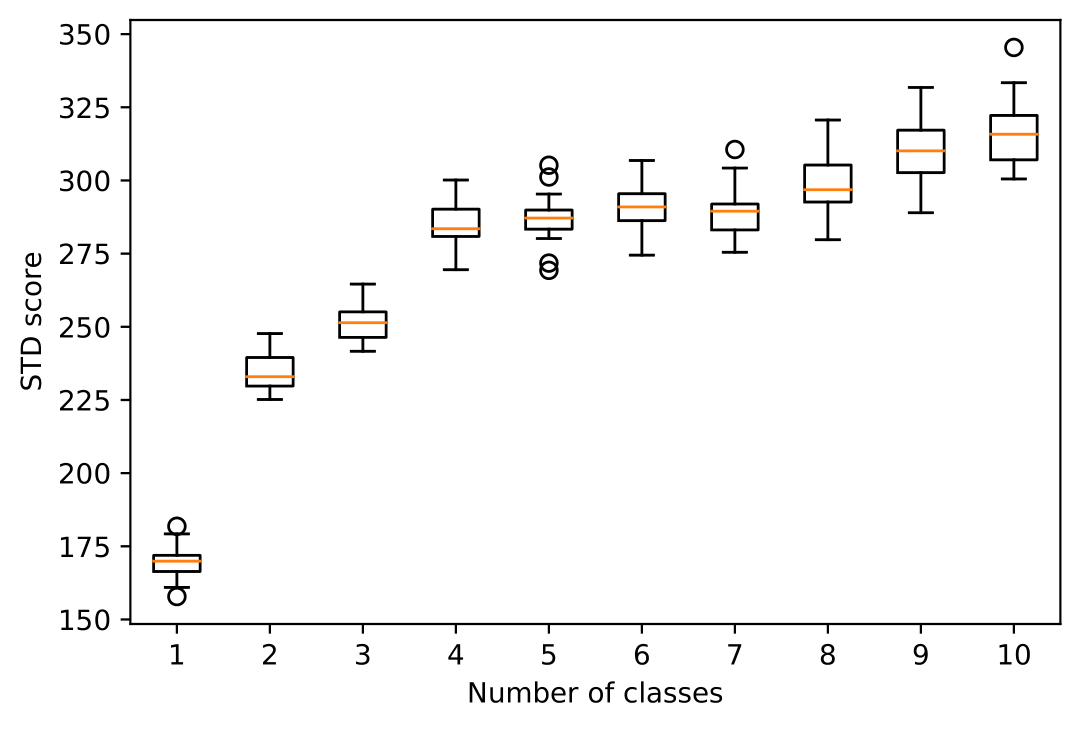}\\
(e) Evolution of STD on MNIST
\label{Fig:STDMnist}
\end{minipage}
\begin{minipage}{.3\textwidth}
\centering
\includegraphics[scale=0.4]{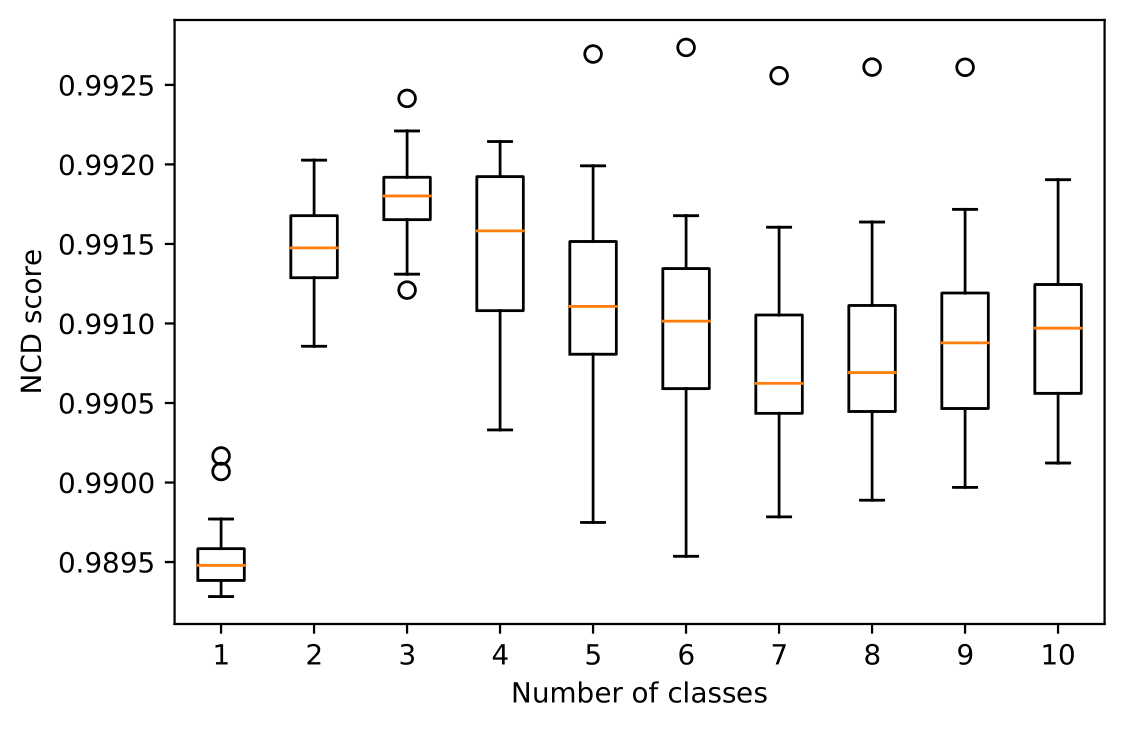}\\
(f) Evolution of NCD on MNIST
\label{Fig:NCDMnist}
\end{minipage}
\vspace{1em}
\caption{Evolution of the diversity scores for input sets from Cifar-10 and MNIST. Each boxplot shows the distribution of diversity scores of 20 input sets of size 100.}
    \label{fig:RQ1}
\vspace{1em}
\end{figure*}

Algorithm~\ref{Algorithm-RQ1} describes at a logical level our experiment's procedure to answer RQ1. This procedure aims to increase the actual diversity of the content of image sets in a controlled manner and observe whether diversity metrics are sensitive to such changes. 
Given a certain dataset, we started by randomly selecting the first class $C_i$ from the dataset in our experiment (Line 1).  Then, we randomly sampled, with replacement, 20 input sets of size 100. Each of these input sets was sampled from the same class $C_i$ (Lines 2-4). We measured the diversity scores for each initial input set (Lines 5-7). For each such input set, we incrementally increased the number of classes it covered by replacing some of its inputs with new ones from a new class $C_{k \neq i}$ while maintaining a uniform distribution across classes inside the samples. To do so, for each initial input set $Fset$, we randomly selected another class $C_{k}$ that we wanted to add (Lines 8 and 10) and randomly selected new inputs from the class $C_{k}$ as a $Newset$ (Line 10-11). 
We randomly kept about $100/k$ inputs (k is the number of selected classes) of each existing class in $Fset$ (Line 12) and merged their inputs in $Fset$ with the newly selected ones in $Newset$ (Line 13). Finally, we measured the diversity scores for each input set (Lines 15-17) and repeated the process until we included images from all  classes in the selected dataset (Line 9).  
The distribution of the diversity scores that are related to each metric using boxplots is depicted in Figure~\ref{fig:RQ1}. Each boxplot illustrates the distribution of the diversity scores of 20 input sets of size 100, each with the same number of classes.

\begin{algorithm}[t!]

\DontPrintSemicolon
  \KwInput{C: set of $n$ classes in the dataset ($C\xleftarrow{} \{c_{1},c_{2},...,c_{n}\}$)}

  \KwOutput{$GDs,STDs, NCDs$}

    $c_{i} \xleftarrow{} RandomClassSelect(1, C)$\;
    \For{$j $\ in $\ \{1,2,...,20\}$}{
    $k \xleftarrow{} 1$ \;
    $Fset \xleftarrow{} RandomInputSelect(100, c_{i})$\;

    $GDs \xleftarrow{} GD(Fset)$\;
    $STDs \xleftarrow{} STD(Fset)$\;
    $NCDs\xleftarrow{} NCD(Fset)$\;
    $C \xleftarrow{} C\setminus \{c_{i}\}$\;
   
    \For{$k $\ in $\ \{2,...,c_{n}\}$}{
    $c_{k} \xleftarrow{} RandomClassSelect(1, C)$\;
    
    $NewSet \xleftarrow{} RandomInputSelect(100/k, C_{k})$\;
    $Fset \xleftarrow{} Keep(100/k, Fset)$\;
    $Fset \xleftarrow{} Merge (Fset, Newset )$\;
    $C \xleftarrow{} C\setminus \{c_{k}\}$\;
    $GDs \xleftarrow{} GD(Fset)$\;
    $STDs \xleftarrow{} STD(Fset)$\;
    $NCDs \xleftarrow{} NCD(Fset)$\;
    }

    }
    \KwRet{$GDs, STDs, NCDs$}
\caption{Experimental Procedure for RQ1}
\label{Algorithm-RQ1}
\end{algorithm}

For example, when we evaluated Cifar-10, we selected 20 input sets of size 100. All the selected images inside each input set corresponded to the class \textit{Deer}. For each selected input set, we measured the GD, NCD and STD scores. For each metric, we reported the distribution of the diversity scores related to these samples using boxplots, as depicted in Figures~\ref{fig:RQ1}.a, ~\ref{fig:RQ1}.b and ~\ref{fig:RQ1}.c.
We then increased the number of classes inside each sample by randomly replacing 50 images in \textit{Deer} with new images from the \textit{Truck} class. Each input set contained equal distribution of \textit{Deer} and \textit{Truck} images. We reported again the distribution of the diversity scores using boxplots. We repeated the process until we reached a total number of 10 classes inside the selected samples, while maintaining a uniform distribution across classes inside the input sets at each sampling iteration.
As shown in Figure~\ref{fig:RQ1}, we observed that GD outperforms NCD and STD as it exhibits a monotonic increase when increasing the number of classes inside the input sets. As shown in Figures~\ref{fig:RQ1}.a and~\ref{fig:RQ1}.d, the more diverse the input sets, the higher the GD in all datasets and models that we have evaluated. We observed a similar but noisier trend in STD. Using the example of STD scores for MNIST (Cf. Figure~\ref{fig:RQ1}.e), we observe that these scores slightly decrease for samples embedding seven classes. A similar observation can be made in Cifar-10 when going from nine to ten classes (Cf. Figure~\ref{fig:RQ1}.b).   
\newline Surprisingly, we found that NCD scores do not increase when input sets become more diverse. We also observed that this diversity metric has low variability in the generated scores. As shown in Figures~\ref{fig:RQ1}.c and ~\ref{fig:RQ1}.f, the range of the calculated mean NCD scores for the different input sets in the experiment is between 0.9895 and 0.9913. We should note that we have tried other types of features in our experiments with NCD to further assess the reliability of this metric in evaluating diversity. For this purpose, we followed one of the recommendations of Cohen \textit{et al.}~\cite{cohen2014normalized} and Cilibrasi \textit{et al.}~\cite{cilibrasi2005clustering} and calculated the NCD scores of the input sets based on the raw images from MNIST. However, we obtained similarly poor results because the NCD score did not consistently increase when input sets became more diverse.
\newline Besides its poor performance in measuring data diversity, we note that NCD is computationally expensive. It took approximately one hour to calculate the NCD score for one input set of size 100, suggesting another limitation regarding its applicability in testing DNN models. We conclude that, in our context, this metric is neither practical nor reliable in measuring data diversity and is therefore excluded from the rest of our study. 
\newline \Revise{Note that the NCD metric's poor results may be due to the combination of feature inputs and compression tools that fail to generate accurate compression distances in our datasets.}
\Revise{We therefore believe that NCD cannot be applied or  function properly without careful selection of image formats and their associated feature representation and the compression tool, which is highly sensitive to these elements. Although we have tried several combinations of the aforementioned configurations based on existing prior works~\cite{cohen2014normalized,cilibrasi2005clustering}, we aim in future work to investigate more combinations of feature images and dedicated compression tools to achieve better results for the NCD metric.}

\begin{tcolorbox}
\textbf{Answer to RQ1:} 
Geometric diversity and STD performed well in measuring actual data diversity in all the studied datasets. This is not the case with NCD, which we excluded from the following experiments.  
\end{tcolorbox}
\subsubsection{RQ2. How does diversity relate to fault detection?}

\begin{table}[ht!]
\color{black}
\centering
\small
\begin{tabular}{|c|c|c||c|c|}
\hline
\hline
Dataset                & Model               & Test Set Size & Spearman & P-value \\ \hline \hline

\multirow{4}{*}{Cifar-10}    & \multirow{4}{*}{\rotatebox{90}{\begin{tabular}[c]{@{}c@{}}12-layer \\ ConvNet \end{tabular}  }}  & 100  & 8\% & 0.53   \\ \cline{3-5} 

                          &                                           & 200  & 17\% & 0.20   \\ \cline{3-5}
                          &                                           & 300  & -4\% & 0.78  \\ \cline{3-5} 
                          &                                           & 400  & -9\% & 0.50  \\ \hline \hline

\multirow{4}{*}{MNIST}    & \multirow{4}{*}{\rotatebox{90}{LeNet-5}}  & 100  & \cellcolor{gray!25}\textbf{-27\%} &  0.04   \\ \cline{3-5} 

                         &                                           & 200  & 4\% & 0.75   \\ \cline{3-5}
                         &                                           & 300  & -0.2\% & 0.98   \\ \cline{3-5} 
                       &                                           & 400  & -10\% & 0.46   \\ \hline 

\end{tabular}
\caption{\color{black} Correlation between faults in subsets and clusters in the entire test set}
\label{Tab:Fault_Data_distributions}
\end{table}

\begin{table}[]
\resizebox{0.449\textwidth}{!}{
\tiny
\begin{tabular}{|c|c|c|c||c|c|}
\hline
\hline
Dataset                    & Model                              & Metric               & Test Set Size & Spearman & P-value \\ \hline \hline
\multirow{45}{*}{Cifar-10} & \multirow{45}{*}{\rotatebox{90}{12-layer ConvNet}} & \multirow{5}{*}{GD}  & 100  & \cellcolor{gray!25}\textbf{29\%} & 0.02 \\ \cline{4-6} 
                          &                                    &                      & 200  & \cellcolor{gray!25}\textbf{32\%} & 0.01   \\ \cline{4-6} 
                          &                                    &                      & 300  & \cellcolor{gray!25}\textbf{25\%} & 0.05 \\ \cline{4-6} 
                          &                                    &                      & 400  & \cellcolor{gray!25}\textbf{31\%} & 0.02 \\ \cline{4-6}
                          &                                    &                      & 1000  & \cellcolor{gray!25}\textbf{29\%} & 0.02 \\
                          \cline{3-6}
                          &                                    & \multirow{4}{*}{STD} & 100  & \cellcolor{gray!25}\textbf{26\%}              & 0.05                 \\ \cline{4-6} 
                          &                                    &                      & 200  &   \cellcolor{gray!25}\textbf{26\%}              &         0.05      \\ \cline{4-6} 
                          &                                    &                      & 300  &  19\%             &     0.14     \\ \cline{4-6} 
                          &                                    &                      & 400  &   21\%            &   0.11             \\ \cline{4-6}  &                                    &                      & 1000  & \cellcolor{gray!25}\textbf{33\%} & 0.01  \\
                          \cline{3-6}
                          &                                    & \multirow{4}{*}{LSC} & 100  & 8\%           & 0.53          \\ \cline{4-6} 
                          &                                    &                      & 200  & 4\%          & 0.74           \\ \cline{4-6} 
                          &                                    &                      & 300  & 0.5\%         & 0.97          \\ \cline{4-6} 
                          &                                    &                      & 400  & 5\%           & 0.70           \\ \cline{4-6} &                                    &                      & 1000  & -5\% & 0.70 \\
                          \cline{3-6}
                          &                                    & \multirow{4}{*}{DSC} & 100  & 2\%           & 0.85           \\ \cline{4-6} 
                          &                                    &                      & 200  & 18\% & 0.18           \\ \cline{4-6} 
                          &                                    &                      & 300  & 3\%           & 0.80          \\ \cline{4-6} 
                          &                                    &                      & 400  & -8\%          & 0.55     \\\cline{4-6} &                                    &                      & 1000  & 24\% & 0.07   \\
                          \cline{3-6} 
                          &                                    & \multirow{4}{*}{\added{NC}} & \color{black}100  & \color{black}-22\%           &   \color{black}0.10         \\ \cline{4-6} 
                          &                                    &                      & \color{black}200  & \color{black}5.3\% &     \color{black}0.69       \\ \cline{4-6} 
                          &                                    &                      & \color{black}300  & \color{black}0.3\% &      \color{black}0.98   \\ \cline{4-6} 
                          &                                    &                      & \color{black}400  & \color{black}22\% &       \color{black}0.10    \\\cline{4-6} &                                    &    & \color{black}1000
                          &            \color{black}14\% &      \color{black}0.28       \\
                          \cline{3-6}
                     
                          &                                    &      \color{black}\multirow{4}{*}{KMNC} &      \color{black}100  & \color{black}0.51\%     &     \color{black}0.97       \\ \cline{4-6} 
                          &                                    &                      & \color{black}200  & \color{black}14\%           &   \color{black}0.29        \\ \cline{4-6} 
                          &                                    &                      & \color{black}300  & \color{black}12\%           &    \color{black}0.35      \\ \cline{4-6} 
                          &                                    &                      & \color{black}400  & \color{black}-4\% &           \color{black}0.76      \\\cline{4-6} &                                    &                      & \color{black}1000            &   \color{black}19\% &  \color{black}0.15  \\
                          \cline{3-6}
                          &                                    & \multirow{4}{*}{\color{black}NBC} & \color{black}100  & \color{black}15\%  &   \color{black}0.27          \\ \cline{4-6} 
                          &                                    &                      & \color{black}200  & \color{black}5.6\% &    \color{black}0.67        \\ \cline{4-6} 
                          &                                    &                      & \color{black}300  & \color{black}7\%   &      \color{black}0.58  \\ \cline{4-6} 
                          &                                    &                      & \color{black}400  & \color{black}-0.6\% &      \color{black}0.96      \\\cline{4-6} &                                    &                      & \color{black}1000  & \color{black}11\% & \color{black}0.40  \\
                          \cline{3-6}
                          &                                    & \multirow{4}{*}{\color{black}TKNC} & \color{black}100  & \color{black}15\%           & \color{black}0.27            \\ \cline{4-6} 
                          &                                    &                      & \color{black}200  & \color{black}\cellcolor{gray!25}\textbf{36\%} &       \color{black}0.005     \\ \cline{4-6} 
                          &                                    &                      & \color{black}300  & \color{black}\cellcolor{gray!25}\textbf{28\%}         &     \color{black}0.031        \\ \cline{4-6} 
                          &                                    &                      & \color{black}400  & \color{black}\cellcolor{gray!25}\textbf{27\%}          &     \color{black}0.04       \\\cline{4-6} &                                    &                      & \color{black}1000  & \color{black}0.2\% & \color{black}0.98   \\
                          \cline{3-6}
                          &                                    & \color{black}\multirow{4}{*}{SNAC} & \color{black}100  & \color{black}16\%           &      \color{black}0.22      \\ \cline{4-6} 
                          &                                    &                      & \color{black}200  & \color{black}5\% &      \color{black}0.70      \\ \cline{4-6} 
                          &                                    &                      & \color{black}300  & \color{black}8\%           &   \color{black}0.52        \\ \cline{4-6} 
                          &                                    &                      & \color{black}400  & \color{black}-2\%          &    \color{black}0.89        \\\cline{4-6} &                                    &                      & \color{black}1000  & \color{black}15\% &  \color{black}0.24 \\
                          \cline{3-6} \hline \hline
\multirow{45}{*}{MNIST}    & \multirow{45}{*}{\rotatebox{90}{LeNet-5}}           & \multirow{4}{*}{GD}  & 100  & \cellcolor{gray!25}\textbf{34\%} & 0.009 \\ \cline{4-6} 
                          &                                    &                      & 200  & \cellcolor{gray!25}\textbf{26\%} & 0.04 \\ \cline{4-6} 
                          &                                    &                      & 300  & \cellcolor{gray!25}\textbf{33\%} & 0.01\\ \cline{4-6} 
                          &                                    &                      & 400  & \cellcolor{gray!25}\textbf{37\%} & 0.004\\ \cline{4-6} &                                    &                      & 1000  & \cellcolor{gray!25}\textbf{35\%} & 0.005 \\
                          \cline{3-6}
                          &                                    & \multirow{4}{*}{STD} & 100  &     6\%          &        0.67       \\ \cline{4-6} 
                          &                                    &                      & 200  &   \cellcolor{gray!25}\textbf{26\%}            &    0.04         \\ \cline{4-6} 
                          &                                    &                      & 300  & \cellcolor{gray!25}\textbf{34\%}              &  0.01              \\ \cline{4-6} 
                          &                                    &                      & 400  &  23\%             &    0.07    \\ \cline{4-6} &                                    &                      & 1000  & 13\% & 0.34  \\
                          \cline{3-6}
                          &                                    & \multirow{4}{*}{LSC} & 100  & 28\% & 0.83 \\ \cline{4-6} 
                          &                                    &                      & 200  & 12\%          & 0.36            \\ \cline{4-6} 
                          &                                    &                      & 300  & 24\%          & 0.07          \\ \cline{4-6} 
                          &                                    &                      & 400  & \cellcolor{gray!25}\textbf{32\%} & 0.01  \\ \cline{4-6} &                                    &                      & 1000  & 19\% & 0.14  \\
                          \cline{3-6}
                          &                                    & \multirow{4}{*}{DSC} & 100  & 3\%          & 0.80           \\ \cline{4-6} 
                          &                                    &                      & 200  & -10\%         & 0.42        \\ \cline{4-6} 
                          &                                    &                      & 300  & 19\%          & 0.16        \\ \cline{4-6} 
                          &                                    &                      & 400  & 30\% & 0.33 \\ \cline{4-6} &                                    &                      & 1000 &   8\% &  0.53 \\
                          \cline{3-6} 
                          
                          &                                    & \color{black}\multirow{4}{*}{NC} & \color{black}100  & \color{black}-7\%           &   \color{black}0.58         \\ \cline{4-6} 
                          &                                    &                      & \color{black}200  & \color{black}2\% &      \color{black}0.90      \\ \cline{4-6} 
                          &                                    &                      & \color{black}300  & \color{black}2\%           &   \color{black}0.85        \\ \cline{4-6} 
                          &                                    &                      & \color{black}400  &  \color{black}\cellcolor{gray!25}\textbf{29\%}          &       \color{black}0.03     \\\cline{4-6} &                                    &                      & \color{black}1000  &  \color{black}\cellcolor{gray!25}\textbf{27\%} & \color{black}0.03  \\
                          \cline{3-6} 
                          &                                    & \color{black}\multirow{4}{*}{KMNC} & \color{black}100  & \color{black}-3\%           &            \color{black}0.83\\ \cline{4-6} 
                          &                                    &                      & \color{black}200  & \color{black}5\% &    \color{black}0.69        \\ \cline{4-6} 
                          &                                    &                      & \color{black}300  & \color{black}-13\%           &       \color{black}0.34    \\ \cline{4-6} 
                          &                                    &                      & \color{black}400  &\color{black} 11\%          &     \color{black}0.40       \\\cline{4-6} &                                    &                      & \color{black}1000  & \color{black}19\% & \color{black}0.14  \\
                          \cline{3-6}
                          &                                    & \color{black}\multirow{4}{*}{NBC} & \color{black}100  & \color{black}2\%           &    \color{black}0.88        \\ \cline{4-6} 
                          &                                    &                      & \color{black}200  & \color{black}-22\% &    \color{black} 0.09       \\ \cline{4-6} 
                          &                                    &                      & \color{black}300  & \color{black}9\%           &    \color{black}0.53       \\ \cline{4-6} 
                          &                                    &                      & \color{black}400  & \color{black}-2\%          &    \color{black}0.87        \\\cline{4-6} &                                    &                      & \color{black}1000  &  \color{black}\cellcolor{gray!25}\textbf{31\%} & \color{black}0.0006  \\
                          \cline{3-6}
                          &                                    & \color{black}\multirow{4}{*}{TKNC} & \color{black}100  & \color{black}27\%           &    \color{black}0.06        \\ \cline{4-6} 
                          &                                    &                      & \color{black}200  & \color{black}22\% &      \color{black}0.08      \\ \cline{4-6} 
                          &                                    &                      & \color{black}300  & \color{black}17\%           &   \color{black}0.19        \\ \cline{4-6} 
                          &                                    &                      & \color{black}400  & \color{black}13\%          &     \color{black}0.31       \\\cline{4-6} &                                    &                      & \color{black}1000  &  \color{black}\cellcolor{gray!25}\textbf{30\%} & \color{black}0.02  \\
                          \cline{3-6}
                          &                                    & \color{black}\multirow{4}{*}{SNAC} & \color{black}100  & \color{black}5\%           &    \color{black}0.72        \\ \cline{4-6} 
                          &                                    &                      & \color{black}200  & \color{black}\cellcolor{gray!25}\textbf{-29\%}  &   \color{black}0.02         \\ \cline{4-6} 
                          &                                    &                      & \color{black}300  & \color{black}3\%           &    \color{black}0.80       \\ \cline{4-6} 
                          &                                    &                      & \color{black}400  & \color{black}-10\%          &     \color{black}0.45       \\\cline{4-6} &                                    &                      & \color{black}1000  &  \color{black}\cellcolor{gray!25}\textbf{32\%} &  \color{black}0.0004 \\
                          \cline{3-6} \hline \hline

\end{tabular}}
\centering
\caption{\color{black} Correlation results between test criteria and DNN faults. The grey boxes refer to statistically significant correlations (p-value $<=0.05)$}
\label{Tab:Correlation1}
\end{table}


\begin{table}[]
\resizebox{0.489\textwidth}{!}{
\tiny
\begin{tabular}{|c|c|c|c||c|c|}
\hline
\hline
Dataset                    & Model                              & Metric               & Test Set Size & Spearman & P-value \\ \hline \hline

\multirow{45}{*}{MNIST}    & \multirow{45}{*}{\rotatebox{90}{LeNet-1}}           & \multirow{4}{*}{GD}  & 100  & \cellcolor{gray!25}\textbf{33\%} & 0.01   \\ \cline{4-6} 
                          &                                    &                      & 200  & \cellcolor{gray!25}\textbf{39\%} & 0.002  \\ \cline{4-6} 
                          &                                    &                      & 300  & \cellcolor{gray!25}\textbf{28\%} & 0.04  \\ \cline{4-6} 
                          &                                    &                      & 400  & \cellcolor{gray!25}\textbf{33\%} & 0.01  \\ \cline{4-6} &                                    &                      & 1000  & \cellcolor{gray!25}\textbf{29\%} & 0.03 \\
                          \cline{3-6}
                          &                                    & \multirow{4}{*}{STD} & 100  &     6\%          &        0.67        \\ \cline{4-6} 
                          &                                    &                      & 200  &   \cellcolor{gray!25}\textbf{26\%}            &    0.04               \\ \cline{4-6} 
                          &                                    &                      & 300  &   20\%         &   0.15             \\ \cline{4-6} 
                          &                                    &                      & 400  &  12\%            &   0.36          \\ \cline{4-6} &                                    &                      & 1000  & 16\% & 0.21  \\
                          \cline{3-6} 
                          &                                    & \multirow{4}{*}{LSC} & 100  & -6\% & 0.62 \\ \cline{4-6} 
                          &                                    &                      & 200  & 23\%          & 0.08     \\ \cline{4-6} 
                          &                                    &                      & 300  & 18\%  & 0.19    \\ \cline{4-6} 
                          &                                    &                      & 400  & 12\%  & 0.35 \\ \cline{4-6} &                                    &                      & 1000  & 16\% & 0.22 \\
                          \cline{3-6}
                          &                                    & \multirow{4}{*}{DSC} & 100  & 13\%          & 0.33           \\ \cline{4-6} 
                          &                                    &                      & 200  & -21\%         & 0.10    \\ \cline{4-6} 
                          &                                    &                      & 300  &   -25\%    &   0.07    \\ \cline{4-6} 
                          &                                    &                      & 400  & 17\%    &   0.19    \\ \cline{4-6} &                                    &                      & 1000  & 6\% & 0.67  \\
                          \cline{3-6}
                          &                                    & \multirow{4}{*}{\added{NC}} & \added{100}  & \added{6\%}           &    \added{0.67}        \\ \cline{4-6} 
                          &                                    &                             & \added{200}  & \added{22\%}           &    \added{0.10}               \\ \cline{4-6} 
                          &                                    &                             & \added{300}  & \added{24\%}           &    \added{0.08}            \\ \cline{4-6} 
                          &                                    &                             & \added{400}  & \added{-6\%}           &    \added{0.67}              \\\cline{4-6} &                                    &                                  & \added{1000}  & \added{3\%}          &    \added{0.82}      \\
                          \cline{3-6} 
                          &                                    & \multirow{4}{*}{\added{KMNC}} & \added{100}  & \added{22\%}           &    \added{0.10}              \\ \cline{4-6} 
                          &                                    &                               & \added{200}  & \added{13\%}           &    \added{0.32}             \\ \cline{4-6} 
                          &                                    &                               & \added{300}  & \added{2\%}           &    \added{0.86}             \\ \cline{4-6} 
                          &                                    &                               & \added{400}  & \added{-3\%}           &    \added{0.84}             \\\cline{4-6} &                                    &                                   & \added{1000}  & \added{19\%}          &    \added{0.15 }   \\
                          \cline{3-6}
                          &                                    & \multirow{4}{*}{\added{NBC}} & \added{100}  & \cellcolor{gray!25}\added{\textbf{-30\%}}           &    \added{0.02}               \\ \cline{4-6} 
                          &                                    &                              & \added{200}  & \cellcolor{gray!25}\added{\textbf{32\%}}           &    \added{0.01}           \\ \cline{4-6} 
                          &                                    &                              & \added{300}  & \added{20\%}           &    \added{0.15}          \\ \cline{4-6} 
                          &                                    &                              & \added{400}  & \added{8\%}           &    \added{0.52}            \\\cline{4-6} &                                    &                                    & \added{1000} & \added{-13\%}           &    \added{0.33 }     \\
                          \cline{3-6}
                          &                                    & \multirow{4}{*}{\added{TKNC}} & \added{100}  & \added{10\%}         &    \added{0.46}              \\ \cline{4-6} 
                          &                                    &                                & \added{200}  & \added{2\%}         &    \added{0.89}            \\ \cline{4-6} 
                          &                                    &                                & \added{300}  & \added{2\%}         &    \added{0.88}          \\ \cline{4-6} 
                          &                                    &                                & \added{400}  & \added{19\%}         &    \added{0.15}            \\\cline{4-6} &                                    &                                      & \added{1000} & \added{13\%}           &    \added{0.31 }   \\
                          \cline{3-6}
                          &                                    & \multirow{4}{*}{\added{SNAC}} & \added{100}  & \cellcolor{gray!25}\added{\textbf{-28\%}}           &    \added{0.03}             \\ \cline{4-6} 
                          &                                    &                               & \added{200}  & \cellcolor{gray!25}\added{\textbf{27\%}}           &    \added{0.03}         \\ \cline{4-6} 
                          &                                    &                               & \added{300}  & \added{25\%}           &    \added{0.07}        \\ \cline{4-6} 
                          &                                    &                               & \added{400}  & \added{4\%}           &    \added{0.76}             \\\cline{4-6} &                                    &                                   & \added{1000} & \added{-13\%}           &    \added{0.32}     \\
                          \cline{3-6} \hline \hline

\multirow{45}{*}{\added{Fashion-MNIST}}    & \multirow{45}{*}{\rotatebox{90}{\added{LeNet-4}}}           & \multirow{4}{*}{\added{GD}} & \added{100}  & \cellcolor{gray!25}\added{\textbf{31\%}}           &    \added{0.02 }\\ \cline{4-6} 
                          &                                    &                             & \added{200}  & \cellcolor{gray!25}\added{\textbf{32\%}}           &    \added{0.01 }               \\ \cline{4-6} 
                          &                                    &                             & \added{300}  & \cellcolor{gray!25}\added{\textbf{30\%}}           &    \added{0.02 }            \\ \cline{4-6} 
                          &                                    &                             & \added{400}  & \cellcolor{gray!25}\added{\textbf{26\%}}           &    \added{0.05 }              \\\cline{4-6} &                                    &                                  & \added{1000}  & \cellcolor{gray!25}\added{\textbf{28\%}}   &    \added{0.03 }      \\
                          \cline{3-6}
                          &                                    & \multirow{4}{*}{\added{STD}} & \added{100}  & \added{10\%}           &    \added{0.48 }        \\ \cline{4-6} 
                          &                                    &                             & \added{200}  & \added{10\%}           &    \added{0.43}               \\ \cline{4-6} 
                          &                                    &                             & \added{300}  & \added{3\%}           &    \added{0.82 }            \\ \cline{4-6} 
                          &                                    &                             & \added{400}  & \added{9\%}           &    \added{0.48 }              \\\cline{4-6} &                                    &                                  & \added{1000}  & \added{18\%}          &    \added{0.18}      \\
                          \cline{3-6} 
                          &                                    & \multirow{4}{*}{\added{LSC}} & \added{100}  & \added{14\%}           &    \added{0.31 }              \\ \cline{4-6} 
                          &                                    &                               & \added{200}  & \added{10\%}           &    \added{0.44 }             \\ \cline{4-6} 
                          &                                    &                               & \added{300}  & \added{11\%}           &    \added{0.42}             \\ \cline{4-6} 
                          &                                    &                               & \added{400}  & \added{9\%}           &    \added{0.48 }             \\\cline{4-6} &                                    &                                   & \added{1000}  & \added{7\%}          &    \added{0.61}   \\
                          \cline{3-6}
                          &                                    & \multirow{4}{*}{\added{DSC}} & \added{100}  & \added{15\%}           &    \added{0.28 }               \\ \cline{4-6} 
                          &                                    &                              & \added{200}  & \added{-19\%}           &    \added{0.14}           \\ \cline{4-6} 
                          &                                    &                              & \added{300}  & \added{-12\%}           &    \added{0.37}          \\ \cline{4-6} 
                          &                                    &                              & \added{400}  & \added{-14\%}           &    \added{0.28}            \\\cline{4-6} &                                    &                                    & \added{1000} & \added{8\%}           &    \added{0.55}     \\
                          \cline{3-6}
                          &                                    & \multirow{4}{*}{\added{NC}} & \added{100}  & \added{-6\%}           &    \added{0.66}        \\ \cline{4-6} 
                          &                                    &                             & \added{200}  & \added{25\%}           &    \added{0.06}               \\ \cline{4-6} 
                          &                                    &                             & \added{300}  & \added{6\%}           &    \added{0.63}            \\ \cline{4-6} 
                          &                                    &                             & \added{400}  & \added{22\%}           &    \added{0.10}              \\\cline{4-6} &                                    &                                  & \added{1000}  & \added{10\%}          &    \added{0.46}      \\
                          \cline{3-6} 
                          &                                    & \multirow{4}{*}{\added{KMNC}} & \added{100}  & \added{18\%}           &    \added{0.18 }              \\ \cline{4-6} 
                          &                                    &                               & \added{200}  & \added{20\%}           &    \added{0.13}             \\ \cline{4-6} 
                          &                                    &                               & \added{300}  & \cellcolor{gray!25} \added{\textbf{36\%}}           &    \added{0.01}             \\ \cline{4-6} 
                          &                                    &                               & \added{400}  & \cellcolor{gray!25}\added{\textbf{27\%}}           &    \added{0.03 }             \\\cline{4-6} &                                    &                                   & \added{1000}  & \added{24\%}          &    \added{0.07 }   \\
                          \cline{3-6}
                          &                                    & \multirow{4}{*}{\added{NBC}} & \added{100}  & \added{-0.2\%}           &    \added{0.99}               \\ \cline{4-6} 
                          &                                    &                              & \added{200}  & \added{2\%}           &    \added{0.88}           \\ \cline{4-6} 
                          &                                    &                              & \added{300}  & \added{15\%}           &    \added{0.24 }          \\ \cline{4-6} 
                          &                                    &                              & \added{400}  & \added{1\%}           &    \added{0.93}            \\\cline{4-6} &                                    &                                    & \added{1000} & \added{25\%}           &    \added{0.06}     \\
                          \cline{3-6}
                          &                                    & \multirow{4}{*}{\added{TKNC}} & \added{100}  & \added{22\%}         &    \added{0.09 }              \\ \cline{4-6} 
                          &                                    &                                & \added{200}  & \added{13\%}         &    \added{0.31 }            \\ \cline{4-6} 
                          &                                    &                                & \added{300}  & \added{18\%}         &    \added{0.18 }          \\ \cline{4-6} 
                          &                                    &                                & \added{400}  & \added{6\%}         &    \added{0.66 }            \\\cline{4-6} &                                    &                                      & \added{1000} & \added{13\%}           &    \added{0.32}   \\
                          \cline{3-6}
                          &                                    & \multirow{4}{*}{\added{SNAC}} & \added{100}  & \added{1\%}           &    \added{0.96 }             \\ \cline{4-6} 
                          &                                    &                               & \added{200}  & \added{2\%}           &    \added{0.86 }         \\ \cline{4-6} 
                          &                                    &                               & \added{300}  & \added{16\%}           &    \added{0.22 }        \\ \cline{4-6} 
                          &                                    &                               & \added{400}  & \added{1\%}           &    \added{0.95 }             \\\cline{4-6} &                                    &                                   & \added{1000} & \added{25\%}           &    \added{0.06}     \\
                          \cline{3-6} \hline \hline

\end{tabular}}
\centering
\caption{\color{black} Correlation results between test criteria and DNN faults. The grey boxes refer to statistically significant correlations (p-value $<=0.05)$}
\label{Tab:Correlation2}
\end{table}

\begin{table}[]
\resizebox{0.489\textwidth}{!}{
\tiny
\begin{tabular}{|c|c|c|c||c|c|}
\hline
\hline
Dataset                    & Model                              & Metric               & Test Set Size & Spearman & P-value \\ \hline \hline

\multirow{45}{*}{\added{Cifar-10}}    & \multirow{45}{*}{\rotatebox{90}{\added{ResNet20}}}           &\multirow{4}{*}{\added{GD}} & \added{100}  & \added{\cellcolor{gray!25}\textbf{26\%}}   & \added{0.05 }\\ \cline{4-6} 
                          &                                    &                             & \added{200}  & \added{\cellcolor{gray!25}\textbf{31\%}}           &    \added{0.01 }               \\ \cline{4-6} 
                          &                                    &                             & \added{300}  & \added{\cellcolor{gray!25}\textbf{37\%}}           &    \added{0.004 }            \\ \cline{4-6} 
                          &                                    &                             & \added{400}  & \added{\cellcolor{gray!25}\textbf{29\%}}           &    \added{0.03 }              \\\cline{4-6} &                                    &                                  & \added{1000}  & \added{\cellcolor{gray!25}\textbf{28\%}}          &    \added{0.03 }      \\
                          \cline{3-6}
                          &                                    & \multirow{4}{*}{\added{STD}} & \added{100}  & \added{16\%}           &    \added{0.22 }        \\ \cline{4-6} 
                          &                                    &                             & \added{200}  & \added{\cellcolor{gray!25}\textbf{28\%}}           &    \added{0.03}               \\ \cline{4-6} 
                          &                                    &                             & \added{300}  & \added{\cellcolor{gray!25}\textbf{34\%}}           &    \added{0.01}            \\ \cline{4-6} 
                          &                                    &                             & \added{400}  & \added{20\%}           &    \added{0.13 }              \\\cline{4-6} &                                    &                                  & \added{1000}  & \added{\cellcolor{gray!25}\textbf{28\%}}          &    \added{0.03 }      \\
                          \cline{3-6} 
                          &                                    & \multirow{4}{*}{\added{LSC}} & \added{100}  & \added{16\%}           &    \added{0.24}              \\ \cline{4-6} 
                          &                                    &                               & \added{200}  & \added{\cellcolor{gray!25}\textbf{33\%}}           &    \added{0.01 }             \\ \cline{4-6} 
                          &                                    &                               & \added{300}  & \added{-10\%}           &    \added{0.43 }             \\ \cline{4-6} 
                          &                                    &                               & \added{400}  & \added{9\%}           &    \added{0.50}             \\\cline{4-6} &                                    &                                   & \added{1000}  & \added{-6\%}          &    \added{0.63 }   \\
                          \cline{3-6}
                          &                                    & \multirow{4}{*}{\added{DSC}} & \added{100}  & \added{-3\%}           &    \added{0.82}               \\ \cline{4-6} 
                          &                                    &                              & \added{200}  & \added{-14\%}           &    \added{0.28}           \\ \cline{4-6} 
                          &                                    &                              & \added{300}  & \added{1\%}           &    \added{0.98 }          \\ \cline{4-6} 
                          &                                    &                              & \added{400}  & \added{9\%}           &    \added{0.48 }            \\\cline{4-6} &                                    &                                    & \added{1000} & \added{18\%}           &    \added{0.16 }     \\
                          \cline{3-6}
                          &                                    & \multirow{4}{*}{\added{NC}} & \added{100}  & \added{12\%}           &    \added{0.38 }        \\ \cline{4-6} 
                          &                                    &                             & \added{200}  & \added{16\%}           &    \added{0.23 }               \\ \cline{4-6} 
                          &                                    &                             & \added{300}  & \added{-6\%}           &    \added{0.64 }            \\ \cline{4-6} 
                          &                                    &                             & \added{400}  & \added{6\%}           &    \added{0.66 }              \\\cline{4-6} &                                    &                                  & \added{1000}  & \added{7\%}          &    \added{0.60}      \\
                          \cline{3-6} 
                          &                                    & \multirow{4}{*}{\added{KMNC}} & \added{100}  & \added{4\%}           &    \added{0.79}              \\ \cline{4-6} 
                          &                                    &                               & \added{200}  & \added{13\%}           &    \added{0.32 }             \\ \cline{4-6} 
                          &                                    &                               & \added{300}  & \added{18\%}           &    \added{0.17 }             \\ \cline{4-6} 
                          &                                    &                               & \added{400}  & \added{\cellcolor{gray!25}\textbf{26\%}}           &    \added{0.05}             \\\cline{4-6} &                                    &                                   & \added{1000}  & \added{\cellcolor{gray!25}\textbf{33\%}}          &    \added{0.01}   \\
                          \cline{3-6}
                          &                                    & \multirow{4}{*}{\added{NBC}} & \added{100}  & \added{13\%}           &    \added{0.31 }               \\ \cline{4-6} 
                          &                                    &                              & \added{200}  & \added{13\%}           &    \added{0.32 }           \\ \cline{4-6} 
                          &                                    &                              & \added{300}  & \added{15\%}           &    \added{0.25 }          \\ \cline{4-6} 
                          &                                    &                              & \added{400}  & \added{12\%}           &    \added{0.37 }            \\\cline{4-6} &                                    &                                    & \added{1000} & \added{7\%}           &    \added{0.61}     \\
                          \cline{3-6}
                          &                                    & \multirow{4}{*}{\added{TKNC}} & \added{100}  & \added{-1\%}         &    \added{0.96}              \\ \cline{4-6} 
                          &                                    &                                & \added{200}  & \added{1\%}         &    \added{0.95 }            \\ \cline{4-6} 
                          &                                    &                                & \added{300}  & \added{16\%}         &    \added{0.23 }          \\ \cline{4-6} 
                          &                                    &                                & \added{400}  & \added{22\%}         &    \added{0.10 }            \\\cline{4-6} &                                    &                                      & \added{1000} & \added{1\%}           &    \added{0.93}   \\
                          \cline{3-6}
                          &                                    & \multirow{4}{*}{\added{SNAC}} & \added{100}  & \added{16\%}           &    \added{0.24}             \\ \cline{4-6} 
                          &                                    &                               & \added{200}  & \added{13\%}           &    \added{0.34}         \\ \cline{4-6} 
                          &                                    &                               & \added{300}  & \added{19\%}           &    \added{0.15 }        \\ \cline{4-6} 
                          &                                    &                               & \added{400}  & \added{11\%}           &    \added{0.40 }             \\\cline{4-6} &                                    &                                   & \added{1000} & \added{11\%}           &    \added{0.41}     \\
                          \cline{3-6} \hline \hline

\multirow{45}{*}{\added{Fashion-MNIST}}    & \multirow{45}{*}{\rotatebox{90}{\added{LeNet-4}}}           & \multirow{4}{*}{\added{GD}} & \added{100}  & \cellcolor{gray!25}\added{\textbf{27\%}} &

\added{0.04 }\\ \cline{4-6} 
                          &                                    &                             & \added{200}  & \cellcolor{gray!25}\added{\textbf{27\%}}           &    \added{0.03}               \\ \cline{4-6} 
                          &                                    &                             & \added{300}  & \cellcolor{gray!25}\added{\textbf{27\%}}           &    \added{0.04}            \\ \cline{4-6} 
                          &                                    &                             & \added{400}  & \cellcolor{gray!25}\added{\textbf{33\%}}           &    \added{0.01}              \\\cline{4-6} &                                    &                                  & \added{1000}  & \cellcolor{gray!25}\added{\textbf{30\%}}          &    \added{0.02}      \\
                          \cline{3-6}
                          &                                    & \multirow{4}{*}{\added{STD}} & \added{100}  & \cellcolor{gray!25}\added{\textbf{27\%}}           &    \added{0.04}        \\ \cline{4-6} 
                          &                                    &                             & \added{200}  & \added{16\%}           &    \added{0.21}               \\ \cline{4-6} 
                          &                                    &                             & \added{300}  & \added{20\%}           &    \added{0.13 }            \\ \cline{4-6} 
                          &                                    &                             & \added{400}  & \added{23\%}           &    \added{0.08}              \\\cline{4-6} &                                    &                                  & \added{1000}  & \added{6\%}          &    \added{0.65}      \\
                          \cline{3-6} 
                          &                                    & \multirow{4}{*}{\added{LSC}} & \added{100}  & \added{-5\%}           &    \added{0.72 }              \\ \cline{4-6} 
                          &                                    &                               & \added{200}  & \added{20\%}           &    \added{0.12 }             \\ \cline{4-6} 
                          &                                    &                               & \added{300}  & \added{10\%}           &    \added{0.45 }             \\ \cline{4-6} 
                          &                                    &                               & \added{400}  & \cellcolor{gray!25}\added{\textbf{27\%}}           &    \added{0.04}             \\\cline{4-6} &                                    &                                   & \added{1000}  & \added{3\%}          &    \added{0.83 }   \\
                          \cline{3-6}
                          &                                    & \multirow{4}{*}{\added{DSC}} & \added{100}  & \added{-17\%}           &    \added{0.20}               \\ \cline{4-6} 
                          &                                    &                              & \added{200}  & \added{-9\%}           &    \added{0.50 }           \\ \cline{4-6} 
                          &                                    &                              & \added{300}  & \added{-5\%}           &    \added{0.69 }          \\ \cline{4-6} 
                          &                                    &                              & \added{400}  & \added{4\%}           &    \added{0.75 }            \\\cline{4-6} &                                    &                                    & \added{1000} & \added{8\%}           &    \added{0.57 }     \\
                          \cline{3-6}
                          &                                    & \multirow{4}{*}{\added{NC}} & \added{100}  & \added{9\%}           &    \added{0.49 }        \\ \cline{4-6} 
                          &                                    &                             & \added{200}  & \added{6\%}           &    \added{0.64 }               \\ \cline{4-6} 
                          &                                    &                             & \added{300}  & \added{-2\%}           &    \added{0.89 }            \\ \cline{4-6} 
                          &                                    &                             & \added{400}  & \added{-6\%}           &    \added{0.67 }              \\\cline{4-6} &                                    &                                  & \added{1000}  & \added{-6\%}          &    \added{0.63 }      \\
                          \cline{3-6} 
                          &                                    & \multirow{4}{*}{\added{KMNC}} & \added{100}  & \added{15\%}           &    \added{0.25 }              \\ \cline{4-6} 
                          &                                    &                               & \added{200}  & \added{22\%}           &    \added{0.09 }             \\ \cline{4-6} 
                          &                                    &                               & \added{300}  & \added{-2\%}           &    \added{0.89 }             \\ \cline{4-6} 
                          &                                    &                               & \added{400}  & \cellcolor{gray!25}\added{\textbf{26\%}}           &    \added{0.05}             \\\cline{4-6} &                                    &                                   & \added{1000}  & \added{7\%}          &    \added{0.61 }   \\
                          \cline{3-6}
                          &                                    & \multirow{4}{*}{\added{NBC}} & \added{100}  & \added{-6\%}           &    \added{0.64 }               \\ \cline{4-6} 
                          &                                    &                              & \added{200}  & \added{13\%}           &    \added{0.31 }           \\ \cline{4-6} 
                          &                                    &                              & \added{300}  & \added{16\%}           &    \added{0.22 }          \\ \cline{4-6} 
                          &                                    &                              & \added{400}  & \added{11\%}           &    \added{0.41 }            \\\cline{4-6} &                                    &                                    & \added{1000} & \added{-6\%}           &    \added{0.64 }     \\
                          \cline{3-6}
                          &                                    & \multirow{4}{*}{\added{TKNC}} & \added{100}  & \added{14\%}         &    \added{0.29 }              \\ \cline{4-6} 
                          &                                    &                                & \added{200}  & \added{4\%}         &    \added{0.77}            \\ \cline{4-6} 
                          &                                    &                                & \added{300}  & \added{10\%}         &    \added{0.46}          \\ \cline{4-6} 
                          &                                    &                                & \added{400}  & \added{9\%}         &    \added{0.50}            \\\cline{4-6} &                                    &                                      & \added{1000} & \added{-16\%}           &    \added{0.23 }   \\
                          \cline{3-6}
                          &                                    & \multirow{4}{*}{\added{SNAC}} & \added{100}  & \added{-6\%}           &    \added{0.64 }             \\ \cline{4-6} 
                          &                                    &                               & \added{200}  & \added{13\%}           &    \added{0.31}         \\ \cline{4-6} 
                          &                                    &                               & \added{300}  & \added{-16\%}           &    \added{0.22 }        \\ \cline{4-6} 
                          &                                    &                               & \added{400}  & \added{11\%}           &    \added{0.41}             \\\cline{4-6} &                                    &                                   & \added{1000} & \added{-6\%}           &    \added{0.64}     \\
                          \cline{3-6} \hline \hline

\end{tabular}}
\caption{\color{black} Correlation results between test criteria and DNN faults. The grey boxes refer to statistically significant correlations (p-value $<=0.05)$}
\label{Tab:Correlation3}
\end{table}

We aimed to investigate whether higher diversity increases the fault detection capability of test sets. We randomly selected, with replacement, 60 samples of size n $\in$ \{100, 200, 300, 400, 1000\}. 
For each sample, we calculated the corresponding diversity scores (GD and STD) and the number of faults \Revise{(i.e. the number of covered clusters of mispredicted inputs)}. We calculated the Spearman correlation~\cite{bonett2000sample} between the diversity scores and the number of faults. The correlation results are reported in Tables~\ref{Tab:Correlation1},~\ref{Tab:Correlation2} and~\ref{Tab:Correlation3} for the different datasets and DNN models. The grey boxes in the table refer to statistically significant correlations (p-value $<=0.05)$. We chose to use the Spearman correlation because it measures the strength of a monotonic correlation between two variables, without making assumptions about the form of the relationship or data distributions~\cite{bonett2000sample}.

\added{Nonetheless, there is a potential confounding factor in the correlation between faults and diversity. If we were to apply clustering in the test dataset, we would expect higher diversity to lead to more clusters. If there is a high similarity in distribution between the entire test dataset and the subset of mispredicted inputs, the correlation between diversity and faults in subsets could be due to the presence of such a confounding factor. To verify this, we analyzed the correlation between the number of clusters in the entire test dataset that were covered and the number of faults (i.e. fault-related clusters) in subsets. A low correlation would indicate that such a threat is unlikely.}

\added{We applied the same HDBSCAN clustering process (section~\ref{Sec:Faults}) to the entire testing dataset to obtain data clusters. \Revise{We then used the previously selected 60 samples of size n $\in$ \{100, 200, 300, 400\} and calculated the number of faults (i.e. the number of covered clusters of only mispredicted inputs) and data clusters (i.e. the number of covered clusters of both correctly predicted and mispredicted inputs) inside each sample. } Finally, we calculated the Spearman correlation between the number of faults and the number of covered data clusters in subsets. We performed this experiment using 12-layer ConvNet (with Cifar-10) and LeNet-5 (with MNIST). The correlation results are reported in Table~\ref{Tab:Fault_Data_distributions}. As shown in the table, we did not find any positive and statistically significant correlation between faults and data clusters. We therefore conclude that there is no confounding factor in our correlation study between diversity and faults, thus giving us more confidence in the cause-effect relationship underlying the observed correlations.} 
\Revise{These results also suggest that correctly predicted inputs belong to separate clusters from fault-related clusters (i.e. clusters of mispredicted inputs) in general. They provide evidence that mispredicted inputs belong to the same cluster and share common characteristics that are different from the ones shared by correctly predicted inputs.}

\begin{table*}[ht!]
\color{black}
\centering
\begin{tabular}{|l|l||c|c|c|c|c|c|c|c|c|}
\hline
\hline
\textbf{Dataset}                & \textbf{Model }              & \textbf{GD} & \textbf{STD} & \textbf{LSC} & \textbf{DSC} & \textbf{NC }& \textbf{KMNC} & \textbf{NBC} & \textbf{TKNC} & \textbf{SNAC} \\ \hline \hline
MNIST       & LeNet-1   & 5/5 & 1/5 & 0/5 & 0/5 & 0/5 & 0/5 & 1/5 & 0/5 & 1/5  \\ \hline
MNIST       & LeNet-5   & 5/5 & 2/5 & 1/5 & 0/5 & 2/5 & 0/5 & 1/5 & 1/5 & 1/5  \\ \hline
Fashion-MNIST    & LeNet-4 & 5/5 & 0/5 & 0/5 & 0/5 & 0/5 & 2/5 & 0/5 & 0/5 & 0/5  \\ \hline
Cifar-10    & 12-layer ConvNet & 5/5 & 3/5 & 0/5 & 0/5 & 0/5 & 0/5 & 0/5 & 3/5 & 0/5  \\ \hline
Cifar-10    & ResNet20  & 5/5 & 3/5 & 1/5 & 0/5 & 0/5 & 2/5 & 0/5 & 0/5 & 0/5  \\ \hline
SVHN        & LeNet-5   & 5/5 & 1/5 & 1/5 & 0/5 & 0/5 & 1/5 & 0/5 & 0/5 & 0/5  \\ \hline \hline

\multicolumn{2}{|c||}{\textbf{Total}}    & \textbf{30/30} & \textbf{10/30} & \textbf{3/30} & \textbf{0/30} & \textbf{2/30} & \textbf{5/30} & \textbf{2/30} & \textbf{4/30} & \textbf{2/30}  \\ \hline \hline
\end{tabular}
\caption{\color{black} Number of positive statistically significant correlations between testing criteria and faults}
\label{Tab:summary}
\end{table*}

In our correlation experiment between diversity and faults, we evaluated a total of 60 configurations related to diversity metrics (6 models \& datasets x 2 diversity metrics x 5 input sizes). 
\Revise{As mentioned in Tables~\ref{Tab:Correlation1},~\ref{Tab:Correlation2} and~\ref{Tab:Correlation3}, we found that GD outperforms STD in terms of fault-revealing capabilities as we observed that there was a positive, statistically significant correlation between GD and faults in all configurations (30/30)}. Furthermore, they were consistent across all the studied models, datasets and input set sizes.
On the other hand, we found that STD had a positive significant correlation with faults in only ten configurations. These results were expected because, in RQ1, GD showed better performance in measuring actual data diversity than STD.
\newline We expected to have a moderate correlation between diversity and faults because we relied on a clustering approach to approximate faults in DNNs (section~\ref{Sec:Faults}). Such correlation is expected to be higher if we have a more straightforward method to identify faults in DNNs. 

Nevertheless, the obtained results clearly indicate that GD can be used to effectively guide DNN testing by devising input sets with maximum diversity to increase their fault-revealing capabilities. Let us recall that GD also has the practical advantage of being black-box, as opposed to state-of-the-art DNN coverage metrics~\cite{pei2017deepxplore,Ma2018DeepGaugeMT,kim2019guiding,gerasimou2020importance}, which require access to the internals of DNN models or their training sets.

\begin{tcolorbox}
\textbf{Answer to RQ2:}  There is a positive and statistically significant correlation between GD and faults in DNNs. GD is more frequently correlated to faults than STD. Consequently, GD should be used as a black-box approach to guide the testing of DNN models.
\end{tcolorbox}

\subsubsection{RQ3.  How  does  coverage  relate  to  fault  detection?} \label{subsec:RQ3}

Similar to the previous section on diversity, in this research question, we aim to study the correlation between state-of-the-art coverage criteria and faults in DNNs. Our goal is to understand how they compare to diversity in this respect.
Based on three factors, we selected the following two coverage criteria: Likelihood-based Surprise Coverage (LSC) and Distance-based Surprise Coverage (DSC)~\cite{kim2019guiding}. First, we retained criteria that were recently published in the literature. We also chose those that (1) have been compared to other coverage criteria, and (2) showed better performance in guiding the testing of DNN models. 

We selected coverage metrics that we could apply and replicate on our models and datasets. The first two factors yielded four coverage metrics: Likelihood-based Surprise Coverage, Distance-based Surprise Coverage, Importance-Driven Coverage (IDC)~\cite{gerasimou2020importance} and Sign-Sign Coverage (SSC) \cite{sun2019structural}. However, we could not apply IDC and SSC on our datasets and models. We got several execution errors\footnote{Authors have been contacted but the execution errors have not been resolved.} when we tried to compute IDC on the 12-layer ConvNet and LeNet models. For SSC, we obtained different results from the original paper~\cite{sun2019structural} when we applied this metric on LeNet-1, and encountered several execution errors in the remaining models. Therefore, we excluded these metrics from our research and only studied LSC and DSC in the correlation between coverage and fault detection in DNNs.  
\added{In addition to this criteria, we included basic and widely used criteria such as Neuron Coverage (NC)~\cite{pei2017deepxplore} and DeepGauge~\cite{Ma2018DeepGaugeMT} coverage metrics. We considered the following metrics related to DeepGauge: k-Multisection Neuron Coverage (KMNC), Neuron Boundary Coverage (NBC), Top-K Neuron Coverage (TKNC) and Strong Neuron Activation Coverage (SNAC). We describe these metrics and their limitations in Section~\ref{Sec:RW}. }

To investigate the relationship between coverage and fault detection, we ran the same experiment as in RQ2 and evaluated the same selected subsets. We calculated the different coverage scores for all subsets. For LSC and DSC, we used the same recommended settings for hyperparameters (e.g. upper bound, lower bound, number of buckets) as in the original paper~\cite{kim2019guiding} and other existing papers in the literature~\cite{yan2020correlations}. We used the same hyperparameters that were recommended in the literature~\cite{Ma2018DeepGaugeMT} for NC, KMNC and TKNC. We used the activation threshold of 10\% for NC and fixed the number of buckets \textit{K} to three for TKNC and 1,000 for KMNC; this is for the different models and datasets in our experiments. We calculated the Spearman correlation between each coverage criterion and the number of faults. The results are reported in Tables~\ref{Tab:Correlation1},~\ref{Tab:Correlation2},~\ref{Tab:Correlation3} and~\ref{Tab:summary}.  

We considered 30 configurations per metric (6 models \& datasets x 5 input sizes). Further, we accounted for a total of 210 configurations related to the coverage criteria (6 models \& datasets  x 7 criteria x 5 input sizes). As depicted in Table~\ref{Tab:summary}, out of 30 different configurations per metric, the distributions of positive, statistically significant correlations to faults are as follows: 5 for KMNC, 4 for TKNC, 3 for LSC, and 2 for NC, NBC and SNAC. DSC, however, did not show any statistically significant correlation with faults in any of the datasets and models.

In general, we conclude that significant correlations between coverage and faults are rare in the configurations of models and datasets that we used. None of the studied coverage metrics consistently showed statistically significant correlations across all the models, datasets and input set sizes. For example, we found that LSC is positively correlated to faults in only three out of 30 configurations related to LeNet-5 and ResNet20. However, we did not find any statistically significant correlation for this metric with LeNet-1, LeNet-4 and 12-layer ConvNet.

\Revise{Our findings raise questions about the usefulness of the selected coverage criteria for enabling effective DNN testing in fault detection.} These results confirm, from a different angle, many recent studies~\cite{li2019structural,chen2020deep,harel2020neuron} that questioned the reliability of such coverage criteria to guide the testing of DNN models. \Revise{A central concern raised by these articles is  whether such coverage metrics relate to the model's behaviour and its decision results}. Our results suggest that this relationship is, at best, weak.

\begin{tcolorbox}
\textbf{Answer to RQ3:} 
\added{In general, significant positive correlations between coverage and faults are rarely based on the configurations and datasets used in our experiments. Coverage metrics are not a good indicator of fault detection.}  
\end{tcolorbox}

\begin{figure*}[ht]
\centering
\includegraphics[scale=0.62]{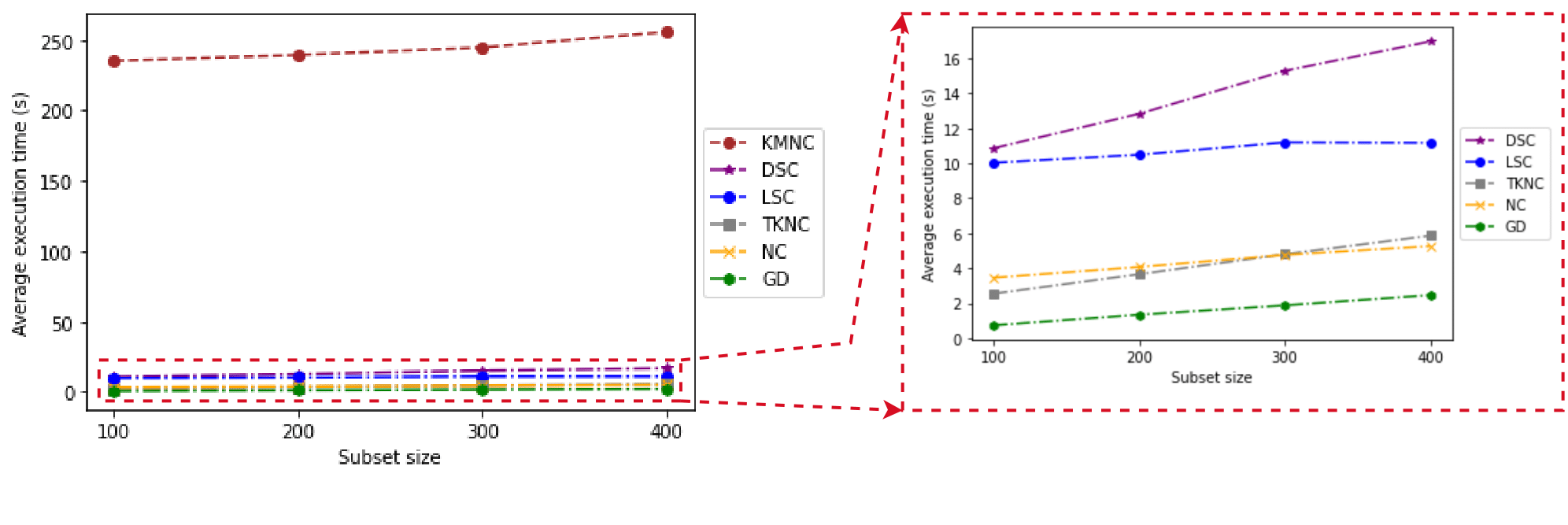}
\centering-\caption{\added{Computation time for diversity and coverage metrics with Cifar-10 and ResNet20}}
\label{fig:timeCIFAR}

\end{figure*}

\begin{figure*}[ht!]
\centering
\includegraphics[scale=0.62]{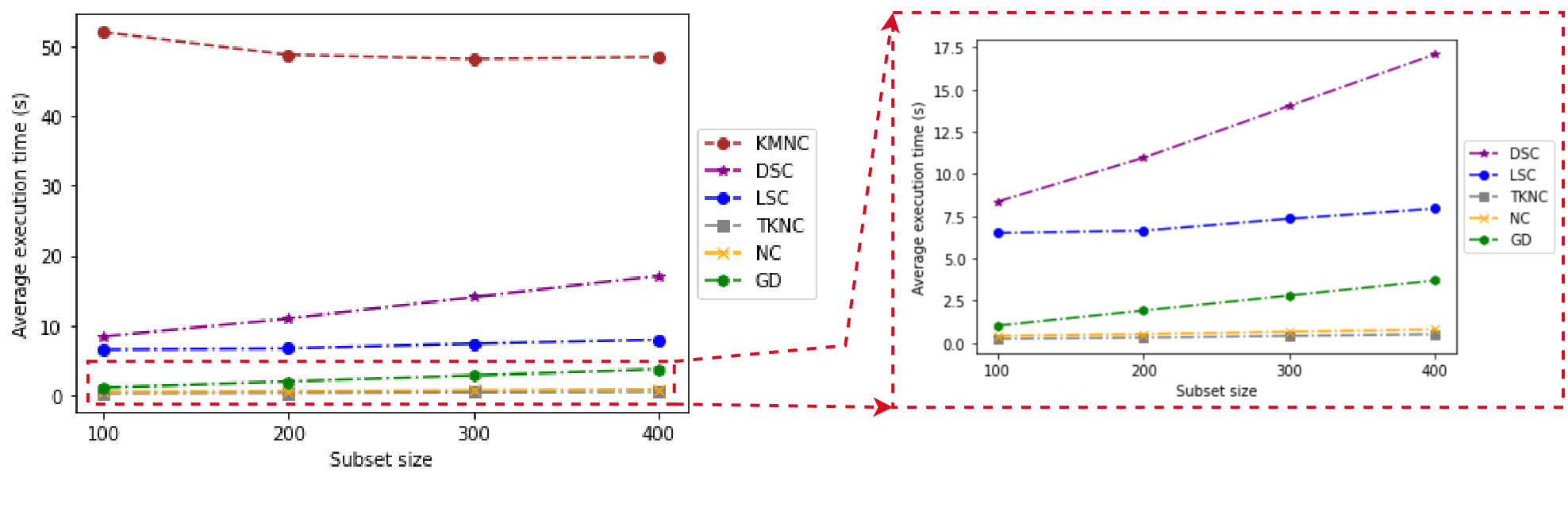}
\caption{\added{Computation time for diversity and coverage metrics with MNIST and LeNet-5}}
\label{fig:timeMNIST}
\end{figure*}

\subsubsection{RQ4. How do diversity and coverage perform in terms of computation time?} We aimed to compare the computation times of the selected diversity metrics and coverage criteria and assessed how they scaled with the sizes of test sets. For this purpose, we randomly selected, with replacement, 60 samples of size n $\in$ \{100, 200, 300, 400\}. We calculated, for each sample, diversity and coverage scores, and measured their respective computation times. Because we found in RQ2 that GD outperforms STD in correlation to faults, in the rest of this work we only used the GD metric to calculate diversity in input sets. For GD, we accounted for the sum of the following two computation times: (1) calculation of diversity based on the extracted features; and (2) the pre-processing time that is required to extract features with the VGG16 model. We report in Figures~\ref{fig:timeCIFAR} and~\ref{fig:timeMNIST} the change in computation times for ResNet20 and LeNet-5 as we increased the sizes of the input sets. We observed that for both diversity and coverage metrics, computation time is linear with test set sizes up to 400.
Both types of metrics are not computationally expensive. For example, the computation time related to diversity and coverage metrics in MNIST and LeNet-5 varies between 0.4 and 49 seconds for samples of size 400. We observed that KMNC, SNAC and NBC have the same computation time. Further, they are the most computationally expensive metrics. For example, KMNC, SNAC and NBC are approximately 25 times more computationally expensive than the other metrics in ResNet20. However, we found that GD generally outperforms most of the coverage metrics in computation time, except for NC and TKNC in LeNet-5. 
We used the Wilcoxon signed-rank test~\cite{woolson2007wilcoxon} to assess the statistical significance of the difference between GD's computation time and the other testing criteria. We found that GD statistically outperformed all coverage metrics in  computation time.
For example, it was three-to-five times faster  to compute GD than LSC and DSC. Although absolute differences are a matter of seconds, such computations, in the context of test selection or minimization, \Revise{may} be performed thousands of times and thus become practically significant.
We studied the distributions of the computation times for diversity and coverage metrics and analyzed their variations. \Revise{Due to space limitations, we included the related boxplots in our replication package~\cite{ReplicationPackage}}. We found that the distributions of the computations times related to GD showed less variation than the studied coverage metrics for samples of the same size. For example, GD showed less variation than LSC and DSC for samples of the same size because the GD metric depends on the calculation of the determinant of a fixed-size feature matrix, while LSC and DSC depend on a search mechanism for the nearest inputs in the training set. Search time may vary from one sample to another, and therefore leads to increased variance in computation time.
\newline Because GD is black-box, its computation time only depends on the used dataset and is not affected by the complexity of the DNN model (e.g. number of layers and neurons). In contrast, the computation time of white-box coverage metrics is highly sensitive to such complexity.

\begin{tcolorbox} \textbf{Answer to RQ4:} Both diversity and coverage metrics are not computationally expensive (seconds). However, GD significantly outperforms coverage criteria. In application contexts, such as test case selection and minimization, and based on searches in which we perform many test set evaluations, this difference can become practically significant. \end{tcolorbox}

\subsubsection{RQ5. How does diversity relate to coverage?}
We aimed to study the relationship between diversity and coverage to assess if diverse input sets increase the coverage of DNN models. Conversely, increasing coverage should, in theory, increase diversity. Although the results of previous research questions make it unlikely for such correlations to be strong, this needed to be investigated. 

We ran the same experiment as in RQ2 and RQ3 and used the same selected subsets. For each subset, we calculated the geometric diversity and coverage scores and measured the Spearman correlation between each pair of diversity and coverage metrics. We evaluated a total of 175 configurations (1 diversity metric x 7 coverage criteria x 5 models \& datasets x 5 test set sizes). We include all the results in Appendix~\ref{Appendix:Corr-div_coverage}, and they are available online~\cite{ReplicationPackage}. 
Out of 175 configurations, only 13 correlations were positive and statistically significant. For example, the only positive correlation (46\%) between GD and LSC was for input sets of size 400 from Cifar-10 using ResNet-20. Furthermore, the only statistically significant correlation (-35\%) between GD and DSC was for input sets of size 300 from Fashion-MNIST using LeNet-4. For NC, we found only three statistically significant correlations in three different models (LeNet-4, LeNet-5 and ResNet-20). Although we found six statistically significant correlations between KMNC and GD, these results were not consistent across all models and input set sizes. Additionally, the only statistically significant correlation (48\%) between GD and NBC was for input sets of size 1,000 from MNIST using LeNet-5. Finally, for TKNC and SNAC, we found only two statistically significant correlations for each metric with GD. To summarize, most configurations (159) show no significant correlations between diversity and coverage metrics, which suggests that, in general, diversity and coverage in DNN models are not correlated. In other words, diverse input sets do not necessarily increase the coverage of DNN models and higher coverage does not systematically lead to higher diversity. These results are also consistent with our previous observations in RQ3 and RQ4, where we found that while geometric diversity is correlated to fault detection, coverage is not.

\begin{tcolorbox}
\added{\textbf{Answer to RQ5:} In general, for most configurations there is no significant correlation between diversity and coverage in DNN models. }

\end{tcolorbox}

\section{Discussion and Recommendations}\label{Sec:Discussions}

We should note that our correlation results between testing criteria \Revise{(diversity and coverage metrics)} and faults are consistent across different datasets and DNN models. Based on our experiments, we show that studying the diversity of the features embedded in test input sets is more reliable as a basis to guide DNN testing than considering the coverage of their hidden neurons. We show that geometric diversity is potentially more effective than existing coverage metrics in guiding the testing of DNN models. \Revise{This metric requires neither knowledge of the model under test nor access to the training set. Further, it does not require execution of the test input nor reliance on the output of the DNN model under test. It is therefore a practical, black-box approach that can be used to guide the testing of DNN models. Although the results are encouraging, we only investigated geometric diversity with DNN models using images as input. Further experiments should be conducted to evaluate the performance on other input data types, such as audio and text data. We will therefore explore appropriate feature extraction models to represent new data types with feature vectors used by our diversity metrics (mainly diversity and STD metrics, because NCD supports, by default, any data type). Because diversity metrics are black-box and do not depend on the type of DNN model, we also aim in future work to consider other DNN models for regression and multi-classification tasks to further generalize our results.}

In our experiments, we were surprised to see only a few significant correlations between coverage and faults across all the models and datasets we evaluated.

We selected both widely used coverage criteria in the literature and the best coverage metrics in published results and reproducibility (section~\ref{subsec:RQ3}).
Nevertheless, coverage showed poor performance as an indicator of detected faults in DNNs. 
In traditional software, one of the potential reasons for the effectiveness of coverage criteria is that they rely on the logical structure of the system's source code. However, the decision logic in DNNs is not explicit, which makes the definition and usage of coverage criteria more challenging in DNNs. Also, in traditional software, relying on diverse test cases tends to increase code coverage and the fault-detection capabilities of test suites~\cite{biagiola2019diversity,biagiola2017search}. In contrast, we show that in DNN testing diverse test input sets do not lead to increased DNN coverage but, at least for geometric diversity, lead to more fault detection in the DNN model. 

Furthermore, traditional software systems and DNNs are fundamentally different with respect to the notion of fault and their detection. Given a test input, in general, we detect faults in software by comparing the actual test output to the expected output. If there is a mismatch, we consider this to be a failure, and we can debug the system using various fault localization techniques~\cite{zou2019empirical,pearson2017evaluating,wong2013dstar} to identify faulty statement(s). However, in DNN models, the notion of fault is elusive because of the black-box nature of DNN models. If the DNN model mispredicts an input, we consider this to be a failure, but debugging and localizing faults in the DNN causing such failure is challenging because there is no explicit and interpretable decision logic. This is also why DNNs are usually fixed through retraining~\cite{fahmy2021supporting}. 
Because it is common for many mispredicted inputs to be caused by the same problems in the DNN model~\cite{fahmy2021supporting}, and because we cannot directly identify root causes, we relied on a clustering-based approach to group similar mispredicted inputs and therefore relied on the number of these clusters to approximate fault-counting in our experiments. Our clustering relied on a density-based clustering algorithm that grouped similar mispredicted inputs based on their (image) features and their misprediction behaviour (pairs of actual and mispredicted classes). \Revise{Our fault estimation approach is therefore not ``\textit{complete}'' because we only considered faults with a sufficient number of observed mispredictions to be grouped into a cluster. The others were considered noisy inputs by our clustering approach. In other words, we obtained a good approximation of commonly occurring faults, which underestimates the total number of faults because we do not account for noisy inputs. Because it is more important for testing approaches to detect faults, leading to more frequent mispredictions, this is practically acceptable. As described in Table~\ref{tab:Cluster} and to reinforce this point, the number of noisy inputs is very small compared to the total number of mispredicted inputs. 
We acknowledge that although our retraining-based validation and evaluation results show promise, they only indirectly validate our fault model since there is no direct way to check its fault estimation accuracy. More research is needed to investigate alternative ways to enable fault detection comparisons in experiments involving DNN models. }

\Revise{Our study of the computation time of diversity and coverage metrics was generic and did not target a specific DNN testing scenario (e.g. selection, generation, minimization). However, this was intentional, as we wish to provide general insights into the computational complexity of coverage and diversity metrics. We showed that both types of metrics are not computationally intensive and that GD is generally three-to-five times faster to compute than the studied coverage metrics. However, whether such differences practically matter and to what extent depends on how frequently they are computed in a given application context. Some coverage-based test selection approaches entail computing the coverage score only once for each test input and selecting the test inputs with the highest coverage score. In contrast, in a typical diversity-based test selection approach, where the goal is to select a set of diverse test inputs, the GD score may be computed many times for selected subsets when, for example, the GD metric is used to drive a search algorithm. Finally, we aim in future work to further investigate the computation time of coverage and diversity metrics when used in specific DNN testing scenarios, such as test set selection, generation and minimization. }

\Revise{To summarize, before using any testing criteria to support a particular test scenario for DNNs (e.g. test selection, minimization and generation), one should investigate the correlation between the targeted testing criteria and faults. This is our main motivation in this work, where we investigate the relationship between testing criteria (coverage and diversity metrics) and faults in DNNs.  The stronger the correlation between testing criteria and fault detection, the better. The practical implications of our results suggest that one should not rely on coverage, as currently defined, to guide DNN testing if the objective is to detect as many faults as possible. Alternatively, we show that geometric diversity has strong potential as an alternative. It outperforms existing coverage metrics in fault-revealing capability, applicability (as it is black-box) and computation time. We therefore recommend investigating its practical use in testing DNNs to guide the selection, minimization or generation of test input sets. }

\section{Threats to Validity} \label{sec:Threats}
In this section, we discuss the different threats to the validity of our study and describe how we mitigated them. \\
\noindent{\textbf{Internal threats to validity}} concern the causal relationship between the treatment and the outcome.  
We reimplemented three diversity metrics because their source code was unavailable (GD and STD) or not applicable on our datasets (NCD). Consequently, an internal threat to validity might be related to our implementations. To mitigate this threat, we carefully checked our code and its conformance to the original papers in which they were published. We also verified the correctness of our implementation of the NCD metric by comparing our results with an existing implementation\footnote{https://github.com/simonpoulding/DataGenerators.jl} that supported the calculation of the NCD score only for pairs of images or \textit{txt} files. In \textbf{\textit{RQ1}}, we tested, through a controlled experiment, the reliability of the selected diversity metrics in measuring actual data diversity and excluded the metrics that failed the test.

As we were targeting black-box diversity metrics, we needed to rely on a feature extraction model to build our feature matrix. Therefore, an internal threat to validity might be caused by a low-quality representation of inputs. To mitigate this threat, we relied on VGG16, which is one of the most-used, accurate, state-of-the-art feature extraction models. Furthermore, this DNN model has been pre-trained on the extremely large ImageNet dataset, which contains over 14 million labelled images belonging to 22,000 categories. 
Further, the configuration of the different hyperparameters in our study may induce additional internal threats to validity. We mitigate this threat in two ways: (1) for coverage metrics hyperparameters, we made use of the original papers' hyperparameter values~\cite{kim2019guiding} for each dataset and model that we used; and (2) for fault estimation hyperparameters (clustering), we tried more than 500 configurations related to HDBSCAN and UMAP for each of the datasets and models that we evaluated. To reduce potential bias, we evaluated the configurations' results by using two clustering evaluation metrics (section~\ref{Sec:Faults}) and by visualizing heatmaps.

A final internal threat to validity is related to randomness when sampling test inputs. We addressed this issue by repeating such sampling multiple times while considering different input set sizes and different datasets and models.

\noindent{\textbf{Construct threats to validity}} concern the relation between the theory and the observations. 
To study the effectiveness of a given test criterion in guiding DNN testing, we relied on a clustering-based approach to estimate detected faults in DNNs. It is a potential threat to construct validity because this estimate may not be sufficiently accurate. If that is the case, correlations with diversity and coverage might appear weaker than they actually are. Alternatively, relying on misprediction rates is, as previously discussed, misleading, because in practice, many similar mispredicted inputs typically result from the same problems in the DNN model. Accounting for numerous redundant test inputs would blur our correlation analysis, an effect we observed in our study. Further, we relied on a density-based clustering algorithm that is capable of grouping similar inputs in clusters of arbitrary shapes, as opposed to other types of clustering algorithms, such as k-means and hierarchical clustering, which assume that clusters are convex. Next, we clustered similar mispredicted inputs based on their (image) features and misprediction behaviour, thus relying on what semantically distinguishes images. Finally, we quantitatively and qualitatively assessed the obtained clusters to group test inputs with similar characteristics.  

\noindent{\textbf{Reliability threats to validity}} concern the replicability of our study results. We relied on publicly available models and datasets and provided all the materials required to replicate our study results online~\cite{ReplicationPackage}. This includes the set of all selected samples in the experiments and the different configurations we used for all the selected testing criteria. 

\noindent{\textbf{Conclusion threats to validity}} concern the relation between the treatment and the outcome. We relied on the Spearman correlation because it does not rely on  assumptions about the data set distributions or on the shapes of the relationships, except for the latter being monotonic. 

\noindent{\textbf{External threats to validity}} concern the generalizability of our study. We mitigated this threat by using \added{four} large datasets and \added{five} widely used and architecturally different DNN models. Further, in each of our experiments, we evaluated many samples and input set sizes. The selected coverage metrics may not be representative of all existing coverage criteria. However, we selected the best metrics based on their published results and our ability to reproduce their results.

\section{Related Work}
\label{Sec:RW}

\Revise{The work presented in this paper relies on concepts related to test diversity, black-box testing and model coverage in DNNs. In this section, we provide an overview of existing coverage metrics for DNN models. We also describe existing work on black-box DNN testing and studies making use of diversity to guide testing of DNNs and traditional software systems.} 

\subsection{Test Coverage Criteria for DNNs} 
Several coverage metrics have been proposed in the literature. 
The first attempt was carried out by Pei \textit{et al}.~\cite{pei2017deepxplore} who proposed the Neuron Coverage (NC) metric for test inputs, which is defined as the proportion of activated neurons (neurons whose activation value is above the defined threshold) over all neurons when all available test inputs are supplied to a DNN. However, several studies~\cite{sun2019structural,sekhon2019towards} have shown that 100\% neuron coverage is easy to achieve with a small set of inputs, and consequently is going to limit the applicability of such metric when testing DNNs.

Ma \textit{et al}.~\cite{Ma2018DeepGaugeMT} proposed DeepGauge, a set of DNN coverage metrics. They introduced K-Multisection Neuron Coverage (KMNC), Neuron Boundary Coverage (NBC) and Strong Neuron Activation Coverage (SNAC). KMNC partitions the ranges of neuron activation values into K buckets based on training inputs and counts the number of total covered buckets by a given test input set. NBC measures the ratio of corner-case regions that have been covered. Corner-case regions correspond to the activation values that are beyond the activation ranges observed during training. SNAC measures how many upper corner-cases have been covered. Upper corner-cases correspond to neuron activation values that are greater than the activation ranges observed during training. The authors showed that input tests generated by adversarial methods increase coverage in terms of their metrics. However, they did not study how these metrics relate to DNN mispredictions using natural inputs. 

Inspired by the MC/DC test coverage in traditional software testing, Sun \textit{et al}.~\cite{sun2019structural} proposed four coverage metrics that account for the causal relationship between neurons in neighbouring layers of a DNN model. These metrics were used to guide the generation of test inputs using adversarial methods to test the robustness of DNN models. 

Kim \textit{et al}.~\cite{kim2019guiding} proposed two coverage criteria called Likelihood-based Surprise Coverage (LSC) and Distance-based Surprise Coverage (DSC). These criteria are based on an analysis of how surprising test sets are given the training set. LSC uses Kernel Density Estimation (KDE)~\cite{wand1994kernel} to estimate the likelihood of seeing a test input during the training phase. DSC relies on the calculation of Euclidean distances between vectors that correspond to (1) the neurons' activation values in inputs from the test set, and (2) the neurons' activation values in inputs from the training set. They argue that an input set that covers a wide and diverse range of surprise values is preferable to test and retrain a DNN model. They show that their metrics are correlated with existing coverage criteria \cite{pei2017deepxplore,Ma2018DeepGaugeMT} when the diversity of inputs is increased. However, our study shows that there is no strong correlation between surprise adequacy coverage and diversity by using only natural inputs. We also show that there is no strong correlation between these coverage metrics and faults in DNNs. Another study conducted by Chen \textit{et al.}~\cite{chen2020deep} showed similar results with respect to DNN misprediction rates when using only natural inputs. 

Gerasimou \textit{et al}.~\cite{gerasimou2020importance} proposed the Importance-Driven Coverage (IDC) criterion to focus on the coverage of the most influential neurons in DNN predictions. 
They argue that IDC is sensitive to adversarial inputs and achieves higher values when applied to input sets that comprise diverse inputs. They also evaluated DeepGauge~\cite{kim2019guiding} and surprise adequacy coverage criteria~\cite{Ma2018DeepGaugeMT} in their experiments and observed that IDC shows a similar increase in these coverage criteria when evaluated with test sets augmented with adversarial inputs. 

\Revise{Byun et al.~\cite{byun2021black} have recently proposed Manifold Combination Coverage (MCC), a black-box coverage metric for testing DNN models based on projecting test inputs onto a manifold space using a Variational AutoEncoder (VAE). This metric relies on manifold learning~\cite{bengio2013representation} that compresses a high-dimensional input space into a lower-dimensional manifold space. It then measures the coverage of test inputs within a subset in the manifold space to assess test thoroughness~\cite{byun2021black}. The authors compared the misprediction-revealing effectiveness and retraining efficacy of MCC and state-of-the-art white-box coverage metrics. They found that the misprediction-revealing effectiveness of MCC is similar to that of the selected white-box coverage metrics. 
Further, MCC failed to outperform the white-box coverage metrics in retraining effectiveness.
Given the reported performance of MCC compared to white-box coverage metrics, we did not consider this recent work in our empirical analysis.}

Despite active research on DNN coverage, several recent articles have questioned the usefulness of coverage criteria to guide the testing of DNN models~\cite{li2019structural,dong2019there,chen2020deep}. For example, Li \textit{et al}.~\cite{li2019structural} studied a number of structural coverage criteria and discussed their limitations in fault detection capabilities in DNN models. Their experiments found no strong correlation between coverage and the number of misclassified inputs in a natural test set. Furthermore, Dong \textit{et al}.~\cite{dong2019there} found that retraining DNN models with new input sets that improve coverage does not increase the robustness of the model to adversarial attacks. 

Our work on diversity metrics is orthogonal to existing research regarding DNN test coverage. \Revise{Most of the state-of-the-art coverage metrics require full access to the internals of DNN state or training data, both of which are often not available to testers in practical contexts}. Thus, in our approach we focused on black-box diversity metrics, and aimed to provide guidance to assess test suites or select test cases for DNNs. 

State-of-the-art coverage criteria have been largely validated with artificial inputs that have been generated based on adversarial methods~\cite{pei2017deepxplore,Ma2018DeepGaugeMT,sun2019structural,kim2019guiding,gerasimou2020importance}. However, their relationship to (often unrealistic) adversarial inputs does not imply they relate to the fault detection capability of natural test input sets. Li \textit{et al}.~\cite{li2019structural} argue that adversarial inputs are distributed pervasively over the divided space defined by existing coverage criteria. On the other hand, misclassified natural inputs are distributed sparsely, making their detection difficult when using such coverage criteria~\cite{li2019structural}. Existing studies~\cite{li2019structural,chen2020deep} have failed to find a significant correlation between coverage and the number of misclassified inputs in a test set. Consequently, coverage criteria may be ineffective at guiding DNN testing to increase the fault-detection capability of natural input sets. Furthermore, existing studies~\cite{pei2017deepxplore,Ma2018DeepGaugeMT,sun2019structural,kim2019guiding,gerasimou2020importance} have used the number of mispredicted inputs to study the effectiveness of coverage criteria to support DNN testing. However, as previously discussed, simply comparing mispredictions is highly misleading because many test inputs may (and usually do) fail due to the same causes in the DNN model. To address this problem, we approximated faults (i.e. common misprediction causes) by relying on a clustering strategy and by studying the correlation between test criteria (i.e. coverage and diversity) and faults instead of misprediction rates. 
\Revise{\subsection{Black-box DNN Testing}}
\Revise{In this section we describe black-box testing approaches for DNN models because the focus of this paper is on black-box metrics and diversity metrics, and on their association with detected faults in DNN models. Feng \textit{et al.}~\cite{feng2020deepgini} proposed DeepGini, a black-box test selection approach that prioritizes test inputs along with higher uncertainty scores. Intuitively, a test input is likely to be misclassified by a DNN if the model is uncertain about the classification and outputs similar probabilities for each class~\cite{feng2020deepgini}. They found that DeepGini outperforms random and coverage-based test selection approaches to reveal misclassifications. However, this approach is only applicable to classification problems and cannot be used for regression tasks.}
\Revise{A recent study by Gao \textit{et al.}~\cite{gao2022adaptive}, developed concurrently to our work, proposed Adaptive Test Selection (ATS), a method based on uncertainty scores and distribution of the output probability vectors of test inputs in DNN models.}
\Revise{The selection is guided by a fault pattern coverage score that is computed using the output probability vectors of the DNN under test. They introduce a mapping from the output domain of the DNN model to a set of intervals in local domains to describe the fault pattern of a given test input or test set. They select test inputs that both cover different fault patterns~\cite{gao2022adaptive} and have higher uncertainty scores. Similar to DeepGini, this approach can only be used for classification problems. 
Empirical results show that ATS outperforms coverage and uncertainty-based test selection methods (including DeepGini~\cite{feng2020deepgini}) in misprediction-revealing capability. They also studied the effectiveness of the proposed approach in finding diverse mispredicted inputs. Consequently, they introduced the concept of fault types by looking at pairs of actual test input labels and labels predicted by the DNN model under test. Inputs that have different pairs of actual and predicted labels are assumed to correspond to different types of faults. However, as opposed to our work, this proposed method for counting distinct faults was not validated. We relied on a more fine-grained fault estimation approach that is based on clustering mispredicted inputs and accounts for their labels (actual and mispredicted) as well as their features. }

\Revise{Li \textit{et al.}~\cite{li2019boosting} proposed two black-box metrics called Cross Entropy-based Sampling (CES) and Confidence-based Stratified Sampling (CSS) for DNN test set minimization. These metrics are used to guide the selection of a small set of test inputs that can accurately estimate the entire testing dataset. The authors show that their approach outperforms random sampling and requires only about half the labeled test inputs to achieve the same level of accuracy as the whole testing set. The authors also report that CSS does not perform well on poorly trained DNNs because the confidence values it produces cannot be trusted~\cite{li2019boosting}.}

\Revise{These black-box testing approaches and their underlying metrics are conceptually different from our black-box diversity metrics. They are model-dependent because they rely on DNN outputs (output diversity) and uncertainty assessments. From a practical standpoint, this implies that all inputs must first be executed to be selected based on their diversity, which is a strong practical impediment in many application contexts, for example when working with large models and large databases of unlabeled inputs. Further, as mentioned above, these diversity metrics cannot be trusted when the model is not accurate. In contrast, our diversity metrics are model-independent and based on an analysis of the diversity of input features, thus requiring no model execution.
Finally, these black-box testing approaches focus on specific testing scenarios, while our study is generic and focuses on investigating fundamental and pervasive assumptions related to the relationship between testing criteria and faults in DNNs.}

\vspace{1em}        

\subsection{Diversity in Testing} 
In this section, we describe existing work that relied on diversity to test DNNs and regular software. 

\noindent\textbf{Diversity in DNN Testing.} A recent study by Langford and Cheng~\cite{langford2021enki} proposed Enki, a DNN input-generation approach based on evolutionary search~\cite{eiben2003introduction}. Their objective is to diversify image transformation types and to generate new inputs from existing ones to test and retrain DNN models. They started by evolving an archive of image transformation types that have a diverse impact on the DNN model. Given a subset of synthetic inputs generated with a certain image transformation type, the diversity of the impact was evaluated against three elements: (1) the F1-score of the DNN model when applied on the subset; (2) the neuron coverage score~\cite{pei2017deepxplore}; and (3) the neuron's activation pattern~\cite{langford2021enki}. After building the final Enki archive containing the most diverse image transformation types, they (1) tested the DNN models using synthetic inputs generated with the identified image transformation types, and (2) studied the accuracy of the DNNs by retraining them with such synthetic training data. They also compared their results with random input generation and DeepTest~\cite{DeepTestRef}. They concluded that Enki outperformed these two input generation approaches, and  reported that testing DNNs with their generated data led to the lowest DNN model accuracy. They also reported that retraining DNNs with their generated data increased the accuracy of DNN models.  

What differentiates our work from Enki is that Enki provides a search-based approach to diversify image transformation types. Its goal is to minimize the model accuracy and then use it to guide the generation of synthetic inputs to test and retrain DNNs. In contrast, our approach investigated ways to measure diversity in natural test input sets and compared the best diversity metric with state-of-the-art coverage criteria to guide DNN testing to maximize fault detection. Such diversity metrics can then be used for multiple purposes such as test suite assessment and guidance for selection, minimization and generation. Our focus on faults, as opposed to accuracy, aimed to find test inputs whose mispredictions resulted from distinct root causes. For practical reasons, as already discussed and as opposed to Enki, we intentionally devised an approach that is black-box and did not rely on internal information about the model or its training set.    

\noindent\textbf{Diversity in Software Testing.} Input and output diversity has been investigated to support different aspects related to traditional software testing. Since executing similar test cases tends to exercise similar parts of the source code, this is likely to lead to revealing the same faults in the system under test. Therefore, relying on diverse test cases should increase the exploration of the fault space and thus increase fault detection rates~\cite{cartaxo2011use,hemmati2015prioritizing,de2018improving}.

Feldt \textit{et al.}~\cite{feldt2016test} proposed Test Set Diameter (TDSm), a diversity-based test case selection strategy. The approach uses the NCD metric to measure the diversity of test inputs. They applied their approach on four systems and concluded that diverse test input sets increase code coverage. Finally, they show that test sets with larger NCD scores exhibit better fault-detection capabilities.

Hemmati \textit{et al.}~\cite{hemmati2013achieving} conducted an empirical study on similarity-based test selection techniques for test cases generated from state machine models. They studied and compared over 320 variants that relied on different similarity metrics and selection algorithms. Based on their experiments, they found that the best test-selection technique used the Gower-Legendre similarity function~\cite{xu2005survey} and applied a (1+1) Evolutionary Algorithm~\cite{droste2002analysis} to select tests with minimum pairwise similarity and thus maximized the diversity of the selected test cases. They further showed that such similarity-based test selection configuration outperformed random selection and coverage-based techniques in fault detection rates and computational cost.

Biagiola \textit{et al.}~\cite{biagiola2019diversity} introduced a web test generation algorithm that produces and selects candidate test cases that are executed in the browser based on their diversity. They showed that their test generation technique achieved higher code coverage and fault detection rates when compared to state of-the-art, search-based web test generators~\cite{biagiola2017search,mesbah2011invariant}.

Our objectives in this paper are similar to these studies, but in the context of DNN testing. 
As several studies have shown the effectiveness of diversity metrics in guiding the testing of software systems, we investigated its usefulness in testing DNN models. We compared the performance of existing diversity metrics with state-of-the-art DNN coverage criteria in terms of their fault detection capabilities and computational cost.

\section{Conclusion}
\label{Sec:Conclusion}

In this paper, we studied the effectiveness of input diversity metrics in guiding the testing of DNN models. We focused on DNN models using images as inputs, as they are  common in many systems. Our motivation is to provide a black-box mechanism that does not rely on DNN internal information or training data to assess test sets. This requirement aims to make our approach more applicable in the many practical contexts where such information is not (easily) accessible. \Revise{We also do not rely on output diversity because this approach would require executing the model with all inputs and would be affected by poor model accuracy.} Further, we compare the results achieved by white-box coverage criteria defined for DNNs with those achieved by black-box input diversity. 

To this end, we selected and adapted three input diversity metrics and, by means of a controlled experiment, evaluated their capability to measure actual input diversity. We selected the best metrics and analyzed their association with fault detection in DNNs using \added{four datasets and five DNN models}. 
Because simply comparing mispredictions is highly misleading, as many test inputs fail for the same reasons, we relied on a clustering-based approach to group similar mispredicted inputs and thus estimated faults based on the number of such clusters. We further selected the best state-of-the-art coverage criteria based on published results and our ability to reproduce such results. We studied the associations of the selected coverage criteria to both diversity and fault detection.

Based on our experiments, we found that the best diversity metric is geometric diversity and that, though there is still room for improvement (e.g. fault estimation in DNNs, investigating other diversity metrics and feature extraction models), it is a far better surrogate metric than the investigated coverage criteria in terms of their relationship to fault detection. This metric outperforms these coverage criteria in correlation to detected faults and computational time. We therefore recommend investigating the use of geometric diversity as a black-box metric to guide the testing of DNN models using images as inputs. We aim to extend our work by studying the application of input diversity in supporting different testing applications such as test set selection, minimization and generation. We also intend to investigate alternatives for DNN fault estimation.

\section*{Acknowledgements}

We are grateful to Kim \textit{et al.}~\cite{kim2019guiding} for their help and support to replicate the surprise adequacy coverage results.  
This work was supported by a research grant from General Motors as well as the Canada Research Chair and Discovery Grant programs of the Natural Sciences and Engineering Research Council of Canada (NSERC).


\bibliographystyle{IEEEtran}
\bibliography{main.bib} 

\newpage

\appendices
\section{Correlation Results Between Geometric Diversity and Coverage Metrics}
\label{Appendix:Corr-div_coverage}
Note that the grey boxes in all the following tables refer to statistically significant correlations (p-value $<=0.05)$

\begin{table}[ht]
\resizebox{0.489\textwidth}{!}{
\tiny
\color{black} 
\begin{tabular}{|c|c|c|c||c|c|}
\hline
\hline
Dataset                    & Model                              & Metric               & Test Set Size & Spearman & P-value \\ \hline \hline
\multirow{35}{*}{Cifar-10} & \multirow{35}{*}{\rotatebox{90}{12-layer ConvNet}} & \multirow{4}{*}{GD \& LSC} & 100  & 13\%           & 0.45          \\ \cline{4-6} 
                          &                                    &                      & 200  & -2\%          & 0.92           \\ \cline{4-6} 
                          &                                    &                      & 300  & 1\%         & 0.97         \\ \cline{4-6} 
                          &                                    &                      & 400  & -10\%           & 0.58          \\ \cline{4-6} &                                    &                      & 1000  & -12\% & 0.53 \\
                          \cline{3-6}
                          &                                    &
         
                          \multirow{4}{*}{GD \& DSC} & 100  & -20\%   &     0.27       \\ \cline{4-6} 
                          &                                    &                      & 200  & -11\% &     0.55     \\ \cline{4-6} 
                          &                                    &                      & 300  & 11\%           & 0.49          \\ \cline{4-6} 
                          &                                    &                      & 400  & -17\%          & 0.36     \\\cline{4-6} &                                    &                      & 1000  & -7\% & 0.73   \\
                          \cline{3-6}
                          & 
                          & \multirow{4}{*}{GD \& NC} & 100  & -30\%           &   0.09         \\ \cline{4-6} 
                          &                                    &                      & 200  & 12\% &     0.49       \\ \cline{4-6} 
                          &                                    &                      & 300  & 10\%           &  0.56        \\ \cline{4-6} 
                          &                                    &                      & 400  & 2\%          &  0.90          \\\cline{4-6} &                                    &                      & 1000  & 2\% & 0.91  \\
                          \cline{3-6} 
                          &                                    & \multirow{4}{*}{GD \& KMNC} & 100  & \cellcolor{gray!25}\textbf{44\%}     &     0.01       \\ \cline{4-6} 
                          &                                    &                      & 200  & 15\%           &   0.40        \\ \cline{4-6} 
                          &                                    &                      & 300  & 15\%           &     0.37      \\ \cline{4-6} 
                          &                                    &                      & 400  & 9\%      &   0.63  \\\cline{4-6} &                                    &                      & 1000  & 12\%  & 0.54  \\
                          \cline{3-6}
                          &                                    & \multirow{4}{*}{GD \& NBC} & 100  & 31\%  &   0.08          \\ \cline{4-6} 
                          &                                    &                      & 200  & 6\% &    0.75       \\ \cline{4-6} 
                          &                                    &                      & 300  & 17\%           &    0.31      \\ \cline{4-6} 
                          &                                    &                      & 400  & -10\%          &       0.62    \\\cline{4-6} &                                    &                      & 1000  & -28\% & 0.14  \\
                          \cline{3-6}
                          &                                    & \multirow{4}{*}{GD \& TKNC} & 100  & -19\%           & 0.30           \\ \cline{4-6} 
                          &                                    &                      & 200  & 17\% &       0.33     \\ \cline{4-6} 
                          &                                    &                      & 300  & -4\%           &    0.83      \\ \cline{4-6} 
                          &                                    &        
                          
                          & 400  & 7\%          &   0.69        \\\cline{4-6}
                          &                                    &                      & 1000  & 7\% &  0.71 \\
                          \cline{3-6}
                          &                                    & \multirow{4}{*}{GD \& SNAC} & 100  & \cellcolor{gray!25}\textbf{34\%}         &      0.05      \\ \cline{4-6} 
                          &                                    &                      & 200  & 6\% &      0.75      \\ \cline{4-6} 
                          &                                    &                      & 300  & 13\%           &   0.44        \\ \cline{4-6} 
                          &                                    &                      & 400  & -13\%         &       0.47     \\\cline{4-6} &                                    &                      & 1000  & -28\% &  0.14 \\

                          \cline{3-6} \hline
\end{tabular}}
\centering
\caption{Correlation results between GD and coverage metrics using 12-layer ConvNet and Cifar-10. }
\label{Tab:Correlation6-cov-div}
\end{table}

\begin{table}[hb]
\resizebox{0.489\textwidth}{!}{
\tiny
\color{black} 
\begin{tabular}{|c|c|c|c||c|c|}
\hline
\hline
Dataset                    & Model                              & Metric               & Test Set Size & Spearman & P-value \\ \hline \hline

\multirow{35}{*}{MNIST}    & \multirow{35}{*}{\rotatebox{90}{LeNet-1}}           & \multirow{4}{*}{GD \& LSC}&	100	&	-6\%	&	0.78	 \\ \cline{4-6} 
&    &    &	200	&	-12\%	&	0.57	 \\ \cline{4-6} 
&    &    &	300	&	-6\%	&	0.78	 \\ \cline{4-6} 
&    &    &	400	&	15\%	&	0.30	 \\ \cline{4-6} 
&    &    &	1000	&	-4\%	&	0.80	 \\  \cline{3-6}
&    &  \multirow{4}{*}{GD \& DSC} &	100	&	-27	\%	&	0.17	 \\ \cline{4-6} 
&    &    &	200	&	-10\%	&	0.64	 \\ \cline{4-6} 
&    &    &	300	&	-27\%	&	0.17	 \\ \cline{4-6} 
&    &    &	400	&	25\%	&	0.09	 \\ \cline{4-6} 
&    &    &	1000	&	-14\%	&	0.41	 \\ \cline{3-6} 
&    &  \multirow{4}{*}{GD \& NC} &	100	&	18\%	&	0.38	 \\ \cline{4-6} 
&    &    &	200	&	-15\%	&	0.47	 \\ \cline{4-6} 
&    &    &	300	&	3\%	&	0.89	 \\ \cline{4-6} 
&    &    &	400	&	-3\%	&	0.85	 \\ \cline{4-6} 
&    &    &	1000	&	-30\%	&	0.08	 \\ \cline{3-6} 
&    &  \multirow{4}{*}{GD \& KMNC} &	100	&	-15\%	&	0.47	 \\ \cline{4-6} 
&    &    &	200	&	-16\%	&	0.45	 \\ \cline{4-6} 
&    &    &	300	&	8\%	&	0.68	 \\ \cline{4-6} 
&    &    &	400	&	28\%	&	0.06	 \\ \cline{4-6} 
&    &    &	1000	&	6\%	&	0.71	  \\ \cline{3-6} 
&    &  \multirow{4}{*}{GD \& NBC} &	100	&	6\%	&	0.78	 \\ \cline{4-6} 
&    &    &	200	&	6\%	&	0.79	 \\ \cline{4-6} 
&    &    &	300	&	-9\%	&	0.65	 \\ \cline{4-6} 
&    &    &	400	&	15\%	&	0.30	 \\ \cline{4-6} 
&    &    &	1000	&	-15\%	&	0.39	 \\ \cline{3-6} 
&    &  \multirow{4}{*}{GD \& TKNC} &	100	&	7\%	&	0.73	 \\ \cline{4-6} 
&    &    &	200	&	11\%	&	0.61	 \\ \cline{4-6} 
&    &    &	300	&	29\%	&	0.14	 \\ \cline{4-6} 
&    &    &	400	&	21\%	&	0.16	 \\ \cline{4-6} 
&    &    &	1000	&	21\%	&	0.23	 \\ \cline{3-6} 
&    &  \multirow{4}{*}{GD \& SNAC} &	100	&	-6\%	&	0.78	 \\ \cline{4-6} 
&    &    &	200	&	7\%	&	0.75	 \\ \cline{4-6} 
&    &    &	300	&	-14\%	&	0.47	 \\ \cline{4-6} 
&    &    &	400	&	17\%	&	0.26	 \\ \cline{4-6} 
&    &    &	1000	&	-13\%	&	0.45	 \\ \cline{3-6} 
 \hline 
                          
\end{tabular}}
\centering
\caption{Correlation results between GD and coverage metrics using LeNet-1 and MNIST.}
\label{Tab:Correlation5-cov-div}
\end{table}


\begin{table}[hb]
\resizebox{0.489\textwidth}{!}{
\tiny
\color{black} 
\begin{tabular}{|c|c|c|c||c|c|}
\hline
\hline
Dataset                    & Model                              & Metric               & Test Set Size & Spearman & P-value \\ \hline \hline

\multirow{35}{*}{MNIST}    & \multirow{35}{*}{\rotatebox{90}{LeNet-5}}           & \multirow{4}{*}{GD \& LSC}&	100	&	3\% &	0.88	 \\ \cline{4-6} 
&    &    &	200	&	25\% &	0.17	 \\ \cline{4-6} 
&    &    &	300	&	26\% &	0.26	 \\ \cline{4-6} 
&    &    &	400	&	-22\% &	0.23	 \\ \cline{4-6} 
&    &    &	1000	&	-7\% &	0.74	 \\  \cline{3-6}
&    &  \multirow{4}{*}{GD \& DSC} &	100	&	13\% &	0.52	 \\ \cline{4-6} 
&    &    &	200	&	-10\% &	0.59	 \\ \cline{4-6} 
&    &    &	300	&	-22\% &	0.34	 \\ \cline{4-6} 
&    &    &	400	&	5\% &	0.81	 \\ \cline{4-6} 
&    &    &	1000	&	-1\% &	0.95	 \\ \cline{3-6} 
&    &  \multirow{4}{*}{GD \& NC} &	100	&	-22\% &	0.28	 \\ \cline{4-6} 
&    &    &	200	&	-11\% &	0.57	 \\ \cline{4-6} 
&    &    &	300	&	-4\% &	0.86	 \\ \cline{4-6} 
&    &    &	400	&	-34\% &	0.06	 \\ \cline{4-6} 
&    &    &	1000	&	\cellcolor{gray!25}\textbf{39\%} &	0.05	 \\ \cline{3-6} 
&    &  \multirow{4}{*}{GD \& KMNC} &	100	&	-1\% &	0.94	 \\ \cline{4-6} 
&    &    &	200	&	20\% &	0.29	 \\ \cline{4-6} 
&    &    &	300	&	4\% &	0.88	 \\ \cline{4-6} 
&    &    &	400	&	5\% &	0.79	 \\ \cline{4-6} 
&    &    &	1000	&	-31\% &	0.11	  \\ \cline{3-6} 
&    &  \multirow{4}{*}{GD \& NBC} &	100	&	-27\% &	0.17	 \\ \cline{4-6} 
&    &    &	200	&	-1\% &	0.95	 \\ \cline{4-6} 
&    &    &	300	&	-4\% &	0.86	 \\ \cline{4-6} 
&    &    &	400	&	-14\% &	0.46	 \\ \cline{4-6} 
&    &    &	1000	&	\cellcolor{gray!25}\textbf{48\%} &	0.01	 \\ \cline{3-6} 
&    &  \multirow{4}{*}{GD \& TKNC} &	100	&	4	\% &	0.84	 \\ \cline{4-6} 
&    &    &	200	&	24\% &	0.18	 \\ \cline{4-6} 
&    &    &	300	&	-17\% &	0.46	 \\ \cline{4-6} 
&    &    &	400	&	-2\% &	0.90	 \\ \cline{4-6} 
&    &    &	1000	&	3\% &	0.90	 \\ \cline{3-6} 
&    &  \multirow{4}{*}{GD \& SNAC} &	100	&	-18\% &	0.38	 \\ \cline{4-6} 
&    &    &	200	&	-1\% &	0.95	 \\ \cline{4-6} 
&    &    &	300	&	-5\% &	0.81	 \\ \cline{4-6} 
&    &    &	400	&	-21\% &	0.25	 \\ \cline{4-6} 
&    &    &	1000	&	\cellcolor{gray!25}\textbf{41\%} &	0.03	 \\ \cline{3-6} 

 \hline 
                          
\end{tabular}}
\centering
\caption{Correlation results between GD and coverage metrics using LeNet-5 and MNIST.}
\label{Tab:Correlation-4-cov-div}
\end{table}

\begin{table}[h]
\resizebox{0.489\textwidth}{!}{
\tiny
\color{black} 
\begin{tabular}{|c|c|c|c||c|c|}
\hline
\hline
Dataset                    & Model                              & Metric               & Test Set Size & Spearman & P-value \\ \hline \hline                
\multirow{35}{*}{Fashion-MNIST}  & \multirow{35}{*}{\rotatebox{90}{LeNet-4}}  & \multirow{4}{*}{GD \& LSC} &	100	&	0.2\%	& 0.99	 \\ \cline{4-6} 
&    &    &	200	&	28\%	&	0.17	 \\ \cline{4-6} 
&    &    &	300	&	19\%	&	0.29	 \\ \cline{4-6} 
&    &    &	400	&	21\%	&	0.20	 \\ \cline{4-6} 
&    &    &	1000	&	2\%	&	0.94	 \\  \cline{3-6}
&    &  \multirow{4}{*}{GD \& DSC} &	100	&	17\%	&	0.38	 \\ \cline{4-6} 
&    &    &	200	&	-7\%	&	0.73	 \\ \cline{4-6} 
&    &    &	300	&	\cellcolor{gray!25}\textbf{-35\%}	&	0.05	 \\ \cline{4-6} 
&    &    &	400	&	-2\%	&	0.90	 \\ \cline{4-6} 
&    &    &	1000	&	2\%	&	0.93	 \\ \cline{3-6} 
&    &  \multirow{4}{*}{GD \& NC} &	100	&	\cellcolor{gray!25}\textbf{-37\%}	&	0.04	 \\ \cline{4-6} 
&    &    &	200	&	0.3\%	&	0.99	 \\ \cline{4-6} 
&    &    &	300	&	-5\%	&	0.77	 \\ \cline{4-6} 
&    &    &	400	&	23\%	&	0.16	 \\ \cline{4-6} 
&    &    &	1000	&	-14\%	&	0.49	 \\ \cline{3-6} 
&    &  \multirow{4}{*}{GD \& KMNC} &	100	&	23	\%	&	0.22	 \\ \cline{4-6} 
&    &    &	200	&	30\%	&	0.14	 \\ \cline{4-6} 
&    &    &	300	&	33\%	&	0.06	 \\ \cline{4-6} 
&    &    &	400	&	\cellcolor{gray!25}\textbf{64\%}	&	0.0001	 \\ \cline{4-6} 
&    &    &	1000	&	6\%	&	0.76	  \\ \cline{3-6} 
&    &  \multirow{4}{*}{GD \& NBC} &	100	&	-25	\%	&	0.19	 \\ \cline{4-6} 
&    &    &	200	&	-31\%	&	0.13	 \\ \cline{4-6} 
&    &    &	300	&	12\%	&	0.51	 \\ \cline{4-6} 
&    &    &	400	&	-22\%	&	0.19	 \\ \cline{4-6} 
&    &    &	1000	&	-21\%	&	0.31	 \\ \cline{3-6} 
&    &  \multirow{4}{*}{GD \& TKNC} &	100	&	-1	\%	&	0.95	 \\ \cline{4-6} 
&    &    &	200	&	\cellcolor{gray!25}\textbf{41\%}	&	0.04	 \\ \cline{4-6} 
&    &    &	300	&	-3\%	&	0.85	 \\ \cline{4-6} 
&    &    &	400	&	24\%	&	0.15	 \\ \cline{4-6} 
&    &    &	1000	&	11\%	&	0.61	 \\ \cline{3-6} 
&    &  \multirow{4}{*}{GD \& SNAC} &	100	&	-25\%	&	0.19	 \\ \cline{4-6} 
&    &    &	200	&	-31\%	&	0.13	 \\ \cline{4-6} 
&    &    &	300	&	12\%	&	0.52	 \\ \cline{4-6} 
&    &    &	400	&	-21\%	&	0.20	 \\ \cline{4-6} 
&    &    &	1000	&	-19\%	&	0.35	 \\ \cline{3-6} 

 \hline

\end{tabular}}
\centering
\caption{Correlation results between GD and coverage metrics using LeNet-4 and Fashion-MNIST.}
\label{Tab:Correlation3-cov-div}
\end{table}

\begin{table}[h]
\resizebox{0.489\textwidth}{!}{
\tiny
\color{black} 
\begin{tabular}{|c|c|c|c||c|c|}
\hline
\hline
Dataset                    & Model                              & Metric               & Test Set Size & Spearman & P-value \\ \hline \hline

\multirow{35}{*}{Cifar-10}    & \multirow{35}{*}{\rotatebox{90}{ResNet-20}} & \multirow{4}{*}{GD \& LSC}&	100	&	19	\%	&	0.31	 \\ \cline{4-6} 
&    &    &	200	&	14\%	&	0.55	 \\ \cline{4-6} 
&    &    &	300	&	39\%	&	0.11	 \\ \cline{4-6} 
&    &    &	400	&	\cellcolor{gray!25}\textbf{46\%}	&	0.02	 \\ \cline{4-6} 
&    &    &	1000	&	-26\%	&	0.18	 \\  \cline{3-6}
&    &  \multirow{4}{*}{GD \& DSC} &	100	&	4	\%	&	0.85	 \\ \cline{4-6} 
&    &    &	200	&	-8\%	&	0.74	 \\ \cline{4-6} 
&    &    &	300	&	5\%	&	0.84	 \\ \cline{4-6} 
&    &    &	400	&	-10\%	&	0.65	 \\ \cline{4-6} 
&    &    &	1000	&	-3\%	&	0.86	 \\ \cline{3-6} 
&    &  \multirow{4}{*}{GD \& NC} &	100	&	5	\%	&	0.81	 \\ \cline{4-6} 
&    &    &	200	&	26\%	&	0.25	 \\ \cline{4-6} 
&    &    &	300	&	-2\%	&	0.94	 \\ \cline{4-6} 
&    &    &	400	&	12\%	&	0.58	 \\ \cline{4-6} 
&    &    &	1000	&	\cellcolor{gray!25}\textbf{-38\%}	&	0.04	 \\ \cline{3-6} 
&    &  \multirow{4}{*}{GD \& KMNC} &	100	&	19	\%	&	0.3	 \\ \cline{4-6} 
&    &    &	200	&	39\%	&	0.08	 \\ \cline{4-6} 
&    &    &	300	&	17\%	&	0.51	 \\ \cline{4-6} 
&    &    &	400	&	\cellcolor{gray!25}\textbf{41\%}	&	0.05	 \\ \cline{4-6} 
&    &    &	1000	&	28\%	&	0.15	  \\ \cline{3-6} 
&    &  \multirow{4}{*}{GD \& NBC} &	100	&	24	\%	&	0.20	 \\ \cline{4-6} 
&    &    &	200	&	12\%	&	0.60	 \\ \cline{4-6} 
&    &    &	300	&	-4\%	&	0.87	 \\ \cline{4-6} 
&    &    &	400	&	-9\%	&	0.67	 \\ \cline{4-6} 
&    &    &	1000	&	-18\%	&	0.35	 \\ \cline{3-6} 
&    &  \multirow{4}{*}{GD \& TKNC} &	100	&	2	\%	&	0.9	 \\ \cline{4-6} 
&    &    &	200	&	18\%	&	0.43	 \\ \cline{4-6} 
&    &    &	300	&	10\%	&	0.70	 \\ \cline{4-6} 
&    &    &	400	&	\cellcolor{gray!25}\textbf{58\%}	&	0.003	 \\ \cline{4-6} 
&    &    &	1000	&	-7\%	&	0.73	 \\ \cline{3-6} 
&    &  \multirow{4}{*}{GD \& SNAC} &	100	&	17	\%	&	0.35	 \\ \cline{4-6} 
&    &    &	200	&	11	\%	&	0.62	 \\ \cline{4-6} 
&    &    &	300	&	4	\%	&	0.89	 \\ \cline{4-6} 
&    &    &	400	&	-3	\%	&	0.90	 \\ \cline{4-6} 
&    &    &	1000	&	-14	\%	&	0.47	 \\ \cline{3-6} 
 \hline

\end{tabular}}
\centering
\caption{Correlation results between GD and coverage metrics using ResNet-20 and Cifar-10.}
\label{Tab:Correlation2-cov-div}
\end{table}

\begin{table}[t]
\resizebox{0.489\textwidth}{!}{
\tiny
\color{black} 
\begin{tabular}{|c|c|c|c||c|c|}
\hline
\hline
Dataset     & Model                            & Metric               & Test Set Size & Spearman & P-value \\ \hline \hline

\multirow{35}{*}{SVHN}    & \multirow{35}{*}{\rotatebox{90}{LeNet-5}}           & \multirow{4}{*}{GD \& LSC}&	100	&	15\% &	0.37	 \\ \cline{4-6} 
&    &    &	200	&	27\% &	0.09	 \\ \cline{4-6} 
&    &    &	300	&	2\% &	0.91	 \\ \cline{4-6} 
&    &    &	400	&	15\% &	0.46	 \\ \cline{4-6} 
&    &    &	1000	&	34\% &	0.06	 \\  \cline{3-6}
&    &  \multirow{4}{*}{GD \& DSC} &	100	&	-16	\% &	0.33	 \\ \cline{4-6} 
&    &    &	200	&	9\% &	0.58	 \\ \cline{4-6} 
&    &    &	300	&	19\% &	0.28	 \\ \cline{4-6} 
&    &    &	400	&	14\% &	0.47	 \\ \cline{4-6} 
&    &    &	1000	&	-14\% &	0.45	 \\ \cline{3-6} 
&    &  \multirow{4}{*}{GD \& NC} &	100	&	-8\% &	0.61	 \\ \cline{4-6} 
&    &    &	200	&	4\% &	0.79	 \\ \cline{4-6} 
&    &    &	300	&	30\% &	0.08	 \\ \cline{4-6} 
&    &    &	400	&	-28\% &	0.14	 \\ \cline{4-6} 
&    &    &	1000	&	-12\% &	0.52	 \\ \cline{3-6} 
&    &  \multirow{4}{*}{GD \& KMNC} &	100	&	\cellcolor{gray!25}\textbf{44\%} &	0.001	 \\ \cline{4-6} 
&    &    &	200	&	\cellcolor{gray!25}\textbf{66\%} &	0.002	 \\ \cline{4-6} 
&    &    &	300	&	23\% &	0.19	 \\ \cline{4-6} 
&    &    &	400	&	-2\% &	0.9	 \\ \cline{4-6} 
&    &    &	1000	&	\cellcolor{gray!25}\textbf{46\%} &	0.01	  \\ \cline{3-6} 
&    &  \multirow{4}{*}{GD \& NBC} &	100	&	9\% &	0.58	 \\ \cline{4-6} 
&    &    &	200	&	5\% &	0.77	 \\ \cline{4-6} 
&    &    &	300	&	-19\% &	0.27	 \\ \cline{4-6} 
&    &    &	400	&	26\% &	0.18	 \\ \cline{4-6} 
&    &    &	1000	&	28\% &	0.12	 \\ \cline{3-6} 
&    &  \multirow{4}{*}{GD \& TKNC} &	100	&	-3\% &	0.83	 \\ \cline{4-6} 
&    &    &	200	&	1\% &	0.95	 \\ \cline{4-6} 
&    &    &	300	&	-22\% &	0.20	 \\ \cline{4-6} 
&    &    &	400	&	-15\% &	0.43	 \\ \cline{4-6} 
&    &    &	1000	&	18\% &	0.34	 \\ \cline{3-6} 
&    &  \multirow{4}{*}{GD \& SNAC} &	100	&	9\% &	0.58	 \\ \cline{4-6} 
&    &    &	200	&	5\% &	0.77	 \\ \cline{4-6} 
&    &    &	300	&	-19\% &	0.27	 \\ \cline{4-6} 
&    &    &	400	&	26\% &	0.18	 \\ \cline{4-6} 
&    &    &	1000	&	28\% &	0.12	 \\ \cline{3-6} 
 \hline

\end{tabular}}
\centering
\caption{Correlation results between GD and coverage metrics using LeNet-5 and SVHN.}
\label{Tab:Correlation1-cov-div}
\end{table}

\cleardoublepage
\section{AUTHORS’ BIOGRAPHIES}

\begin{wrapfigure}{l}{25mm} 
\includegraphics[width=1in,height=1.25in,clip,keepaspectratio]{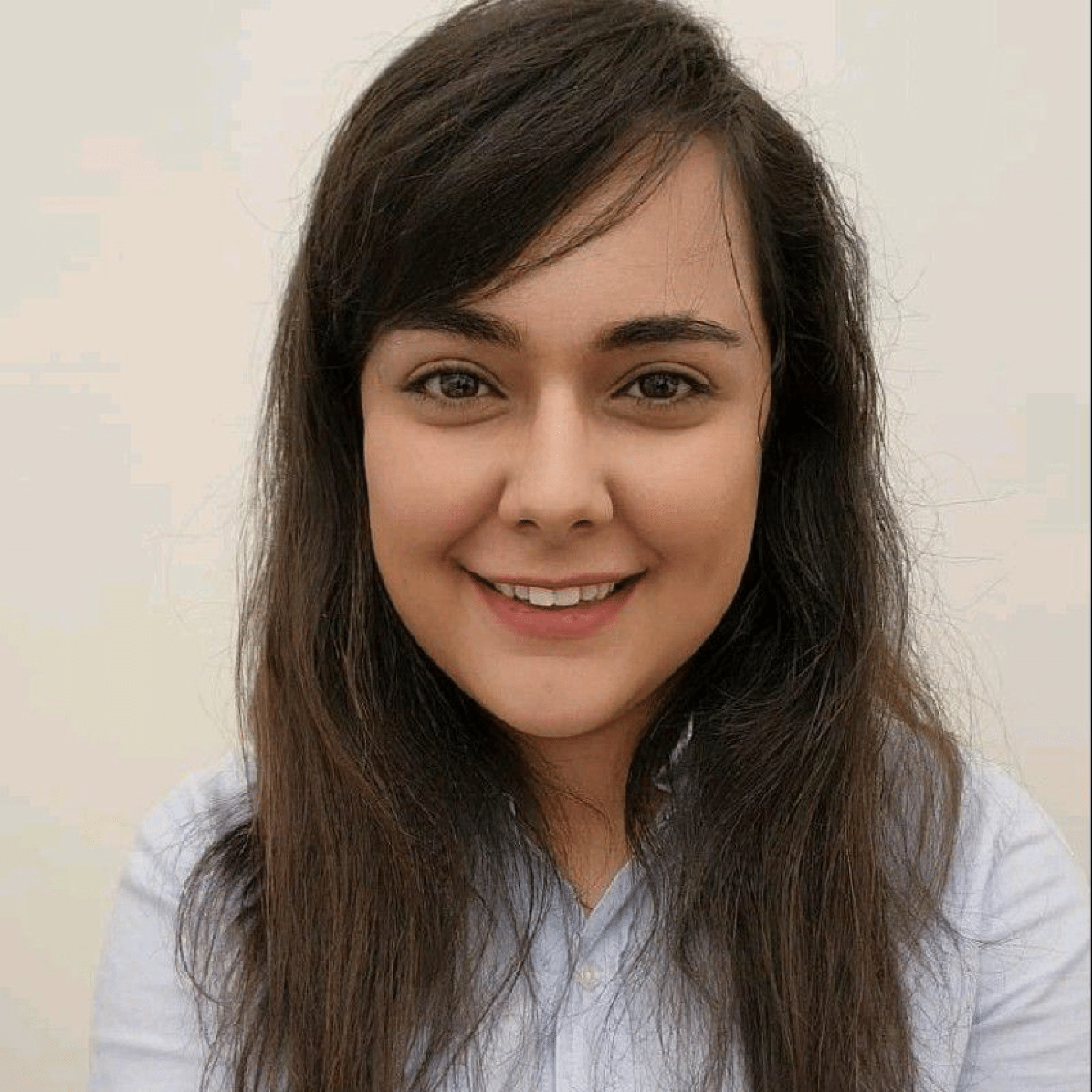}
\end{wrapfigure}\par
\textbf{Zohreh Aghebabaeyn} is a PhD student at the School of EECS at the University of Ottawa and a member of the Nanda Lab. She gained practical experience during her internship at the research and development lab in General Motors, USA.
She received several academic awards, including a PhD admission scholarship, an international doctoral scholarship from University of Ottawa, and an honourable award of admission to the master's program in computer science at Amirkabir University of Technology in Iran. She was also ranked the third-best student among all computer science students at Amirkabir University in 2017. Her research interests include testing and verification of machine learning-based systems and empirical software engineering.\\ \par

\begin{wrapfigure}{l}{25mm} 
\includegraphics[width=1in,height=1.25in,clip,keepaspectratio]{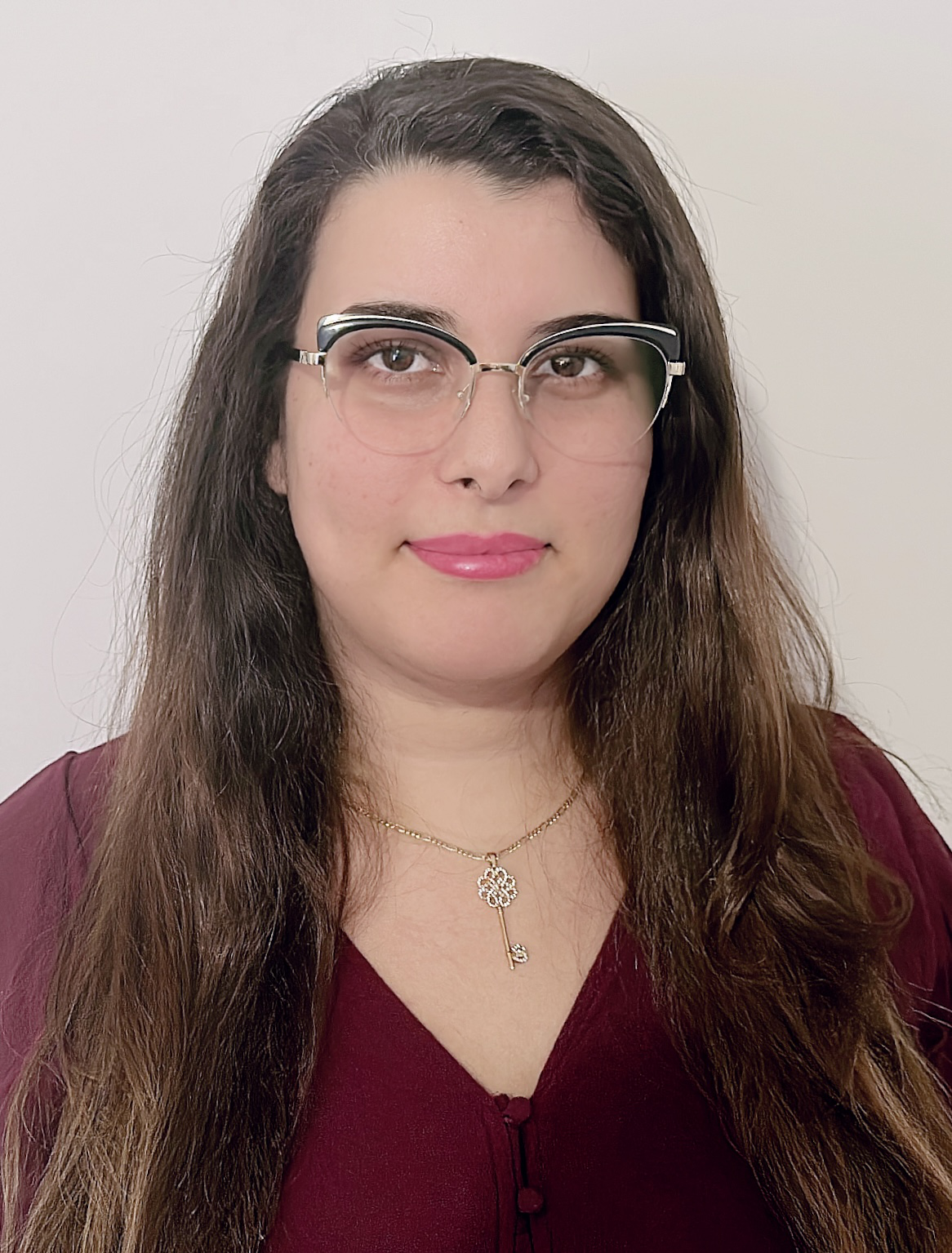}
\end{wrapfigure}\par
\textbf{Manel Abdellatif} received her PhD in computer science from Polytechnique Montréal, Canada (2021). She is a faculty member at École de Technologie Supérieure, Canada. She was a postdoctoral fellow at the School of EECS, University of Ottawa (2022). She received her Master's degree in Information Technology from École de Technologie Supérieure (2016) and earned her Bachelor's degree from École Nationale d'Ingénieurs de Tunis (2013). She served as a program committee member and a reviewer in several journals and conferences. Her research interests include testing machine learning-based systems, service computing, and empirical software engineering.  \\ \par

\begin{wrapfigure}{l}{25mm} 
\includegraphics[width=1in,height=1.25in,clip,keepaspectratio]{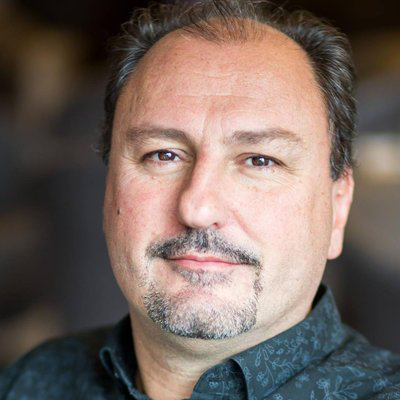}
\end{wrapfigure}\par
\textbf{Lionel Briand} is professor of software engineering and has shared appointments between (1) School of Electrical Engineering and Computer Science, University of Ottawa, Canada and (2) The SnT centre for Security, Reliability, and Trust, University of Luxembourg. He is the head of the SVV department at the SnT Centre and a Canada Research Chair in Intelligent Software Dependability and Compliance (Tier 1). He has conducted applied research in collaboration with industry for more than 25 years, including projects in the automotive, aerospace, manufacturing, financial, and energy domains. He is a fellow of the IEEE and ACM. He was also granted the IEEE Computer Society Harlan Mills award (2012), the IEEE Reliability Society Engineer-of-the-year award (2013), and the ACM SIGSOFT Outstanding Research Award (2022) for his work on software testing and verification. More details can be found on: http://www.lbriand.info. \\ \par

\begin{wrapfigure}{l}{25mm} 
\includegraphics[width=1in,height=1.25in,clip,keepaspectratio]{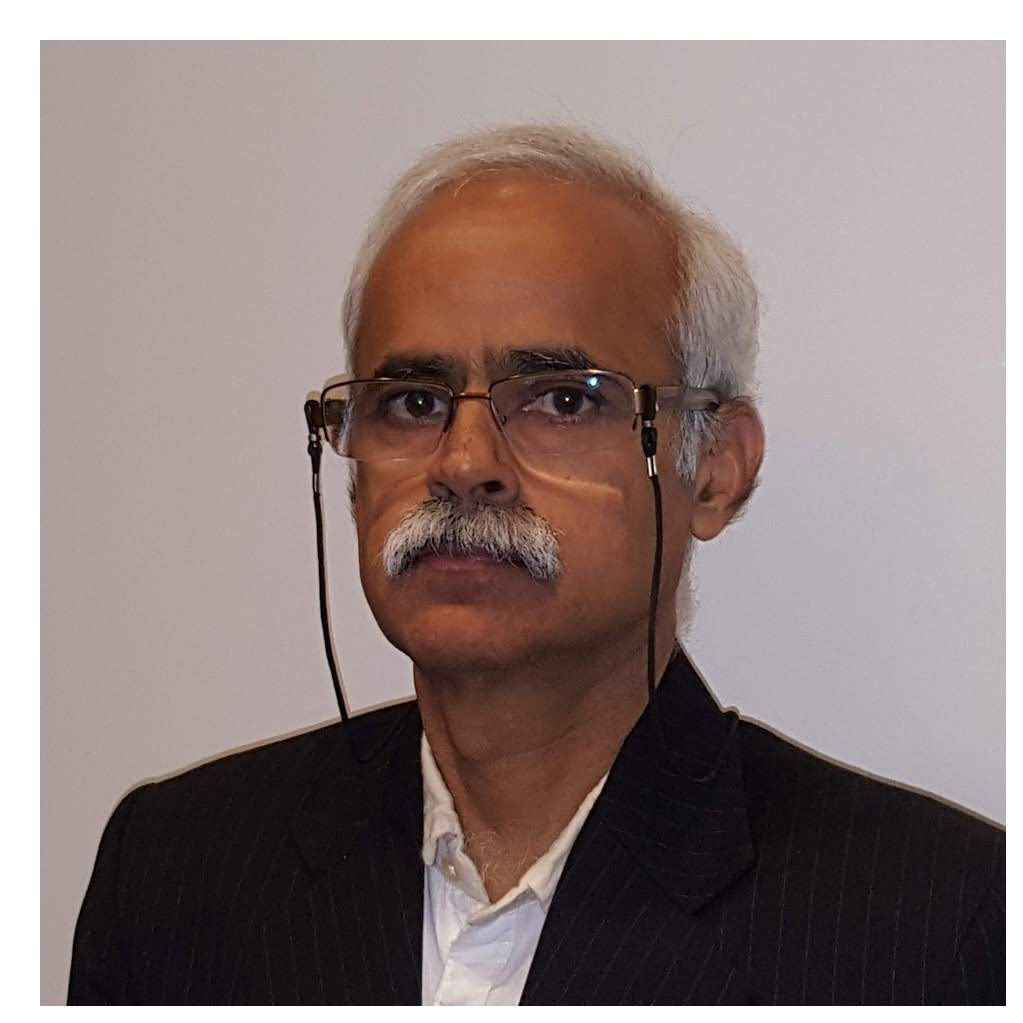}
\end{wrapfigure}\par
\textbf{Ramesh S}, Senior Technical Fellow,  has been with General Motors Research and Development (R\&D) for more than 15 years conducting and leading advanced research projects in the areas of model-based development of embedded systems and software, rigorous verification and validation and more recently AI/ML based systems. Prior to joining GM R\&D, he was a full professor in the department of Computer Science and Engineering at Indian Institute of Technology Bombay, India where he co-founded a Centre for Formal Design and Verification of Software. He has published more than 125 research papers in International Journals and Conferences and author many patents in the areas of modeling, analysis and verification of embedded systems and software. He has been on the program committees of several international research conferences and on the editorial boards of journals. He is leading an USCAR committee and serving as an expert in ISO and SAE committees for developing guidelines for AI/ML based systems. \\ \par

\begin{wrapfigure}{l}{25mm} 
\includegraphics[width=1in,height=1.25in,clip,keepaspectratio]{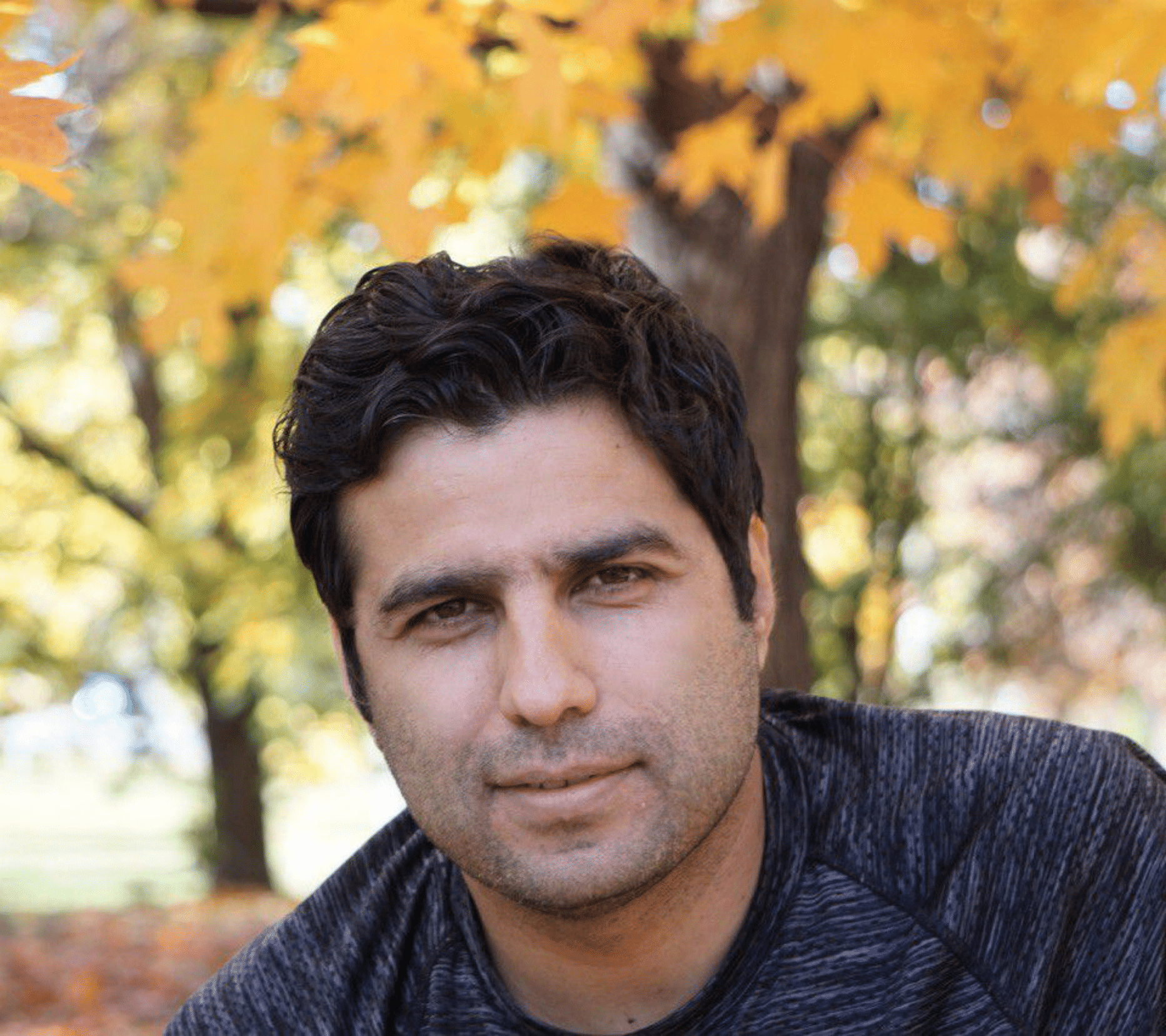}
\end{wrapfigure}\par
\textbf{Mojtaba Bagherzadeh} is a highly experienced Software Engineer with a proven track record of success in both industry and academia. He currently works as a software engineer at Cisco Systems and has previously worked as a software developer at IBM and a startup company. He has contributed to this research during his tenure as a postdoctoral researcher at the University of Ottawa. He obtained his PhD in Computer Science from Queen’s University, Canada, in 2019. His research interests include testing and debugging of machine learning-based systems, model-driven engineering, software testing, and empirical software engineering.\par

\end{document}